\documentclass[superscriptaddress, twocolumn, pre, longbibliography]{revtex4-2}
\usepackage{libertine}
\usepackage{color, amsfonts, amsmath, graphicx, tikz, amssymb, mathtools}
\usepackage{amsthm}
\usepackage[libertine, cmintegrals, bigdelims, vvarbb]{newtxmath}
\usepackage{appendix}
\usepackage{comment}
\usepackage{upgreek}
\definecolor{linkcolor}{rgb}{0,0,0.6}

\definecolor{lgreen} {RGB}{180,210,100}
\definecolor{dblue}  {RGB}{20,66,129}
\definecolor{jblue}  {RGB}{20,50,100}
\definecolor{nblue}  {RGB}{0,120,200}
\definecolor{dgreen} {RGB}{78,138,21}
\definecolor{ngreen} {RGB}{98,158,31}
\definecolor{lred}   {RGB}{220,0,0}
\definecolor{nred}   {RGB}{224,0,0}
\usepackage[pdftex,colorlinks=true,
	pdfstartview = FitV,
	linkcolor    = red,
	citecolor    = blue,
	urlcolor     = blue,	
	hyperindex   = true,
	hyperfigures = false]{hyperref}
\usepackage{cleveref}

\usepackage{array}
\newcolumntype{P}[1]{>{\centering\arraybackslash}p{#1}}

\patchcmd{\subsubsection}{\itshape}{\bfseries}{}{}
\usepackage{orcidlink}
\usepackage{multirow}
\usepackage{natbib}
\usepackage{oplotsymbl}

\begin{document}

\title{Stochastic Thermodynamics of Non-reciprocally Interacting Particles and Fields}
\author{Atul Tanaji Mohite\,\orcidlink{0009-0004-0059-1127}}
\email{atul.mohite@uni-saarland.de}
\affiliation{Department of Theoretical Physics and Center for Biophysics, Saarland University, Saarbrücken, Germany}

\author{Heiko Rieger\,\orcidlink{0000-0003-0205-3678}}
\affiliation{Department of Theoretical Physics and Center for Biophysics, Saarland University, Saarbrücken, Germany}
\begin{abstract}
Nonreciprocal interactions that violate Newton's law `actio=reactio' are ubiquitous in nature and are currently intensively investigated in active matter, chemical reaction networks, population dynamics, and many other fields.  An outstanding challenge is the thermodynamically consistent formulation of the underlying stochastic dynamics that obeys local detailed balance and allows for a rigorous analysis of the stochastic thermodynamics of non-reciprocally interacting particles. Here, we present such a framework for a broad class of active systems and derive by systematic coarse-graining exact expressions for the macroscopic entropy production. Four independent contributions to the thermodynamic dissipation can be identified, among which the energy flux sustaining vorticity currents manifests the presence of non-reciprocal interactions. Then, Onsager’s non-reciprocal relations, the fluctuation-response relation, the fluctuation relation and the thermodynamic uncertainty relations for non-reciprocal systems are derived. Finally, we demonstrate that our general framework is applicable to a plethora of active matter systems and chemical reaction networks and opens new paths to understand the stochastic thermodynamics of non-reciprocally interacting many-body systems. 
\end{abstract}
\date{\today}
\maketitle
\section{Introduction}
\begin{figure*}[]
\centering
\includegraphics[width=\textwidth]{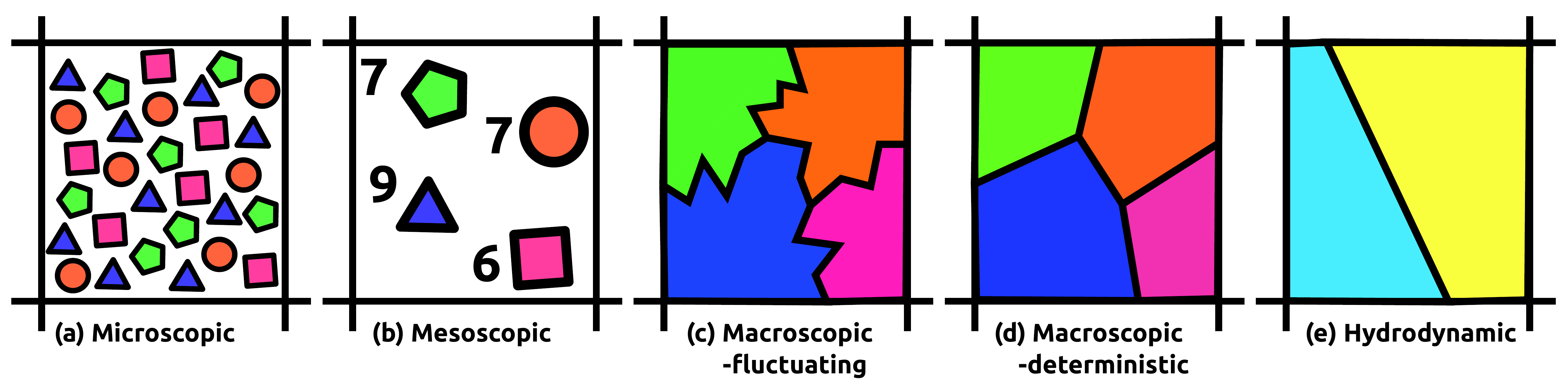}
\caption{ Scheme of various levels of coarse-graining considered in this paper, microscopic, mesoscopic, macroscopic-fluctuating, macroscopic-deterministic and hydrodynamic. (a) Microscopic particles are confined to the lattice site (denoted by the black $\#$). Four different types of particles are illustrated: triangle (blue), square (pink), pentagon (green) and, circle (orange). (b/c) The mesoscopic/macroscopic description is obtained for the fluctuating particle number/density. (d) The straight interface between the density fields represents the suppression of the fluctuation and convergence to the mean-field dynamical description. (e) Cyan and Yellow denote the relevant hydrodynamic order parameter that does not necessarily preserve the microscopic properties. }
\label{fig:CG_diagram}
\end{figure*}
The discovery of fluctuation relations has shed new light on the classical field of thermodynamics \cite{schnakenberg_1976,Bochkov_1977,Bochkov_1979,Evans_1993,Evans_1994,Gallavotti_1995,Jarzynski_1997,Jarzynski_1997_pre,Sekimoto_1997,Sekimoto_1998,Kurchan_1998,Crooks_1999,Maes_1999,Lebowitz_1999,Tasaki_2000,Crooks_2000,Maes_2003,Seifert_2005,sekimoto} and established the refined framework of stochastic thermodynamics (ST) \cite{seifert_2012,Shiraishi_2023_book}.  
Stochastic thermodynamics provides a framework to define the first and second laws of thermodynamics for a single stochastic transition \footnote{We adhere to the convention of using a stochastic transition instead of a stochastic trajectory, since a stochastic trajectory can be visualized as a collection of sequential stochastic transitions. The Markovian property of the transitions ensures this equivalence, provided that the initial probability distributions (boundary terms) are the same. We clarify this notion to avoid confusion with the trajectory-based approach used to compute the information-theoretic definition of entropy production. The information-theoretic approach is not necessarily thermodynamically consistent if each microscopic transition is not properly resolved or if the transition rates do not satisfy the local detailed balance condition.}. This is achieved by defining the energy, work, heat, and entropy production for a stochastic transition, so that these thermodynamic quantities have probability distributions. Fluctuations vanish in the thermodynamic limit, where the resulting stochastic thermodynamic quantities converge to the deterministic thermodynamic description, thereby recovering classical thermodynamics. Stochastic thermodynamics applies to small systems with inherent stochastic fluctuations in the presence of externally imposed driving of control parameters. The validity of this framework relies on a key constraint of time-scale separation, which results in the Markovian property for the transitions: the states of the stochastic process comprise the slow degrees of freedom, while the collective fast degrees of freedom form the thermal environment (bath). The Local Detailed Balance (LDB) condition establishes the equivalence between the forward and backward transition rates for the system and the thermodynamic transition cost supported by the environment \cite{Katz_1983,Katz_1984}, ensuring thermodynamic consistency for all transitions.

Newton's third law of motion states that for every action force, there is an equal and opposite reaction force. However, multiple real-world counter-examples to Newton's third law exist. Systems that violate Newton's third law are called \textit{non-reciprocal} \cite{Ivlev_2015}. In particular, a multitude of biological systems provides frontiers for studying and understanding the implications of such violations. For instance, toy models for ecological systems \cite{Goel_1971,Rieger_1989_jpa,Reichebach_2007_prl,Knebel_2013,Knebel_2015,Amir_2015,Bunin_2017,Geiger_2018,knebel_2020,Yoshida_2022,Yoshida_2021}, stochastic systems \cite{Tang_2021,Mahault_2022,Nelson_2024}, neural networks \cite{Sompolinsky_1986,Kree_1987,Crisanti_1988,Sompolinsky_1988,Sommers_1988,Rieger_1988,Rieger_1989}, mixtures of non-reciprocal particles \cite{Loos_2020}, birds with vision cones \cite{Durve_2018}, solids with odd elasticity \cite{Scheibner_2020}, particles breaking action-reaction symmetry \cite{Soto_2014,Agudo_canaljo_2019}, chiral active particles \cite{Kreienkamp_2022,Knezevic_2022}, non-reciprocal flocking models \cite{Fruchart2021,martin2023nr_aim,NR-TSAIM,Random-NR-Flocking,Multi-NR-Vicsek1,Multi-NR-Vicsek2,NR-Vicsek-Chaos}, non-reciprocal frustration \cite{Hanai_2023_nonreciprocal}, and systems described by non-reciprocal Cahn-Hilliard equations \cite{You_2020,Saha_2020,saha_2022effervescent,Frohoff_2023,Frohoff_2021,Frohoff_2021_nonreciprocal,brauns_2023_nonreciprocal}.  
Recent developments in understanding the microscopic dynamics of non-reciprocal systems have revealed more sophisticated behaviors and phases absent in reciprocal systems, such as traveling waves, oscillations, spiral waves, and effervescence \cite{Fruchart2021,martin2023nr_aim,You_2020,Saha_2020,saha_2022effervescent,Frohoff_2023,Frohoff_2021,Frohoff_2021_nonreciprocal,brauns_2023_nonreciprocal}. The non-reciprocal phases have been further elucidated by delineating the underlying topological properties \cite{Fruchart2021}.  

Active matter and non-reciprocal systems are conventionally analyzed using hydrodynamic descriptions \cite{Hohenberg_1977,Cross_1993}, which successfully recover the phases observed in the microscopic dynamics \cite{martin2023nr_aim}. Deriving hydrodynamic equations of motion requires identifying the relevant degrees of freedom, termed \textit{order parameters}, with the help of symmetry arguments: a \textit{top-down approach} \cite{Hohenberg_1977,Cross_1993}. Subsequently, fluctuations of the order parameters can be incorporated using more sophisticated exact \textit{coarse-graining} techniques \cite{Kawasaki_1994,Dean_1996}. This procedure is usually effective and reliable for understanding system dynamics, phases, and phase transitions.  

However, for microscopic systems, the hydrodynamic description fails both quantitatively and qualitatively in the small particle number regime. For example, the onset of the phase transition predicted by the hydrodynamic description differs from that predicted by the microscopic description \cite{Solon2013}. Moreover, it does not correctly predict the order of the phase transition, which is usually addressed by introducing ad-hoc noise corrections \cite{Solon2013}. The hydrodynamic description also fails to adequately describe and interpret the thermodynamic implications at the macro/meso-scale because of coarse-graining. The effective hydrodynamic description does not resolve microscopic dissipation at the hydrodynamic scale. In particular, microscopic states with different thermodynamic properties are coarse-grained into a single order parameter. In addition, the mapping between microscopic and macroscopic/mesoscopic control parameters is less transparent. We aim to construct the macrostate dynamics using a \textit{bottom-up approach}, implementing the coarse-graining procedure only to an order at which the microscopic dissipation is both qualitatively and quantitatively preserved, thereby avoiding a subsequent coarse-graining to obtain an effective description in terms of relevant order parameters.

The coarse-graining scheme is illustrated in \cref{fig:CG_diagram}. Microscopic particles with different dynamic and thermodynamic properties exist. Lumping together microstates with identical thermodynamic and physical properties forms a mesostate: the particle number. The macroscopic limit corresponds to taking the limit of a large number of particles per site. It defines the scaling from the intensive density order parameter to the extensive particle number. The densities are fluctuating stochastic fields, indicated by the rough edges. The thermodynamic limit suppresses macroscopic fluctuations, leading to reaction-diffusion dynamics for deterministic density fields. In comparison, the top-down construction of an effective hydrodynamic description fails to track microscopic thermodynamic dissipation both qualitatively and quantitatively.  

Despite significant progress in understanding the dynamics of active matter on different scales, their implications for the thermodynamics of non-reciprocal systems remain elusive. For reciprocal active matter systems, stochastic thermodynamics has been formulated for active Brownian particles \cite{Pietzonka_2018,Speck_2023,bebon_2024_mips_thermodynamics} and for flocking models \cite{atm_2024_flocking_thermo}, and fluctuation relations have been derived \cite{Mandal_2017,Dabelow_2019}. To quantify thermodynamic irreversibility, information-theoretic measures have been utilized, for instance, for the Active Ising model \cite{Tu2021} and the non-reciprocal Cahn-Hilliard equation \cite{Suchanek_2023_prl,Alston_2023,Suchanek_2023_pre_epr,Suchanek_2023_pre_pt}, which have been proven to be good statistical measures for computing lower bounds on thermodynamic entropy production \cite{Kawai_2007,Horowitz_2009,Barato_2013,Gaveau_2014,Gaveau_2014_entropy,Martinez_2019,Roldan_2021}. However, the information-theoretic definition of irreversibility does not necessarily coincide with thermodynamic entropy production \cite{Kawai_2007,Horowitz_2009,Barato_2013,Gaveau_2014,Gaveau_2014_entropy,Martinez_2019,Roldan_2021}. The information-theoretic measure can be defined without the state transitions satisfying local detailed balance (LDB), thus lacking thermodynamic consistency \cite{Tu2021,Suchanek_2023_prl,Alston_2023,Suchanek_2023_pre_epr}.  

In this paper, we combine the three main motifs mentioned so far: Stochastic Thermodynamics, Coarse-graining, and Non-reciprocal Interacting Particles. We aim to formulate a microscopic Markov jump process description for interacting non-reciprocal particles, which enables a thermodynamically consistent definition of their dynamics and thermodynamics. Subsequently, we aim to obtain a coarse-grained mesoscopic/macroscopic description that preserves microscopic thermodynamic dissipation. We ensure this by properly identifying the LDB condition at the mesoscale/macroscale. This allows us to obtain thermodynamically consistent dynamics for the time evolution of the mesostate/macrostate, ensuring that thermodynamics for non-reciprocal systems is defined at these scales. This procedure establishes the connection between microscopic and macroscopic/mesoscopic control parameters, enabling qualitative and quantitative equivalence between dynamics and thermodynamic quantities. Importantly, this extends the applicability of stochastic thermodynamics to coarse-grained many-body descriptions.  

{Throughout this work, the definition of a `microscopic' non-reciprocally interacting particle refers to the `minimal' experimentally relevant observation scale, where an `effective' non-reciprocity is observed. This non-reciprocity may arise from asymmetric interactions with the surrounding medium, complex internal dynamics, or the cognitive capabilities of living systems, corresponding to active matter models, chemical reaction networks, and population dynamics models, respectively. 
{
For example, two particles of type ``$1$" and ``$2$" of different shapes and materials submerged in a medium interact asymmetrically with the surrounding environment, which generates effective asymmetric non-reciprocal forces between them \cite{Soto_2014,Agudo_canaljo_2019}. Although composed of atoms and molecules that may satisfy Newton's third law at the smallest scales, ``$1$" and ``$2$" are treated as `effective' particles exhibiting non-reciprocal interactions due to their asymmetric interactions with the surrounding medium. We treat such an `effective non-reciprocal particle' as the fundamental microscopic object of our study. The notion of a `microscopic particle' refers to such an independent degree of freedom.
}
{
Throughout this work, the term `microscopic' means: it has not intended as `elemental' in the sense of irreducible, rather as the effective small-scale description of the system’s independent units (degree of freedom).
}
Questions such as `How does such non-reciprocity originate, and what is the energetic cost to generate it?' are \textit{not} the aim of this work. Rather, we focus on the open question: given the abundant existence of `effective' non-reciprocal particles, is it possible to formulate a thermodynamically consistent microscopic theoretical description of their dynamics? We focus on a broad class of lattice models with non-reciprocal interactions, for which we introduce a thermodynamically consistent dynamics. Then it is a well-posed question to derive the stochastic thermodynamics for it, but we intended this as a purely theoretical framework without any concrete experimental realization in mind.}  

We define a class of thermodynamically consistent lattice models for non-reciprocally interacting microscopic particles coupled to an external driving reservoir \cite{schnakenberg_1976,Lebowitz_1999,seifert_2012}. 
We achieve this by formulating the LDB condition for the dynamics of microscopic non-reciprocal particles. 
{In stochastic thermodynamics, thermodynamic consistency refers specifically to the LDB condition \cite{Katz_1983,Katz_1984,Pietzonka_2018}, the LDB condition for the microscopic system ensures thermodynamic consistency of these models; we adhere to this convention throughout this paper.} 
The LDB for the microscopic system ensures thermodynamic consistency of these models. This allows us to define the microscopic entropy production rate (EPR) for externally driven non-reciprocal particles using a Markov Jump Process (MJP) with dynamics governed by the Master equation \cite{schnakenberg_1976,Lebowitz_1999,seifert_2012}. Further, we define a coarse-grained description (mesoscopic/macroscopic) by lumping microstates with identical dynamical and thermodynamic properties \cite{atm_cg_nr_2024}. The mesoscopic/macroscopic scale is defined for particle number/density as a coarse-grained state. The coarse-grained description is chosen flexibly depending on system-specific requirements. We employ the Doi-Peliti exact coarse-graining method, which preserves microscopic discreteness in the coarse-grained description \cite{atm_cg_nr_2024}.  

Next, we formulate transitions between coarse-grained states using the LDB in a thermodynamically consistent way. This enables us to quantify and decompose coarse-grained state dissipation using stochastic thermodynamics for any non-reciprocal system. In addition, the dynamic and thermodynamic equivalence of microscopic and coarse-grained systems is ensured. The fluctuating dynamics of coarse-grained states generalize Macroscopic Fluctuation Theory (MFT) for non-reciprocal systems \cite{Bertini_2015}. We obtain a non-quadratic dissipation function, which plays an important role in the exact quantification of the EPR, particularly for far-from-equilibrium systems \cite{atm_2024_var_epr,atm_2025_var_epr_derivation}. We decompose the EPR into its linearly independent contributions through an orthogonal decomposition, consisting of four components: relaxation of the reciprocal interaction energy functional, non-reciprocal interaction, coupling to the external reservoir, and external driving work. We show that non-reciprocal phase transitions are equivalent to dynamical phase transitions and are characterized by different scaling regimes for the non-reciprocal EPR, which physically correspond to sustaining dissipative vorticity currents.  

Finally, we demonstrate non-reciprocal and non-equilibrium analogs of reciprocal systems. For instance, irreversible thermodynamic relations: Onsager's non-reciprocal relations, fluctuation-response relations, higher-order non-linear responses, and stochastic thermodynamic relations: fluctuation relations, thermodynamic kinetic uncertainty relations, and information thermodynamics, which are generalized to macroscopic systems. In the deterministic macroscopic limit, we formulate irreversible thermodynamics, Arrhenius transition state theory (LDB), and reaction-diffusion systems (dynamics) for interacting non-reciprocal density fields \cite{Turing_1952,Hohenberg_1977,Cross_1993}. We elaborate on the application of our framework using a few prototypical models. Overall, our framework extends the applicability of stochastic thermodynamics for non-reciprocal and externally driven systems across different observation scales.
\section{Microscopic Description}\label{sec:microscopic}
%
%
{
We consider a multi-particle system that comprises several particle types and resides in a discretized lattice cells (the sites of the lattice ): a regular lattice with lattice constant $l$ and continuous time. We set $l = 1$ throughout this paper unless we explicitly specify the limit $l \to 0$. Our definition of lattice cell seems analogous to the `subvolume' defined in Refs.\cite{Hattne_2005,Elf_2004} or the `voxel' defined in Ref.\cite{Gillespie_2014}. We measure the infinitesimal timestep in units of the fastest microscopic transition among the set of all possible transitions, so that both reactive and diffusive transitions defined subsequently are relevant. Because of this self-adjusting formulation, our setup incorporates reaction-dominated and diffusion-dominated regimes in a specific limit of observation timescale. This preserves the physical effects associated with all microscopic transitions and avoids any assumption associated with timescale separation between transitions \cite{atm_2024_var_epr,atm_2025_var_epr_derivation}. The number of particles of type $i$ at lattice site index $\#$ is denoted by $N_i^\#$. The configuration of the lattice is denoted by $\{N\}$, whose dimension is determined by the lattice size and the number of particle types. The volume of the $d$-dimensional lattice space is denoted by $\mathcal{V}$ and is equal to the number of lattice sites. The faster degrees of freedom form the environment, which provides the thermodynamic cost for all microstate changes of the system, consisting of the slower degrees of freedom \cite{Katz_1983,Katz_1984,seifert_2012}.
}
%
%
%
%
%
%
\subsection{Microscopic Boltzmann weight}\label{sec:microscopic_model}
\subsubsection{Theoretical formulation} 
\begin{figure*}[t!]
\centering
\includegraphics[width=0.9\textwidth]{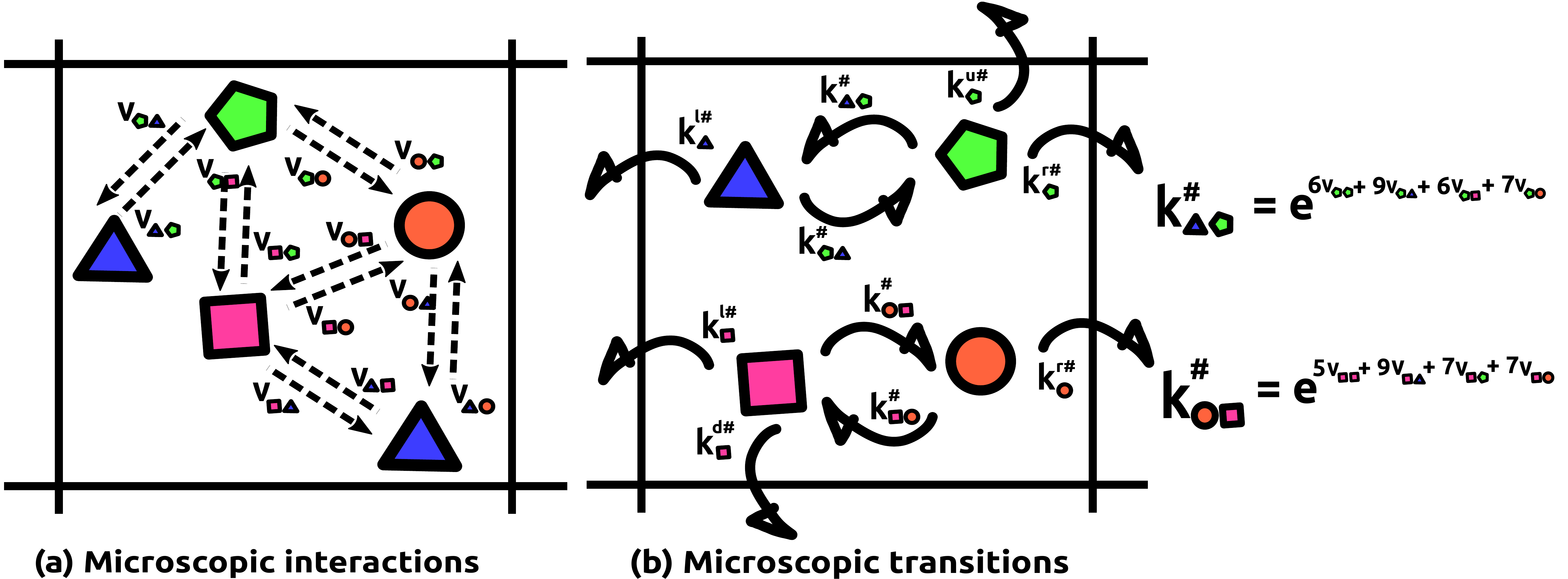}
\caption{(a) Illustration of the microscopic interactions between particle types, denoted by $v_{ij}$, confined to the lattice site $\#$, with $v_{ij} \neq v_{ji}$. (b) All possible transitions are indicated by curved harpoons. The exact expressions for two transition rates $k_{\gamma\gamma'}$ in terms of $v_{ij}$ for the microscopic lattice configuration in \cref{fig:CG_diagram}\textcolor{red}{(a)} are demonstrated. Throughout this work, superscript $\#$ and subscript $i$ denote the lattice site and particle-type index, respectively. For microscopic interaction coefficients between particle types, two indices are needed; the convention `left–right' denotes `influenced–influencer' particle. Reactive or diffusive transitions correspond to changes in particle type or lattice location, respectively. For transitions, the `right-to-left' notation denotes `initial-to-final' microstate. Latin letters $i, j$ refer to state-space properties requiring the particle type specification, e.g., $v_{ij}, v_{ij}^r, v_{ij}^{nr}, \epsilon_i^\#, \epsilon_i^r, f_i^{nr}$. Greek letters $\gamma, \gamma'$ refer to transition-space properties specifying particle type changes, e.g., $\Delta_{\gamma\gamma{'}}^{\#}, f_{\gamma\gamma{'}}^{ch}, k_{\gamma\gamma{'}}^{\#}$. This distinction clarifies the difference between state-space and transition-space quantities. }
\label{fig:particle_interactions}
\end{figure*}

The microscopic interaction energy experienced by the type $i$ particle due to the type $j$ particle is denoted by $v_{ij}$. $v_{ij} < 0$ implies that the type $i$ particle is attracted towards the type $j$ particle, while $v_{ij} > 0$ indicates a repulsive interaction. Analogously, $v_{ij}$ represents the thermodynamic cost, supported by the environment, to place a type $i$ particle in the presence of a type $j$ particle, see \cref{fig:particle_interactions}\textcolor{red}{(a)}. The violation of the actio=reactio symmetry implies $v_{ij} \neq v_{ji}$ microscopically. $v_{ij} = v_{ji}$ indicates that the particles are reciprocal and respect the actio=reactio symmetry. To separate and quantify the actio=reactio symmetry preserving and breaking components, we define the reciprocal $v_{ij}^r = (v_{ij} + v_{ji})/2$ and non-reciprocal $v_{ij}^{nr} = (v_{ij} - v_{ji})/2$ interaction energies. $v_{ij}^r$ and $v_{ij}^{nr}$ correspond to the symmetry-preserving and symmetry-breaking parts, respectively, which fixes `the orthogonal reference gauge' such that $v_{ij}^{r} = v_{ji}^r$ and $v_{ij}^{nr} = -v_{ji}^{nr}$ by construction.       

$\epsilon_i^\#$ is microscopic Boltzmann weight of a type $i$ particle at lattice site $\#$, quantifying the total thermodynamic cost of placing the particle at the site. The reciprocal and non-reciprocal components of the microscopic Boltzmann weight $\epsilon_i^\#$ of the type $i$ particle at $\#$ are denoted as $\epsilon_i^r$ and $f_i^{nr}$, respectively \footnote{ Despite the equivalence of $\epsilon_i^\#$ to a chemical potential, we avoid calling so. Rather, it appears in \cref{eq:microscopic_transition_rate} as the argument in the exponential of the transition rates for a single particle. If all particles are reciprocal, the usual physical meaning of the chemical potential is recovered. However, in principle, what we refer to as the `Boltzmann weight' throughout this work is mathematically and physically equivalent to the `chemical potential', but defined for a non-reciprocal and far-from-equilibrium system. This remark clarifies the nomenclature. },
%
%
{
such that, 
\begin{equation}\label{eq:microscopic_boltzmann_weight}
    \epsilon_i^\# = \epsilon_i^r + f_i^{nr},
\end{equation}
is decomposed into its reciprocal and non-reciprocal components,
}
\begin{equation}\label{eq:microscopic_reciprocal_interaction_energy_particle}
\begin{split}
    \epsilon_i^r = \beta \sum_{j \neq i} v_{ij}^r N_j^\# + \beta v_{ii}^r \left( N_i^\# - 1 \right),\hspace{0.5cm}
    f_{i}^{nr} = \beta \sum_{j} v_{ij}^{nr} N_j^\#.
\end{split}    
\end{equation}
We drop the lattice index $\#$ for $\epsilon_i^r$ and $f_i^{nr}$ for brevity, but it is understood that these quantities are evaluated with the appropriate lattice $\#$ and type $i$ indexes. Here, $N_i^\#-1$ accounts for the self-interaction of the particle. $\beta$ is the inverse temperature, and the Boltzmann constant is set to unity throughout this work. $\epsilon_i^r$ and $f_{i}^{nr}$ correspond to the actio=reactio symmetry preserving and breaking terms, respectively, and quantify the total reciprocal and non-reciprocal interaction energy experienced by the type $i$ particle due to all other particles. $\epsilon_i^r$ and $f_{i}^{nr}$ are local properties of each particle. For non-interacting ideal particles, $\epsilon_i^\# = 0$, since $v_{ij} = 0, \forall j$. \Cref{table:microscopic_single_particle_notation} summarizes the notation for microscopic systems.

We examine the possibility of representing the locally defined microscopic Boltzmann weight as a global property of the lattice, namely the total interaction energy $\varepsilon^{int}$. This is possible only if $v_{ij} = v_{ji}$, which is satisfied by $v_{ij}^r$ and violated by $v_{ij}^{nr}$. Hence, the reciprocal part of the microscopic Boltzmann weight can be expressed as the microscopic interaction energy functional $\varepsilon^{int}$ of the lattice,
\begin{equation}\label{eq:microscopic_energy_functional}
\begin{split}
    \varepsilon^{int} = \frac 1 2 \beta \sum_{\#, \: i} v_{ii}^{r} N_i^\# \left( N_i^\# - 1 \right) + \beta \sum_{\#, \: i \neq j} v_{ij}^{r} N_i^\# N_j^\#.
\end{split}
\end{equation}
In the absence of non-reciprocity, i.e., $v_{ij}^{nr} = 0, \forall i,j$, the system satisfies the Boltzmann distribution generated by $\varepsilon^{int}$. A detailed discussion on thermodynamics follows.

Equation \cref{eq:microscopic_energy_functional} does not necessarily represent the equilibrium interaction energy functional. Consider a decomposition $v_{ij} = v_{ij}^{eq} + v_{ij}^{neq}$, where $v_{ij}^{eq}$ is derived from the equilibrium energy functional of the form \cref{eq:microscopic_energy_functional} and $v_{ij}^{neq}$ is the remaining microscopic non-equilibrium interaction energy attributed to the violation of Newton's third law. This gives $v_{ij}^{r} = v_{ij}^{eq} + (v_{ij}^{neq} + v_{ji}^{neq})/2$ and $v_{ij}^{nr} = (v_{ij}^{neq} - v_{ji}^{neq})/2$, so that $v_{ij}^r$ and $v_{ij}^{nr}$ incorporate the symmetric and anti-symmetric parts of $v_{ij}^{neq}$, respectively. In conclusion, even though the microscopic physical interaction rules are governed by $v_{ij}, v_{ji}, v_{ij}^{eq}, v_{ij}^{neq}$, the underlying geometric and thermodynamic structure is captured by $v_{ij}^{r}, v_{ij}^{nr}$, and $|\varepsilon^{int}| \geq |\varepsilon^{eq}|$ gives a tighter bound on the conservative interaction energy functional.

Moreover, this analysis shows that the decomposition of $v_{ij}$ and $v_{ji}$ (gauge fixing) is not necessarily unique, unless `the orthogonal reference gauge' is imposed. Further, only `the orthogonal reference gauge' leads to an observation scale-invariant decomposition of actio-reactio symmetry conserving and breaking terms. Consequently, it also yields the physically correct thermodynamic dissipation of the system. The decomposition of $\epsilon_i^\#$ into $\epsilon_i^r$ and $f_i^{nr}$ is equivalent to the Helmholtz-Hodge decomposition of $\epsilon_i^\#$ into a conservative gradient force $\epsilon_i^r$ derived from $\varepsilon^{int}$ and a divergence-free non-conservative force $f_i^{nr}$.
\subsubsection{Illustration of the model setup}
%
%
{We illustrate the significance of the concept of Boltzmann weights
in non-reciprocal systems with a minimal model, namely the 
sheep–dog dynamics. Here, $v_{sd}$ represents the energy cost for an external agent to place a sheep on the lattice when a dog is already present. Since the sheep repels the dog, let us assume that 5 units of energy are required to implement this operation experimentally, implying $v_{sd} = 5$ units. Similarly, $v_{ds}$ is the energy cost for an external agent to place a dog on the lattice when a sheep is already present. Because the dog is attracted to the sheep, this operation releases energy. This calibration is equivalently achieved by performing the inverse operation of removing the dog from the lattice when both sheep and dog are present, which requires $-v_{ds}$ units of energy (note the sign convention due to defining the cost for the inverse operation). Let us assume that 3 units of energy are needed to remove the dog from the lattice in the presence of the sheep. Therefore, $v_{ds} = -3$ units, signifying an attractive interaction of the dog towards the sheep. All parameters of our 
theoretical model, namely the microscopic interaction energies $v_{ij}$ for different particle types, are to be interpreted in this way. }
\subsection{Dynamics}\label{sec:microscopic_dynamics}
\subsubsection{Single particle microstate transition rates}

The particle can change its type or its lattice location, which are defined as the reactive and diffusive transitions, respectively. A reactive transition from type $\gamma'$ to $\gamma$ at the lattice site $\#$ is indicated by
$\Delta_{\gamma \gamma'}^\#$, resulting in $\{N\} \to \{N + \Delta_{\gamma \gamma'}^\# \}$, see \cref{fig:particle_interactions}\textcolor{red}{(b)}. Similarly, a diffusive transition of a type $i$ particle from the lattice site $\#$ in the direction $\vec{\mathcal{D}}$ is indicated by $\Delta_{i}^{\vec{\mathcal{D}}\#}$, resulting in $ \{N\} \to \{N + \Delta_{i}^{\vec{\mathcal{D}}\#} \}$. 
%
%
{
The set of forward and backward transitions is consistently defined between the initial $\{N\}$ and final lattice configurations $\{N + \Delta_{\gamma \gamma'}^\# \}$ and $\{N + \Delta_{i}^{\vec{\mathcal{D}}\#} \}$, respectively, for reactive and diffusive transitions. For example, let us consider a single reactive transition of type $1$ to type $2$ with a single lattice site, then, the forward transition has to be consistently defined from the initial lattice configuration $\{N\} = (N_1, N_2)$ to the final lattice configuration $\{N + \Delta_{21} \} = ( N_1-1, N_2 + 1 )$ and the other way around for the backward transition.
}
The transition rates of reactive ($k_{\gamma' \gamma}^{\#}$) and diffusive ($k_{ i }^{ \vec{\mathcal{D}} \# }$) jumps are constructed using $\epsilon_i^\#$:
\begin{equation}\label{eq:microscopic_transition_rate}
\begin{split}
    k_{\gamma' \gamma}^{\#} = d_{\gamma' \gamma} e^{\epsilon_\gamma^{\#}}, \hspace{0.5cm}
    k_{ i }^{ \vec{\mathcal{D}} \# } = d_i^{\mathcal{D}} e^{\epsilon_i^{\#}}.
\end{split}    
\end{equation}
%
{The thermodynamically consistent exponential parameterization of $k_{\gamma' \gamma}^{\#}$ and $k_{i}^{\vec{\mathcal{D}}\#}$ in $\epsilon_i^\#$ allows one to visualize the thermodynamic meaning of $\epsilon_i^\#$. For non-interacting ideal particles, $\epsilon_i^\# = 0$; therefore, $k_{\gamma' \gamma}^\# = d_{\gamma' \gamma}$ and $k_i^{\vec{\mathcal{D}}\#} = d_i^{\mathcal{D}}$ are the equilibrium transition rates for non-interacting ideal particles, which also quantify the strength of the transition rates in \cref{eq:microscopic_transition_rate}. They satisfy $d_{\gamma \gamma'} = d_{\gamma' \gamma}$ and $d_i^{\mathcal{D}} = d_i^{\mathcal{D}^{-1}}$.}  

The LDB condition for the forward and backward microscopic transitions constructed using \cref{eq:microscopic_transition_rate} reads:  
\begin{equation}\label{eq:microscopic_local_detailed_balance_final}
\begin{split}
    \frac{k_{\gamma \gamma'}^\#}{k_{\gamma' \gamma}^\#} 
    = e^{ \epsilon_{\gamma'}^\# - \epsilon_{\gamma}^\# + f_{\gamma \gamma'}^{ch}}, \hspace{0.5cm}
    \frac{ k_{ i }^{ \vec{\mathcal{D}} \# } } { k_{ { i } }^{ (\vec{\mathcal{D} } \#)^{-1} } }
    = e^{ \epsilon_{i}^\# - \epsilon_{i}^{\vec{\mathcal{D}} \#} + \vec{\mathcal{D}} \cdot \vec{f}_{i}^{sp} }.
\end{split}
\end{equation}
%
{\Cref{eq:microscopic_local_detailed_balance_final} expresses the microscopic LDB for non-reciprocal particles in a more familiar form compared to its original definition in \cref{eq:microscopic_transition_rate}. However, the time-symmetric part of the transition rates cancels out in \cref{eq:microscopic_local_detailed_balance_final} due to the relations $d_{\gamma' \gamma} = d_{\gamma\gamma'}$ and $d_i^{\mathcal{D}} = d_i^{\mathcal{D}^{-1}}$. 
}

The environment supports the thermodynamic cost for the change in the microscopic Boltzmann weight. Compared to \cref{eq:microscopic_transition_rate}, we have modified \cref{eq:microscopic_local_detailed_balance_final} by introducing a non-conservative external driving force $f_{\gamma \gamma'}^{ch}$ and $\vec{f}_{i}^{sp}$ “\textit{along the transition}”, which correspond to chemical and self-propulsion driving. 
{
In accordance with the fundamental assumption of stochastic thermodynamics, under the timescale separation between the system's slow degrees of freedom and the environment's fast degrees of freedom, the thermodynamic cost of such external constant non-conservative driving of the system is supported by the environment \cite{Katz_1983,Katz_1984,seifert_2012}. Here, the environment is physically/experimentally realized by an externally coupled reservoir of a very large number of particles. External driving refers to particle self-propulsion driven by an external reservoir, like Janus particles that are driven by the Hydrogen Peroxide contained in the surrounding liquid in active colloid experiments. Similarly, the reactive transition between particles are driven by enzymes in the surrounding liquid. This self-propulsion or chemical driving mechanism is generally assumed to be independent of the interactions between and reactions of the system particles. Throughout this work, the convention `externally non-conservatively driven' refers to such a `thermodynamic reservoir'. Therefore, the term external means `from outside the system', and externally driven means a reaction or diffusive movement fueled by large external reservoirs which are not explicitly modeled, as sketched in \cref{fig:particle_interactions}(b).
} 

{
Since, an external coupling to a reservoir is a property of a transition, $f_{\gamma\gamma'}^{ch}$ and $\vec{f}_{i}^{sp}$ are introduced in \cref{eq:microscopic_local_detailed_balance_final}. Nevertheless, they could have been introduced directly in \cref{eq:microscopic_transition_rate}, using exponential tilts $e^{f_{\gamma\gamma'}^{ch}/2}$ and $e^{-f_{\gamma\gamma'}^{ch}/2}$ for the forward and backward reactive transitions (or $e^{\vec{\mathcal{D}} \cdot \vec{f}_{i}^{sp} / 2}$ and $e^{-\vec{\mathcal{D}} \cdot \vec{f}_{i}^{sp} / 2}$ for the forward and backward diffusive transitions). Note that the non-reciprocal non-conservative forces $f_i^{nr}$ act in the state space, whereas chemical or self-propulsion non-conservative driving is associated with transitions between states and lies in the transition space. This fundamental physical distinction between the two is clear: non-reciprocity requires the minimal notion of \textit{two states} to define a dissipative current, whereas driven discrete-state processes require \textit{three states and three transitions} \cite{schnakenberg_1976}.}  

We use shorthand notations $\Delta_{\gamma \gamma'}^\#(*) = (*)_{\gamma'}^\# - (*)_{\gamma}^\#$ and $\Delta_{i}^{\vec{\mathcal{D}}\#}(*) = (*)_{i}^\# - (*)_{i}^{\vec{\mathcal{D}}\#}$ for the change in $\epsilon_i^\#, \epsilon_i^r$, and $f_i^{nr}$ due to a transition. We introduce a shorthand notation $\Delta$ to refer to a transition, applied as a diffusive or reactive transition depending on the context.
{Therefore, using \cref{eq:microscopic_boltzmann_weight,eq:microscopic_reciprocal_interaction_energy_particle},  $\Delta_{\gamma \gamma'}^\# \epsilon = \epsilon_{\gamma'}^\# - \epsilon_{\gamma}^\#$ and $\Delta_{i}^{\vec{\mathcal{D}}\#} \epsilon = \epsilon_{i}^\# - \epsilon_{i}^{\vec{\mathcal{D}}\#}$. The corresponding reciprocal non-reciprocal decompositions, $\Delta_{\gamma \gamma'}^\# \epsilon = \Delta_{\gamma \gamma'}^\# \epsilon^r + \Delta_{\gamma \gamma'}^\# f^{nr}$ and $\Delta_{i}^{\vec{\mathcal{D}}\#} \epsilon = \Delta_{i}^{\vec{\mathcal{D}}\#} \epsilon^r + \Delta_{i}^{\vec{\mathcal{D}}\#} f^{nr}$.}
The set of all reactive transitions is denoted by $\{\Delta_{\gamma \gamma'}^\#\}$, and the set of all diffusive transitions is denoted by $\{\Delta_i^{\vec{\mathcal{D}}\#}\}$. Four such possibilities exist for a two-dimensional square lattice: upward, downward, leftward, and rightward. Microscopic interactions and transitions are shown in \cref{fig:particle_interactions}.  

{To demonstrate the connection of \cref{eq:microscopic_transition_rate} to the Glauber-dynamics \cite{Glauber_1963} $k_{21}^\# = e^{\epsilon_1^\#} / ( e^{\epsilon_1^\#} + e^{\epsilon_2^\#} )$ and $k_{12}^\# = e^{\epsilon_2^\#} / ( e^{\epsilon_1^\#} + e^{\epsilon_2^\#} )$ or the Metropolis-dynamics \cite{Metropolis_1953} $k_{12}^\# = \min( 1, e^{ - \epsilon_{1}^\# + \epsilon_{2}^\# } )$ and $k_{21}^\# = \min( 1, e^{ - \epsilon_{2}^\# + \epsilon_{1}^\# } )$, we consider a reactive transition $\Delta_{12}^\#$ between type $1$ and $2$ particles with $f_{12}^{ch} = 0$. Obviously, Glauber-dynamics and Metropolis-dynamics have the exponential form \cref{eq:microscopic_transition_rate}, but with factors $d_{\gamma\gamma'} = 1/(e^{\epsilon_1^\#} + e^{\epsilon_2^\#})$ and $d_{\gamma\gamma'} = 1/\max{(e^{\epsilon_1^\#}, e^{\epsilon_2^\#})}$, respectively,
that does \textit{not} represent the rate for non-interacting 
particles since it depends on $\epsilon_1^\#$ and $\epsilon_2^\#$. The factors non-interacting transition rates $d_{\gamma\gamma'}$ represent the broad range of time-scales for different transitions that are present in generic reaction networks, which requires careful treatment without assumptions, see Refs.\cite{atm_2024_var_epr,atm_2025_var_epr_derivation} for details. This is carefully treated with the definition in \cref{eq:microscopic_transition_rate}, which preserves the different timescales associated with different transitions. Note that the Glauber and Metropolis rates are saturating and therefore advantageous for equilibrium Monte Carlo simulations \cite{Metropolis_1953,Glauber_1963}, but they do not reflect a realistic dynamics of non-equilibrium systems. In any case the time-symmetric components of the transition rates become irrelevant after taking the ratio of transition rates for the LDB in \cref{eq:microscopic_local_detailed_balance_final}. The dynamics defined by \cref{eq:microscopic_transition_rate} can straightforwardly be simulated by the exact Gillespie algorithm \cite{Gillespie_1977}.}
\subsubsection{Multi-particle microstate transition rates.}
Particles with identical type and lattice index have the same dynamic and thermodynamic properties. Hence, $k_{\gamma\gamma'}^{\#}$ are lumped together to give multi-particle transition rates 
$k_{\gamma \gamma'}^{\#}({\{ N \}}) = N_{\gamma'}^{\#} k_{\gamma \gamma'}^{\#}$ and 
$k_{ i }^{ \vec{\mathcal{D}} \#}({\{ N \}}) = N_i^\# k_{ i }^{ \vec{\mathcal{D}} \#}$.
The LDB for the multi-particle transition rates (reactive ($k_{\gamma \gamma'}^{\#}({\{ N \}})$) and diffusive ($k_{ { i } }^{\vec{\mathcal{D}} \#} (\{ N \})$)) is:
\begin{equation}\label{eq:microscopic_multi_particle_local_detailed_balance_final}
\begin{split}
    \frac{k_{\gamma \gamma'}^{\#}({\{ N \}})}{k_{\gamma' \gamma}^{\#}({\{ N + \Delta_{\gamma \gamma'}^\#\}})}
    & = e^{ 
    - \Delta_{\gamma \gamma'}^\# s^b
    + \Delta_{\gamma \gamma'}^\# \epsilon 
    + f_{\gamma \gamma'}^{ch} 
    },  
    \\
    \frac{k_{ { i } }^{\vec{\mathcal{D}} \#} (\{ N \}) }{k_{i}^{ (\vec{\mathcal{D}} \#)^{-1} } ( \{ N + \Delta_{i}^{\vec{\mathcal{D}}\#} \} ) } 
    & = e^{ - \Delta_i^{\vec{\mathcal{D}} \#} s^b + \Delta_i^{\vec{\mathcal{D}} \#} \epsilon 
    + \vec{\mathcal{D}} \cdot \vec{f}_{i}^{sp} }.
\end{split}
\end{equation}
Here, $\Delta s^b$ incorporates the change in the microscopic Boltzmann entropy, defined as a global lattice property \cite{Spinney_2013}:
\begin{equation}\label{eq:microscoipic_boltzmann_entropy}
\begin{split}
    s^{b} = \ln{ \left[\frac{ \left(\sum_{i, \#} N_i^\# \right)! }{\prod_{i,\#} N_i^\# !} \right] }.
\end{split}    
\end{equation}
It quantifies the statistical degeneracy of the microstate and is a system property defined for the lattice. Importantly, $\Delta_{\gamma \gamma'}^\# \epsilon^r = \Delta_{\gamma \gamma'}^\# \varepsilon^{int}$ and $\Delta_{i}^{\vec{ \mathcal{D}} \#} \epsilon^r = \Delta_{i}^{\vec{ \mathcal{D}} \#} \varepsilon^{int}$, as satisfied by the reciprocal microscopic Boltzmann weight. We introduce the notation $\Delta_{\gamma \gamma'}^\# \varepsilon^{int} = \varepsilon^{int}(\{N\}) - \varepsilon^{int}(\{N+\Delta_{\gamma \gamma'}\})$, which represents a change of a function of the lattice state, a global property; this applies similarly to $s^{b}$. In contrast, the change in the non-reciprocal microscopic Boltzmann weight $\Delta_{\gamma \gamma'}^\# f^{nr}$ cannot be represented as a change of a global function of the lattice state. This results in a fundamental difference between the boundary ($\Delta_{\gamma \gamma'}^\# \varepsilon^{int}$ and $- \Delta_{\gamma \gamma'}^\# s^b$) and the bulk terms ($\Delta_{\gamma \gamma'}^\# f^{nr}$) of the transition affinity.
\subsubsection{Master equation}
The probability of the multi-particle lattice microstate $\{N\}$ at time $t$ is denoted by $P_{ \{ N \} } (t)$. The master equation for the dynamic evolution of the probability $P_{\{N\}}(t)$ for the lattice microstate reads \cite{van_kampen,gardiner}:
\begin{equation}\label{eq:microscopic_master_equation}
\begin{split}
    \partial_t P_{\{N\}}(t)
    & = - \sum_{\{ \Delta_{\gamma \gamma'}^\# \}}
    {j_{\gamma \gamma'}^\# ( \{ N \} )}
    - \sum_{ \{ \Delta_i^{\vec{\mathcal{D}} \#} \} }{ j_i^{{ \vec{\mathcal{D}} \# }} ( \{ N \} ) }. 
\end{split}    
\end{equation}
The reactive and diffusive microscopic transition currents are:
\begin{equation}\label{eq:microscopic_transition_currents}
\begin{split}
    j_{\gamma \gamma'}^\# ( \{ N \} )
    & = \left(
    k_{ \gamma \gamma' }^\#( \{ N \} ) 
    P_{ \{ N \} }
    - k_{ \gamma' \gamma}^\# (\{ N + \Delta_{\gamma \gamma'}^\# \}) 
    P_{ \{ N + \Delta_{\gamma \gamma'}^\# \} } 
    \right),
    \\
    j_i^{{ \vec{\mathcal{D}} \# }} ( \{ N \} )
    & =
    \left( 
    k_i^{ \vec{\mathcal{D}} \# } ( \{ N \} ) P_{ \{ N \} }
    - k_i^{ {( \vec{\mathcal{D}} \# )}^{-1} } ( \{ N + \Delta_i^{ \Vec{\mathcal{D}} \# } \} ) P_{ \{ N + \Delta_i^{ \Vec{\mathcal{D}} \# } \} }  
    \right).
\end{split}    
\end{equation}
\subsection{Thermodynamics}\label{sec:microscopic_thermodynamics}
\subsubsection{Transition affinities}
The stochastic state entropy $s^{state}$ of $\{N\}$ is defined as \cite{Seifert_2005}:
\begin{equation}\label{eq:microscopic_state_entropy}
     s^{state} = -\ln{ \left( P_{ \{N\} } \right) }.
\end{equation}
The total microscopic energy functional $\varepsilon$ incorporates energetic and entropic contributions, $\varepsilon = \varepsilon^{int} - s^{b}$. The stochastic Massieu potential is:
\begin{equation}\label{eq:microscopic_stochastic_massieu_potential}
\begin{split}
    \varepsilon^{stoch} = \varepsilon - s^{state}.
\end{split}    
\end{equation}
Importantly, $\varepsilon^{int}, s^b, s^{state}$ are properties of the lattice; therefore, a cumulative change in them due to multiple stochastic transitions depends only on the initial and final lattice states. This subsequently applies to $\varepsilon$ and $\varepsilon^{stoch}$. In comparison, $\Delta_{i}^{\vec{\mathcal{D}}\#} f^{nr}$ and $\Delta_{\gamma \gamma'}^\# f^{nr}$ are defined locally on the lattice. The stochastic transition entropy is defined as: 
\begin{equation}\label{eq:entropy_production_transition_defination}
\begin{split}
\Delta_{\gamma \gamma'}^\# \sigma
& = \Delta_{\gamma \gamma'}^\# \varepsilon^{int} 
- \Delta_{\gamma \gamma'}^\# s^b 
- \Delta_{\gamma \gamma'}^\# s^{state} 
+ \Delta_{\gamma \gamma'}^\# f^{nr} 
+ f_{\gamma \gamma'}^{ch}, 
\\
\Delta_i^{\vec{\mathcal{D}} \#} \sigma
& = \Delta_{i}^{ \vec{\mathcal{D}} \#} \varepsilon^{int}
- \Delta_{i}^{\vec{\mathcal{D}} \#} s^b 
- \Delta_{i}^{\vec{\mathcal{D}} \#} s^{state}
+ \Delta_{i}^{\vec{\mathcal{D}}\#} f^{nr} 
+ \vec{\mathcal{D}} \cdot \vec{f}_{i}^{sp}. 
\end{split}    
\end{equation}
\Cref{eq:entropy_production_transition_defination} quantifies the thermodynamic cost of each microscopic transition supported by the environment. We adhere to the convention \textit{`forces generate currents'} throughout this paper, where forces refer to the transition affinities, for instance, $\Delta \varepsilon, \Delta s^{state}, \Delta f^{nr}, \vec{f}_{i}^{sp}, f_{\gamma \gamma'}^{ch}$. Transition affinities are categorized into boundary (conservative) terms $(\Delta \varepsilon, \Delta s^{state})$ and bulk (non-conservative) terms $(\Delta f^{nr}, \vec{f}_{i}^{sp}, f_{\gamma \gamma'}^{ch})$, which correspond to relaxation and dissipative currents, respectively \cite{seifert_2012}. Considering a set of consecutive stochastic transitions over observation time $\tau$, the conservative and non-conservative total entropy production due to all transitions is $O(1)$ and $O(\tau)$, respectively. Importantly, the non-conservative forces fundamentally differ in their origin. $\Delta f^{nr}$ and $f_{\gamma \gamma'}^{ch}$ (or $\vec{f}_i^{sp}$) lie in the state-space and the transition-space, respectively. Moreover, $f_i^{nr}$ depends on the local particle occupancy, unlike $f_{\gamma \gamma'}^{ch}$ (or $\vec{f}_i^{sp}$), which are constants.
{
Thus, the non-conservative driving due to non-reciprocal interactions depends on the initial $\{N\}$ and final lattice configurations $\{N + \Delta_{\gamma \gamma'}^\# \}$ and $\{N + \Delta_{i}^{\vec{\mathcal{D}}\#} \}$, respectively, for reactive and diffusive transitions, since $f_i^{nr}$ depends on $\{ N \}$, as highlighted earlier in \cref{sec:microscopic_dynamics}. The non-conservative driving due to non-reciprocal interactions is more dynamic in comparison to fixed chemical or self-propulsion driving forces.
}
\subsubsection{Microscopic EPR}\label{sec:microscopic_EPR}
The microscopic reactive and diffusive mean EPR are defined as $\langle \dot{\sigma}^r \rangle$ and $\langle \dot{\sigma}^d \rangle$, respectively \cite{schnakenberg_1976,seifert_2012}:
\begin{equation}\label{eq:microscopopic_net_reactive_EPR}
\begin{split}
    \langle \dot{\sigma}^r \rangle 
    & = \sum_{ \{\Delta_{\gamma \gamma' }^\#\} }^{\{N\}} j_{\gamma \gamma'}^\# ( \{ N \} )  \ln{\left(
    \frac{ k_{ \gamma \gamma'}^\# (\{ N \}) 
    P_{ \{ N \} } 
    }
    {
    k_{ \gamma' \gamma}^\# (\{ N + \Delta_{\gamma \gamma'}^\# \}) 
    P_{ \{ N + \Delta_{\gamma \gamma'}^\# \} }
    }
    \right)},
    \\
    \langle \dot{\sigma}^d \rangle 
    & = \sum_{\{ \Delta_i^{\Vec{\mathcal{D}} \#} \}}^{\{N\}}
    j_i^{{ \vec{\mathcal{D}} \# }} ( \{ N \} )
    \ln{
    \left(
    \frac
    { k_{ i }^{ \vec{\mathcal{D}} \# } (\{ N \}) 
    P_{ \{ N \} } 
    }
    {
    k_{i}^{ (\vec{\mathcal{D}} \#)^{-1} } (\{ N + \Delta_i^{\Vec{\mathcal{D}} \#}  \}) 
    P_{ \{ N + \Delta_i^{\Vec{\mathcal{D}} \#}  \} }
    }
    \right)}.
\end{split}    
\end{equation}
Here, $\sum_{ \{\Delta_{\gamma \gamma' }^\#\} }^{\{N\}}$ and $\sum_{\{ \Delta_i^{\Vec{\mathcal{D}} \#} \}}^{\{N\}}$ denote the sum over all lattice configurations for the reactive and diffusive transitions, respectively. This satisfies the fundamental definition of mean EPR, namely, force times current: 
$\langle \dot{\sigma}^r \rangle = \sum_{ \{\Delta_{\gamma \gamma' }^\#\} }^{\{N\}} j_{\gamma \gamma'}^\# ( \{ N \} ) \Delta_{\gamma \gamma'}^\# \sigma$ and 
$\langle \dot{\sigma}^d \rangle = \sum_{\{ \Delta_i^{\Vec{\mathcal{D}} \#} \}}^{\{N\}} j_i^{{ \vec{\mathcal{D}} \# }} ( \{ N \} ) \Delta_i^{\vec{\mathcal{D}} \#} \sigma$,
due to the LDB condition \cref{eq:microscopic_multi_particle_local_detailed_balance_final} and the definitions \cref{eq:microscopic_state_entropy,eq:entropy_production_transition_defination}. $\langle \dot{\sigma}^r \rangle > 0$ and $\langle \dot{\sigma}^d \rangle \geq 0$ due to the inequality $(x-y)\ln{\left(x/y\right)} \geq 0$, which leads to the second law of thermodynamics. The total mean microscopic EPR reads $\langle \dot{\sigma} \rangle = \langle \dot{\sigma}^r \rangle + \langle \dot{\sigma}^d \rangle$.
\subsubsection{Conservative and non-conservative decomposition of EPR}\label{sec:microscopic_EPR_consv_n_consv_decomposition}
We reorganize $\langle \dot{\sigma} \rangle$ using \cref{eq:microscopic_master_equation,eq:microscopic_transition_currents,eq:microscopic_multi_particle_local_detailed_balance_final}, obtaining the EPR contributions due to the reciprocal, non-reciprocal, Gibbs entropic, and external non-conservative driving forces.   
\begin{equation}\label{eq:microscopic_net_EPR_final}
\begin{split}
    \langle \dot{\sigma} \rangle 
    & = - d_t \langle \varepsilon \rangle
    + d_t s^{gb}
    + \langle \dot{\sigma}^{nr} \rangle 
    + \langle \dot{\sigma}^{sp} \rangle
    + \langle \dot{\sigma}^{ch} \rangle. 
\end{split}
\end{equation}
Here, $- d_t \langle \varepsilon \rangle$ and $d_t s^{gb}$ are attributed to the rate of change of $\varepsilon^{int} - s^b$ and $s^{state}$ from \cref{eq:entropy_production_transition_defination}. We define the mean microscopic work rate $\langle \dot{w} \rangle = -\sum_{\{N\}} P_{\{N\}} \partial_t \varepsilon^{int}$ due to the explicit time-dependent driving of the control parameters $\{\lambda\}$ of $\varepsilon^{int}$ \cite{sekimoto}. Hence, $d_t \langle \varepsilon \rangle = \sum_{\{N\}} \partial_t P_{\{N\}} \varepsilon^{int} + P_{\{N\}} \partial_t \varepsilon^{int}$ incorporates the stochastic work. The Gibbs entropy and the Gibbs EPR read \cite{Spinney_2013}: 
\begin{equation}\label{eq:microscopic_gibbs_entropy}
\begin{split}
    s^{gb} = -  \sum_{\{N\}} P_{\{N\}} \ln{ \left( P_{\{N\}}  \right) }, \hspace{0.5cm} 
    d_t s^{gb} =  - \sum_{\{N\}} d_t P_{\{N\}} \ln{ \left( P_{\{N\}}  \right) }.
\end{split}    
\end{equation}
Such that $s^{gb} = \langle s^{state} \rangle$ holds. The mean microscopic non-reciprocal EPR is:
\begin{equation}\label{eq:microscopopic_non-reciprocal_reactive_EPR}
    \langle \dot{\sigma}^{nr} \rangle
    = \sum_{\{ \Delta_{\gamma \gamma'}^\# \} }^{\{N\}} 
    j_{\gamma \gamma'}^\#( \{ N \} ) 
    \: \Delta_{\gamma \gamma'}^\# f^{nr}
    + \sum_{\{ \Delta_{i}^{\vec{\mathcal{D}}\#} \} }^{\{N\}} 
    j_{i}^{ \vec{\mathcal{D}} \#}( \{ N \} ) 
    \: \Delta_{i}^{\vec{\mathcal{D}}\#} f^{nr}.
\end{equation}
The mean microscopic non-conservative EPR due to the chemical and self-propulsion driving is:
\begin{equation}\label{eq:microscopopic_non_conservative_EPR}
\begin{split}
    \langle \dot{\sigma}^{ch} \rangle
    = \sum_{\{ \Delta_{\gamma \gamma'}^\# \} }^{\{N\}} 
    j_{\gamma \gamma'}^\#( \{ N \} )  
    f_{\gamma \gamma'}^{ch}, 
    \hspace{0.3cm}
    \langle \dot{\sigma}^{sp} \rangle
    = \sum_{\{ \Delta_{i}^{ \vec{\mathcal{D}}\#} \}}^{\{N\}} 
    j_{i}^{ \vec{\mathcal{D}} \#}( \{ N \} ) 
    \: \vec{\mathcal{D}} \cdot \vec{f}_{i}^{sp}.
\end{split}    
\end{equation}
\Cref{eq:microscopic_net_EPR_final} is the second law of thermodynamics decomposed on the basis of the origin of the forces acting along the transitions. The first two terms are derived from conservative forces. Because their time-integrated EPR is given by the change of a functional defined for the lattice, it depends only on the initial and final states. In contrast, the remaining terms give the EPR due to the non-conservative forces, namely non-reciprocal, self-propulsion, and chemical driving.
\subsubsection{Orthogonal decomposition of the EPR}\label{sec:microscopic_epr_orthogonal_decomosition}
We define the Boltzmann probability distribution $P^{\varepsilon}_{ \{ N \} } (t) = e^{-\varepsilon + \psi_{\varepsilon}}$ with reference energy functional $\varepsilon$. We compute the free energy $\psi_{\varepsilon} = -\ln{(\mathcal{Z}^\varepsilon)}$ in terms of the partition function $\mathcal{Z}^{\varepsilon} = \sum_{\{N\}} e^{-\varepsilon}$. In the absence of non-conservative forces, i.e., $f_i^{nr}, \vec{f}_i^{sp}, f_{\gamma\gamma'}^{ch} = 0$, the system satisfies the Boltzmann distribution. We decompose \cref{eq:microscopic_stochastic_massieu_potential} using $P^{\varepsilon}_{ \{ N \} } (t)$ to \cite{Bergman_1955,Lebowitz_1957,Qian_2001}:
\begin{equation}\label{eq:microscopic_stochastic_massieu_potential_orthogonal}
\begin{split}
    \varepsilon^{stoch} = \psi_\varepsilon + \ln{ \left( \frac {P_{ \{ N \} } (t)}{ P^{\varepsilon}_{ \{ N \} } (t) } \right)}.
\end{split}    
\end{equation}
Comparing \cref{eq:microscopic_stochastic_massieu_potential_orthogonal} to \cref{eq:microscopic_stochastic_massieu_potential}, analogously to \cref{eq:microscopic_state_entropy}, we define the reference state entropy $s^{state}_\varepsilon$,
\begin{equation}\label{eq:microscopic_state_entropy_orthogoonal}
\begin{split}
    s^{state}_\varepsilon = - \ln{ \left( \frac {P_{ \{ N \} } (t)}{ P^{\varepsilon}_{ \{ N \} } (t) } \right)}.
\end{split}    
\end{equation}
$s^{state}_\varepsilon$ quantifies the thermodynamic distance of $P_{ \{ N \} } (t)$ from the reference Boltzmann probability $P^{\varepsilon}_{ \{ N \} } (t)$. The total thermodynamic distance between the probability distributions $P_{ \{ N \} } (t)$ and $P^{\varepsilon}_{ \{ N \} } (t)$ is given by the KL-divergence \cite{Bergman_1955,Lebowitz_1957,Schloegl_1971,Amari_2000_book,Qian_2001,Ge_2009,Ge_2010,Shiraishi_2019}, $D_{\varepsilon}^{KL}(t) = \sum_{ \{N\} } P_{\{N\}}(t) \ln{ \left( {P_{\{N\}}(t) }/{P^{\varepsilon}_{\{N\}}(t) } \right) }$, where, $-D_{\varepsilon}^{KL}(t) = \langle  s^{state}_\varepsilon \rangle$ holds. Analogously to $d_t s^{gb}$, the rate of change of $D_{\varepsilon}^{KL}(t)$ is defined as:
\begin{equation}\label{eq:microscopic_KL_divergence_relaxation_rate}
\begin{split}
   d_t D_{\varepsilon}^{KL}(t) = \sum_{ \{N\} } d_t P_{\{N\}}(t) \ln{ \left( \frac{P_{\{N\}}(t) }{P^{\varepsilon}_{\{N\}}(t) } \right) }.
\end{split}    
\end{equation}
We introduce the orthogonal decomposition of $\langle \dot{\sigma}^{ch} \rangle $ and 
$ \langle \dot{\sigma}^{sp} \rangle $,
\begin{equation}\label{eq:microscopopic_non_conservative_EPR_orthogonal}
\begin{split}
    \langle \dot{\sigma}^{ch} \rangle = \sum_{\{ \Delta_{\gamma \gamma'}^\# \} }^{\{N\}} j_{\gamma \gamma'}^{ch} \: f_{\gamma \gamma'}^{ch},
    \hspace{0.5cm}
    \langle \dot{\sigma}^{sp} \rangle = \sum_{\{ \Delta_{i}^{ \vec{\mathcal{D}}\#} \}}^{\{N\}} 
    j_{i}^{sp} 
    \: \vec{\mathcal{D}} \cdot \vec{f}_{i}^{sp}.
\end{split}    
\end{equation}
Here, $ j_{\gamma \gamma'}^{ch}$ is the anti-symmetric part of $j_{\gamma \gamma'}^\#( \{ N \} )$ under the adjoint transformation $f_{\gamma \gamma'}^{ch} \to -f_{\gamma \gamma'}^{ch}$. Similarly, $ j_{i}^{sp}$ is the anti-symmetric part of $ j_{i}^{ \vec{\mathcal{D}} \#}( \{ N \} ) $ under the adjoint transformation $\vec{f}_{i}^{sp} \to -\vec{f}_{i}^{sp}$. We have dropped the $\#$ and $\{N\}$ index for brevity. The exact expressions for $j_{\gamma \gamma'}^{ch}$ and $ j_{i}^{sp}$ read:
\begin{equation}\label{eq:microscopopic_currents_orthogonal}
\begin{split}
    j_{\gamma \gamma'}^{ch} 
    & = d_{\gamma\gamma'} \left( P_{\{N\}} N_{\gamma'}^{\#} e^{\epsilon_{\gamma'}^\#} + P_{\{N + \Delta_{\gamma \gamma'}^\# \}} N_{\gamma} e^{\epsilon_{\gamma}^\#} \right) \sinh{ \left( \frac{f_{\gamma\gamma'}}{2} \right)},
    \\
    j_{i}^{sp}
    & = d_{i}^{\mathcal{D}} \left( P_{\{N\}} N_{i}^{\#} e^{\epsilon_{i}^\#} + P_{\{N + \Delta_{i}^{ \vec{\mathcal{D}} \#} \}} N_{i}^{\vec{\mathcal{D}}\#} e^{ \epsilon_{i}^{\vec{\mathcal{D}} \#} } \right) \sinh{ \left( \frac{\vec{\mathcal{D}} \cdot \vec{f}_{i}^{sp}}{2} \right)}.
\end{split}    
\end{equation}
$j_{\gamma \gamma'}^{ch} $ and $ j_{i}^{sp}$ in \cref{eq:microscopopic_currents_orthogonal} are the dissipative currents generated by non-conservative driving forces $f_{\gamma \gamma'}$ and $\vec{f}_i^{sp}$, compared to the total reactive/diffusive currents $ j_{\gamma \gamma'}^\#( \{ N \} ) $ and $ j_{i}^{ \vec{\mathcal{D}} \#}( \{ N \} )$ in \cref{eq:microscopopic_non_conservative_EPR} that include currents generated by conservative forces. The total bidirectional current, commonly called traffic, is obtained by replacing $-$ with $+$ in \cref{eq:microscopic_transition_currents}, which also equivalently quantifies the scaled fluctuations resulting from the transitions. The form of \cref{eq:microscopopic_currents_orthogonal} reveals that the amplitude of $j_{\gamma \gamma'}^{ch} $ and $ j_{i}^{sp}$, and subsequently $\langle \dot{\sigma}^{ch} \rangle$ and $\langle \dot{\sigma}^{sp} \rangle$, is proportional to the traffic (scaled fluctuations) in the direction orthogonal to the external non-conservative driving force. 

Utilizing \cref{eq:microscopic_stochastic_massieu_potential_orthogonal,eq:microscopic_state_entropy_orthogoonal,eq:microscopic_KL_divergence_relaxation_rate,eq:microscopopic_non_conservative_EPR_orthogonal}, the second law of thermodynamics \cref{eq:microscopic_net_EPR_final} has the following orthogonal form:
\begin{equation}\label{eq:microscopic_net_EPR_final_orthogonal}
\begin{split}
    \langle \dot{\sigma} \rangle 
    & = 
    - d_t \psi_{\varepsilon}
    - d_t D^{KL}_{ \varepsilon}(t)
    + \langle \dot{\sigma}^{nr} \rangle 
    + \langle \dot{\sigma}^{sp} \rangle
    + \langle \dot{\sigma}^{ch} \rangle. 
\end{split}
\end{equation}
In \cref{eq:microscopic_net_EPR_final_orthogonal}, we have decomposed the EPR into its four linearly independent orthogonal contributions. 
First, $-d_t \psi_{\varepsilon}$ quantifies the rate of change of the free energy attributed to the external driving work required to change the control parameters $\{\lambda\}$ of $\varepsilon$.  
Second, $- d_t D^{KL}_{ \varepsilon}(t)$ quantifies the EPR due to the relaxation towards the Boltzmann distribution $P_{\{N\}}^\varepsilon$, where $-d_t \psi_{\varepsilon}$ and $- d_t D^{KL}_{ \varepsilon}(t)$ are physically interpreted as the boundary terms in the $\{\lambda\}$ and $P_{\{N\}}$ spaces, respectively. Third, $\langle \dot{\sigma}^{nr} \rangle$ quantifies the EPR due to the anti-symmetric forces (and vorticity currents generated by them) between the non-reciprocal particles. Fourth, $\langle \dot{\sigma}^{sp} \rangle + \langle \dot{\sigma}^{ch} \rangle$ quantifies the EPR due to the external non-conservative forces along the transitions, a thermodynamic cost supported by an external chemical or self-propulsion reservoir, which generates the dissipative transition currents \cite{Oono_1998}.

One identifies the non-adiabatic EPR $\langle \dot{\sigma}_{\varepsilon}^{na} \rangle = - d_t D^{KL}_{ \varepsilon}(t)$ and the remaining terms as the housekeeping EPR $\langle \dot{\sigma}_{\varepsilon}^{hs} \rangle$. The orthogonal decomposition is more fundamental because it does not require the existence of a steady state, as $\varepsilon$ is well-defined for any dynamic state. The lower bound on the total EP is obtained using $\langle \dot{\sigma}_{\varepsilon}^{na} \rangle$ \cite{Shiraishi_2019,Amari_2000_book}, and it reads $\langle \sigma_{\varepsilon}^{na} \rangle \geq D^{KL}_\varepsilon(0) - D^{KL}_\varepsilon(\tau) \geq D^{KL}(P(0)||P(\tau))$. Other notable consequences of the orthogonal decomposition have utilized different reference gauges \cite{Espigares_2012,maes_2007,Maes_2014,dechant_2022_geometric_epr,Ding_2022_st_ft,dechant_2022_geometric_epr_cpl,yoshimura_2023_epr_decomposition_nld,nagayama_2023_geometric_epr_rd}. Choosing $P_{\{N\}}^{ss}$ as the reference distribution, the nonadiabatic-housekeeping decomposition of $\langle \dot{\sigma} \rangle$ is detailed in \cref{sec:microscopic_epr_excess_housekeeping_decomosition}.
\subsection{Coarse-graining}\label{sec:cg}
We implement the Doi-Peliti coarse-graining (DPCG) procedure to obtain a coarse-grained description \cite{Doi_1976,Doi_1976_2,Peliti,Peliti_1986,grassberger_1980,rose_1979,Mikhailov_1981,Mikhailov_1981_2,Mikhailov_1985,Cardy_2008,Howard_1995,Wiese_2016,Tauber_2014,Weber_2017}. The DPCG procedure incorporates the fluctuations (discreteness of the microscopic particle nature) owing to the second-quantization approach used. The technical details of the coarse-graining procedure and its physical implications are summarized in Ref. \cite{atm_cg_nr_2024}.
\section{Fluctuating Macroscopic Description}\label{sec:macroscopic}
The set of coarse-grained mesostates is denoted by $\{ N_i^\# \}$. Here, $\{ N_i^\# \}$ is defined over the discrete (lattice) space for the mesoscopic description. Physically, it corresponds to lumping the microstates (particles) with the same dynamic and thermodynamic properties into a single extensive coarse-grained mesostate. It achieves the coarse-graining step from (a) to (b) shown in \cref{fig:CG_diagram}. We define $\Omega = N_{tot}/\mathcal{V}$, where $N_{tot}$ and $\mathcal{V}$ are the total number of particles in the whole lattice and the volume of the lattice (number of lattice sites), respectively. Thus, $\Omega$ quantifies the mean number of particles per lattice site. $N_i^\#$ is $O(\Omega)$ and its fluctuations due to microscopic transitions are $O(1)$ \cite{Touchette_2009,atm_cg_nr_2024}. 

To proceed further with the mesoscopic to macroscopic coarse-graining step, (b) to (c) depicted in \cref{fig:CG_diagram}, we introduce the mesoscopic to macroscopic scaling $N_i^\# = \Omega \rho_i(\vec{x})$  between the macrostate $\rho_i( \vec{x} ) $  and the particle number mesostate $N_i^\#$. Therefore, $\rho_i$ and its fluctuations due to microscopic transitions are $O(1)$ and $O(1/\Omega)$ respectively \cite{Touchette_2009,atm_cg_nr_2024}. By construction, $\Omega \geq 1$, as $\Omega < 1$ contradicts the mesoscopic to macroscopic coarse-graining. The space field $\vec{x}$ tracks the lattice index, which we drop in our notations and its meaning is presumed. Importantly, we use the macrostate convention consistent with the large-deviation theory \cite{Touchette_2009,atm_cg_nr_2024}, where an intensive variable is defined and its fluctuations are suppressed with the large-deviation scaling parameter $\Omega$. Thus, the notion of an intensive variable is defined in the context of fluctuations. The macrostate $\rho_i$ does not necessarily coincide with the convention of the density used in hydrodynamics unless $\Omega = \mathcal{V}$ is chosen. This is fulfilled in the hydrodynamic limit, where the average number of particles per lattice site scales with the system volume. Hence, $N_i^\# = \mathcal{V} \rho_i$ holds. Importantly, this flexible definition of the macrostate $\rho_i$ allows us to define and study the systems on intermediate observation scales between the mesoscopic and hydrodynamic descriptions. $\Omega = 1$ corresponds to the mesoscopic description that uses the particle number as the coarse-grained state. The thermodynamic limit of $\Omega \to \infty$ suppresses macrostate fluctuations and recovers the deterministic limit corresponding to the reaction-diffusion systems, \cref{fig:CG_diagram}\textcolor{red}{(d)}.  
\subsection{Macroscopic Boltzmann weight}
\subsubsection{Energy Functional and Non-conservative Forces}\label{sec:macroscopic_energy_functional}
The macroscopic Boltzmann weight $\mu_i$ of the macrostate $\rho_i$ is identified using the LDB \cite{atm_cg_nr_2024}. $\mu_i$ is the macroscopic Boltzmann weight of the macrostate $\rho_i$, quantifying the total thermodynamic cost of placing a macrostate $\rho_i$.  It consists of an ideal (entropic) part $\mu_i^{id} = \ln{(\rho_i)}$ and a reciprocal interaction (energetic) part $\mu_i^{int} = \sum_j {V}_{ij} \rho_j$, such that $\mu_i = \mu_i^{id} + \mu_i^{int}$. Here, $V_{ij} = \Omega \left( e^{\beta v_{ij}^r + \beta v^{nr}_{ij}} - 1 \right)$ quantifies the contribution to $\mu_i$ due to the presence of $\rho_j$ arising from the interaction between $\rho_i$ and $\rho_j$. $V_{ij}$ serves as the second Virial coefficient for the interacting non-reciprocal fields, such that $V_{ij} \neq V_{ji}$. $\mu_i^{int} < 0$ indicates attractive interactions experienced by $\rho_i$, whereas $\mu_i^{int} > 0$ corresponds to repulsive interactions. Analogously, $\mu_i$ can be decomposed into reciprocal and non-reciprocal macroscopic Boltzmann weights:
\begin{equation}\label{eq:macroscpic_boltzmann_weight}
\begin{split}
    \mu_{i}^r &= \ln{\rho_i} + \sum_j {V}_{ij}^r \rho_j,\\
    {F}^{nr}_{i} &= \sum_j {V}_{ij}^{nr} \rho_j,
\end{split}    
\end{equation}
where $V_{ij}^r$ and $V_{ij}^{nr}$ denote the reciprocal and non-reciprocal interaction coefficients for $\rho_i$ due to $\rho_j$, respectively:
\begin{equation}\label{eq:macroscopic_interaction_coefficeints}
\begin{split}
    {V}_{ij}^r &= \Omega \left( \cosh{\left(\beta v^{nr}_{ij}\right)} e^{\beta v_{ij}^r} - 1 \right),\\
    {V}_{ij}^{nr} &= \Omega \sinh{ \left( \beta v_{ij}^{nr} \right) } e^{\beta v_{ij}^r}.
\end{split}    
\end{equation}
\Cref{table:macroscopic_single_particle_notation} summarizes the notation used for macroscopic interacting fields. Importantly, $V_{ij}^r = V_{ji}^r$ and $V_{ij}^{nr} = -V_{ji}^{nr}$, as required by $v_{ij}^r = v_{ji}^r$ and $v_{ij}^{nr} = -v_{ji}^{nr}$. This ensures that the microscopic `orthogonal gauge fixing' is scale-invariant, guaranteeing a unique and exact decomposition of reciprocal and non-reciprocal contributions at the coarse-grained macroscopic level. Alternative gauge choices would violate scale invariance and lead to physically incorrect decompositions of the EPR.

The non-linear dependence of $V_{ij}^r$ and $V_{ij}^{nr}$ arises from incorporating the Poissonian statistics of the microscopic occupancy variables \cite{atm_cg_nr_2024}. The reciprocal macroscopic weight $\mu_i^r$ is derived from the global macroscopic energy functional $E$, defined such that $\mu_i^r = \delta E/\delta \rho_i$:
\begin{equation}\label{eq:macro_energy_functional}
\begin{split}
    E = \int_{\mathcal{V}} \left[ 
    \frac{1}{2} \sum_{i, j} V_{ij}^r \rho_i \rho_j 
    + \sum_i \rho_i \ln{\left(\frac{ \rho_i }{e}\right)} 
    \right].
\end{split}    
\end{equation}
$E$ quantifies the total reciprocal interaction energy $E^{int}$ and the macroscopic Boltzmann entropic term $S^b$, such that $E = E^{int} - S^b$. Consequently, $\mu_i = \mu_i^r + F_i^{nr}$ represents a macroscopic Helmholtz-Hodge decomposition of $\mu_i$ into symmetry-preserving ($\mu_i^r$) and symmetry-violating ($F_i^{nr}$) components. The macroscopic non-conservative driving forces along reactive and diffusive transitions are related to their microscopic counterparts as $F^{ch}_{\gamma\gamma'} = f_{\gamma\gamma'}^{ch}$ and $\vec{F}^{sp}_{i} = l\vec{f}_{i}^{sp}$. 

\subsubsection{Equilibrium energy functional}\label{sec:mesoscopic_eq_energy_functional}

The functional $E$ is not the equilibrium energy functional $E^{eq}$. Setting $v_{ij}^{nr} = 0$, $\vec{f}_i^{sp} = 0$, and $f_{\gamma \gamma'}^{ch} = 0$ yields $E^{eq} = E(V_{ij}^{eq})$, where $V_{ij}^{eq} = \Omega \left[ e^{\beta v_{ij}^r} - 1 \right]$ is the second Virial coefficient for equilibrium reciprocal interactions. Notably, $E$ satisfies $|E| \geq |E^{eq}|$, with equality only if $v_{ij}^{nr} = 0$ for all $i,j$. Unlike $E^{eq}$, the functional $E$ incorporates non-reciprocal interactions and their effect on symmetric macrostate correlations through the $\cosh(\beta v_{ij}^{nr})$ term in \cref{eq:macroscopic_interaction_coefficeints}. 
{
$E$ is the macroscopic analogue of \cref{eq:microscopic_energy_functional}. Analogously highlighted in \cref{sec:microscopic_model}, $E$ corresponds to the geometric decomposition of reciprocal interactions that generates the conservative reciprocal interactions through the Helmholtz-Hodge decomposition of $\mu_i$; therefore, $E$ serves as a Lyapunov functional for non-reciprocal systems in the absence of external driving forces and vanishing non-reciprocal vorticity currents. The proof follows trivially from the decomposition of macroscopic mean EPR \cref{eq:macroscopic_net_EPR_final,eq:macroscopic_net_EPR_final_orthogonal}, the second law of thermodynamics, and vanishing EPR cost associated with external driving forces and vanishing non-reciprocal vorticity currents. 
}
\subsection{Dynamics}\label{sec:macroscopic_dynamics}
\subsubsection{Local Detailed Balance}
The macroscopic reactive and diffusive transitions are defined as $\Delta_{\gamma \gamma'}: \rho_{\gamma'} \to \rho_{\gamma}$ and $\Delta_{i}^{\vec{\mathcal{D}}}: \rho_i \to \rho_i^{\vec{\mathcal{D}}}$, respectively, where ${\vec{{\mathcal{D}}}}$ quantifies the direction vector for the diffusive transition. 
{Using the macroscopic transition probability measure from the DPFT \cite{atm_cg_nr_2024}, the macroscopic transition rates for the reactive ($\Delta_{\gamma \gamma'}$) and diffusive (${\Delta}_{i}^{\vec{\mathcal{D}}}$) transitions are:
\begin{equation}
\begin{split}\label{eq:macroscopic_transition_rate}
    K_{\gamma\gamma'} &= d_{\gamma\gamma'} e^{\mu_{\gamma'} + \frac 1 2 F_{\gamma\gamma'}^{ch}}, 
    \hspace{0.5cm}
    K_i^{\vec{\mathcal{D}}} = \tilde{d}_i^{\vec{\mathcal{D}}} e^{ \mu_{i} + \frac 1 2 \vec{\mathcal{D}} \cdot \vec{F}_{i}^{sp} }.
\end{split}    
\end{equation}
Here, $\tilde{d}_i^{\vec{\mathcal{D}}} = d_i l^2 \cosh{( \vec{\mathcal{D}} \cdot \vec{F}_i^{sp} / 2 )}$ is the macroscopic diffusion constant along $\vec{\mathcal{D}}$, measured using the microscopic diffusive length scale $l$. The local detailed balance (LDB) constructed from \cref{eq:macroscopic_transition_rate} for the reactive ($\Delta_{\gamma \gamma'}$) and diffusive (${\Delta}_{i}^{\vec{\mathcal{D}}}$) transitions reads \cite{atm_cg_nr_2024}: }
\begin{equation}\label{eq:macroscopic_local_detailed_balance_reactive}
\begin{split}
    \frac{K_{\gamma\gamma'}}{K_{\gamma'\gamma}} &= e^{ \mu_{\gamma'} - \mu_{\gamma} + F_{\gamma \gamma'}^{ch} },
    \hspace{0.5cm}
    \frac{K_i^{\vec{\mathcal{D}}}}{K_i^{(\vec{\mathcal{D}})^{-1}}} 
    = e^{ \mu_{i} - \mu_i^{\vec{\mathcal{D}}} + \vec{\mathcal{D}} \cdot \vec{F}_{i}^{sp} }.
\end{split}
\end{equation}
The LDB constrains the ratio of forward and backward transitions (a dynamic quantity) using $\mu_i$ (a thermodynamic quantity). For $\Delta_{\gamma \gamma'}$, we define the macroscopic affinity $A_{\gamma\gamma'} = \mu_{\gamma'} - \mu_{\gamma} + F_{\gamma \gamma'}$ and its symmetric counterpart $U_{\gamma \gamma'} = \mu_{\gamma'} + \mu_{\gamma}$. Macroscopic and microscopic external chemical driving forces are related by ${F}_{\gamma\gamma'}^{ch} =  {f}_{\gamma \gamma'}^{ch}$. Similarly, for ${\Delta}_{i}^{\vec{\mathcal{D}}}$, we define ${A}_{i}^{\vec{\mathcal{D}} } = \mu_{i} - \mu_{i}^{\vec{\mathcal{D}}} + \vec{\mathcal{D}} \cdot \vec{F}_i^{sp}$ and ${U}_{i}^{\vec{\mathcal{D}}} = 2 \mu_{i}$. Macroscopic and microscopic self-propulsion are related by $\vec{F}_i^{sp} = l \vec{f}_i^{sp}$. 
{\Cref{eq:macroscopic_transition_rate,eq:macroscopic_local_detailed_balance_reactive} are the macroscopic analogs of the microscopic \cref{eq:microscopic_transition_rate,eq:microscopic_local_detailed_balance_final}, respectively. Importantly, \cref{eq:macroscopic_transition_rate} provides a more fundamental formulation than \cref{eq:macroscopic_local_detailed_balance_reactive}, as it does not assume a timescale separation between different transition rates. Since \cref{eq:macroscopic_local_detailed_balance_reactive} omits the time-symmetric part of the transition rates by taking the ratio of forward and backward transitions, it is analogous to the microscopic counterpart discussed previously.}
\subsubsection{Generalized Macroscopic Fluctuating Dynamics}\label{sec:macroscopic_fluctuation_theory}
The deterministic evolution of $\rho_i$ is given by the `most likelyhood path' of the Doi-Peliti action, namely the `Instanton'. The `dominant macrostate fluctuations' are characterized by the local curvature of the `Instanton' \cite{atm_cg_nr_2024}. The stochastic equation of motion (EOM) for $\rho_i$ is \cite{atm_cg_nr_2024}:
\begin{equation}\label{eq:macroscopic_eom_mft}
\begin{split}
    \partial_t \rho_i 
    &= -\nabla \cdot \vec{J}_{i}^{\vec{\mathcal{D}}}
    - \sum_{ i \in \{ \Delta_{\gamma \gamma'} \} } J_{\gamma \gamma'} 
    + \nabla \cdot \left(
    \sqrt{2 B_i^{\mathcal{D}} } 
    \: \vec{ \hat{\xi} }_i^{\mathcal{D}} \right)
    \\
    & \hspace{2cm} + \sum_{ i \in \{ \Delta_{\gamma \gamma'} \} } \sqrt{2 B^{\mathcal{R} }_{\gamma \gamma'} } 
    \: \hat{\xi}^{ \mathcal{R} }_{\gamma \gamma'}.
\end{split}    
\end{equation}
We adopt the convention that outward transition currents from $\rho_i$ ($-J_{\gamma \gamma'}$ and $-\nabla \cdot \vec{J}_{i}^{\vec{\mathcal{D}}}$) are represented with a negative sign. $\hat{\xi}_{\gamma\gamma'}^{\mathcal{R}}$ and $\hat{\xi}_{i}^{\mathcal{D}}$ are standard Gaussian white noises with unit variance and zero mean. The set $\{ \Delta_{\gamma \gamma'} \}$ denotes all macroscopic reactive transitions. The reactive and diffusive transition currents are:
\begin{equation}\label{eq:macroscopic_eom_mean_current_mft}
\begin{split}
    \vec{J}_{i}^{\vec{\mathcal{D}}} 
    & = - D_i^{\vec{{\mathcal{D}}}} \left( \left\{ \rho\right\} \right)
    \nabla^{\vec{\mathcal{D}}} \mu_i
    + \vec{J}_i^{sp}, 
    \\
    J_{\gamma \gamma'} 
    & = 2 D_{\gamma \gamma'} \left( \{ \rho \} \right) 
    \sinh{ \left( \frac 
    { A_{\gamma \gamma'}}{2} \right) }.
\end{split}    
\end{equation}
Here, $\nabla^{\vec{\mathcal{D}}}$ is the gradient along $\vec{\mathcal{D}}$, and similarly, the Laplacian along $\vec{\mathcal{D}}$ is denoted by $\Delta^{\vec{\mathcal{D}}}$. The macroscopic self-propulsion current is $\vec{J}_i^{sp} = 2 d_i l e^{\mu_i} \sinh{ ( \vec{\mathcal{D}} \cdot \vec{F}_i^{sp}/2 ) }$. Choosing the basis vector $\vec{\mathcal{D}} = \{ \parallel, \perp \}$ as parallel and perpendicular to $\vec{F}_i^{sp}$, the simplified form reads $-\nabla \cdot \vec{J}_i^{\vec{\mathcal{D}}} = \tilde{d}_i^{\parallel} \Delta^{\parallel} e^{\mu_i} + \tilde{d}_i^{\perp} \Delta^{\perp} e^{\mu_i} + \nabla^{\parallel} \cdot \vec{J}_i^{sp}$. Here, $\tilde{d}_i^{\parallel} = d_i l^2 \cosh{(F_i^{sp}/2)}$ and $\tilde{d}_i^{\perp} = d_i l^2$, indicating that self-propulsion renormalizes the diffusion coefficients differently in directions parallel and perpendicular to the self-propulsion force. Without loss of generality, we consider the scaled macroscopic diffusion constants $\tilde{d}_i = d_i l^2$ or equivalently fix $l=1$, using the diffusive length scale as the unit of spatial distance. In the continuum limit $l \to 0$, $\tilde{d}_i^\parallel = \tilde{d}_i^\perp = \lim_{l \to 0} d_i l^2$ and $J_i^{sp} = \tilde{d}_i^\parallel e^{\mu_i} f_i^{sp}$ \cite{atm_cg_nr_2024}, so the macrostates lose the directional dependence of the diffusion coefficients. The anisotropic diffusion coefficients, however, have been shown to generate novel phases \cite{Chatterjee_2020,Mangeat_2020,Chatterjee_2022,Karmakar_2023,Woo_2024}, highlighting their importance.

We define the mobility for diffusive and reactive transitions as:
\begin{equation}\label{eq:macroscopic_eom_mobility}
\begin{split}
    D_i^{\vec{\mathcal{D}}}\left( \{ \rho \} \right) = \tilde{d}_i^{\vec{\mathcal{D}}} e^{ \mu_i },
    \hspace{0.5cm}
    D_{\gamma \gamma'} \left( \{ \rho \} \right) 
     & = d_{\gamma \gamma'} e^{ \left( U_{\gamma \gamma'}/2 \right) }.
\end{split}    
\end{equation}
Mobilities characterize the strength of macroscopic transition currents generated by the macroscopic transition affinities. They play the same role as diffusive mobility and transport coefficients for $\vec{J}_i^{\vec{\mathcal{D}}}$ and $\vec{J}_i^{sp}$, respectively. For repulsive interactions experienced by $\rho_i$ (i.e., $\mu_i > 0$), the amplitude of the transition mobilities increases exponentially, signifying escape from a thermodynamically unfavorable state. Similarly, the transition currents are exponentially suppressed for $\rho_i$ experiencing attractive interactions ($\mu_i < 0$). The variances of the reactive and diffusive currents are \cite{atm_cg_nr_2024}:
\begin{equation}\label{eq:eom_mft_variance_currents}
\begin{split}
2B_i^{\vec{\mathcal{D}}} \left( \{ \rho \} \right) 
& = \frac 2 \Omega D_i^{\vec{\mathcal{D}}} \left( \{ \rho \} \right), 
\\
2B_{\gamma \gamma'} \left( \{ \rho \} \right)
& = \frac 2 \Omega D_{\gamma \gamma'} \left( \{ \rho \} \right)
\cosh{ \left( \frac 
{ A_{\gamma \gamma'} }{2} \right) }.    
\end{split}    
\end{equation}
\Cref{eq:eom_mft_variance_currents} is the Einstein-Smoluchowski relation connecting current fluctuations to transition mobility, where $\Omega$ plays a role analogous to the inverse temperature $\beta$ \cite{Touchette_2009,atm_cg_nr_2024}. Importantly, the same mobility controls the mean current in \cref{eq:macroscopic_eom_mean_current_mft} through a hyperbolic relation, which gives rise to the fluctuation-response relation. 

We define the traffic as the sum of the magnitudes of the forward and backward currents \cite{Maes_2020,Vlad_1994_1,Vlad_1994_2,Vlad_1994_3,Ross_1999}. For reactive and diffusive transitions, the traffic reads:   
\begin{equation}\label{eq:macroscopic_traffic}
\begin{split}
    T_{\gamma \gamma'}
    = 2 D_{\gamma \gamma'} \left( \{ \rho \} \right) 
    \cosh{ \left( \frac 
    { A_{\gamma \gamma'} }{2} \right) },
    \hspace{0.7cm}
    T_{i}^{\vec{\mathcal{D}}}  
    = 2 D_{i} \left( \{ \rho \} \right). 
\end{split}    
\end{equation}
Traffic in \cref{eq:macroscopic_traffic} is related to the scaled variance of the transition currents in \cref{eq:eom_mft_variance_currents}: $T_{\gamma \gamma'} = 2 \Omega B_{\gamma \gamma'} \left( \{ \rho \} \right)$ and $T_{i}^{\vec{\mathcal{D}}}  = 2 \Omega B_{i}^{\vec{\mathcal{D}}} \left( \{ \rho \} \right)$. 

{The mean current and traffic (\cref{eq:macroscopic_eom_mean_current_mft,eq:macroscopic_traffic}) exhibit $\sinh()$ and $\cosh()$ forms due to their thermodynamically consistent exponential parametrization in the affinity $A$. However, they have a fundamental thermodynamic meaning independent of exponential parametrization: namely, the mean current and traffic correspond to the time-antisymmetric and time-symmetric components of the non-equilibrium currents \cite{atm_2024_var_epr,atm_2025_var_epr_derivation}. Consequently, they characterize the thermodynamic length and the dynamical activity for the transition, which physically correspond to the total directional transition current and the inverse timescale for the transition \cite{atm_2024_var_epr,atm_2025_var_epr_derivation}. The shortcomings of the Glauber-dynamics and Metropolis-dynamics discussed in \cref{sec:microscopic_dynamics} are better understood with \cref{eq:macroscopic_eom_mft}, where different transitions contribute with different `diffusivity' or `variance', quantified by the traffic. This exactly preserves the different timescales associated with different transitions that contribute to the macrostate $\rho_i$. Therefore, due to the thermodynamically consistent exact coarse-graining procedure \cite{atm_cg_nr_2024}, \cref{eq:macroscopic_eom_mft} serves as a foundational cornerstone for future studies to investigate macroscopic stochastic thermodynamics without strong assumptions on timescale separation.}

Although \cref{eq:macroscopic_eom_mft} is written in the continuum macroscopic limit $l \to 0$, equivalent analogs exist for systems with finite $l$, e.g., lattice gas models with $l=1$. In such cases, the discrete analogs of the Laplacian and gradient operators replace the continuum-space counterparts, similar to reactive transition dynamics \cite{atm_cg_nr_2024}. 
\subsection{Thermodynamics}\label{sec:macroscopic_thermodynamics}
\subsubsection{Macroscopic Thermodynamics}\label{sec:macroscopic_thermodynamics_subsec}
The total mean macroscopic EPR is $\langle \dot{\Sigma} \rangle = \langle \dot{\Sigma}^{\mathcal{D}} \rangle + \langle \dot{\Sigma}^{\mathcal{R}} \rangle + d_t S^{gb}$. It is decomposed into the reactive mean macroscopic bulk EPR ($\langle \dot{\Sigma}^{\mathcal{R}} \rangle$), the diffusive mean macroscopic bulk EPR ($\langle \dot{\Sigma}^{\mathcal{D}} \rangle$), and the boundary term, i.e., the Gibbs EPR ($d_t S^{gb}$):
\begin{equation}\label{eq:macroscopic_diffusive_reactive_gibbs_EPR}
\begin{split}
    \langle \dot{\Sigma}^{\mathcal{D}} \rangle
    & = \int_{\mathcal{V}} \sum_i \langle \vec{J}_i^{\vec{\mathcal{D}}} \cdot \left( - \nabla^{\vec{\mathcal{D}}} \mu_i + \vec{F}_i^{sp} \right) \rangle,
    \\
    \langle \dot{\Sigma}^{\mathcal{R}} \rangle 
    & = \int_{\mathcal{V}} \sum_{\{  \Delta_{\gamma\gamma'} \}} \langle J_{\gamma \gamma'} \left( \mu_{\gamma'} - \mu_{\gamma} + F_{\gamma \gamma'}^{ch} \right) \rangle,
    \\
    d_t S^{gb} & = - \int \mathbb{D}[\{\rho\}] \, d_t \mathcal{P}[\{\rho\}] \ln{( \mathcal{P}[\{ \rho \}] )}.
\end{split}    
\end{equation}
Here, $\mathbb{D}[\{\rho\}]$ denotes the path integral over the macrostate space, and $\mathcal{P}[\{ \rho \}]$ is the probability distribution for the macrostate. The mean EPR in \cref{eq:macroscopic_diffusive_reactive_gibbs_EPR} has the form of a transition current multiplied by the corresponding transition affinity. The transition affinity is obtained using the macroscopic local detailed balance (LDB) condition in \cref{eq:macroscopic_local_detailed_balance_reactive}. The macroscopic Gibbs entropy is defined as $S^{gb} = - \int \mathbb{D}[\{ \rho \}] \mathcal{P}[\{ \rho \}] \ln{( \mathcal{P}[\{ \rho \}] )}$, which is related to the macroscopic state entropy $S^{state}$, defined as \cite{Seifert_2005,seifert_2012}:
\begin{equation}\label{eq:macroscopic_state_entropy}
\begin{split}
    S^{state} = -\ln{(\mathcal{P} [\{\rho\}])},
\end{split}    
\end{equation}
so that $\langle S^{state} \rangle = S^{gb}$ holds.
\subsubsection{Conservative and non-conservative decomposition of the total macroscopic EPR}\label{sec:macroscopic_thermodynamics_conservative_non_c_decomposition}
$\langle \dot{\Sigma} \rangle$ is decomposed into four contributions, analogous to \cref{eq:microscopic_net_EPR_final}:
\begin{equation}\label{eq:macroscopic_net_EPR_final}
\begin{split}
    \langle \dot{\Sigma} \rangle 
    & = - d_t \langle E \rangle
    + d_t S^{gb}
    + \langle \dot{\Sigma}^{nr} \rangle
    + \langle \dot{\Sigma}^{ch} \rangle
    + \langle \dot{\Sigma}^{sp} \rangle . 
\end{split}
\end{equation}
\Cref{eq:macroscopic_net_EPR_final} represents the macroscopic second law of thermodynamics. The term $- d_t \langle E \rangle + d_t S^{gb}$ corresponds to the conservative EPR $\langle \dot{\Sigma}^c \rangle$, so $\Sigma^c$ depends only on the initial and final states. The exact expressions for $\langle \dot{\Sigma}^{ch} \rangle$ and $\langle \dot{\Sigma}^{sp} \rangle$ are:
\begin{equation}\label{eq:macroscopic_chemical_driving_EPR}
\begin{split}
    \langle \dot{\Sigma}^{ch} \rangle 
    = \int_{\mathcal{V}} \sum_{ \{ \Delta_{\gamma\gamma'} \}} \langle J_{\gamma \gamma'} \rangle F^{ch}_{\gamma \gamma'}, 
    \hspace{0.5cm}
    \langle \dot{\Sigma}^{sp} \rangle 
    = \int_{\mathcal{V}} \sum_{ i } \langle \vec{J}_i^{\vec{\mathcal{D}}} \rangle \cdot \vec{F}^{sp}_{i}.
\end{split}
\end{equation}
The non-reciprocal EPR $\langle \dot{\Sigma}^{nr} \rangle$ consists of reactive 
$\langle \dot{\Sigma}_{\mathcal{R}}^{nr} \rangle = \int_{\mathcal{V}} \sum_{ \{ \gamma \gamma' \} } \langle J_{\gamma \gamma'} \left( F^{nr}_{\gamma'} - F^{nr}_{\gamma} \right) \rangle$ and diffusive contributions $\langle \dot{\Sigma}^{nr}_{\mathcal{D}} \rangle = -\int_{\mathcal{V}} \sum_{ i } \langle \vec{J}_i^{\vec{\mathcal{D}}} \cdot \nabla F^{nr}_{i} \rangle$, such that $\dot{\Sigma}^{nr} = \dot{\Sigma}_{\mathcal{R}}^{nr} +  \dot{\Sigma}^{nr}_{\mathcal{D}}$. Using the EOM \cref{eq:macroscopic_eom_mft} leads to $\dot{\Sigma}^{nr} = - \sum_i \int_{\mathcal{V}} F_i^{nr} \: \partial_t \rho_i $. With $V_{ij}^{nr} = -V_{ji}^{nr}$, this further simplifies to:
\begin{equation}\label{eq:macroscopic_epr_non_reciprocal}
\begin{split}
    \langle \dot{\Sigma}^{nr} \rangle 
    &  = \int_{\mathcal{V}} \sum_{i,j} V_{ij}^{nr} \langle \omega_{ij} \rangle.
\end{split}    
\end{equation}
Here, $\omega_{ij} = \rho_i \partial_t \rho_j -  \rho_j \partial_t \rho_i$ defines the macroscopic vorticity between $\rho_i$ and $\rho_j$. $\langle {\Sigma}^{nr} \rangle$ corresponds to the sustained vorticity currents between the macrostates generated by the macroscopic non-reciprocal forces. Importantly, the vorticity currents are defined in the macrostate space, in contrast to the dissipative currents defined in the transition space. Moreover, vorticity currents are defined using two macrostates, unlike chemical reaction networks, which require a transition cycle of three states to formulate a dissipative current \cite{schnakenberg_1976}. From a fundamental perspective, this reveals a sophisticated mechanism of non-equilibriumness in non-reciprocal systems, with no equivalent counterparts in the known literature.

\subsubsection{Non-reciprocal phase transitions are dynamical phase transitions} 
The dynamics of non-reciprocal systems exhibit dynamic phases due to $\mathcal{PT}$ symmetry breaking \cite{Fruchart2021}, for instance, traveling waves and temporal oscillations. Here, we focus on non-reciprocal phase transitions that exhibit a transition from a static to a dynamic phase. A static phase is defined as the $\mathcal{PT}$ symmetry-preserving phase; in contrast, the $\mathcal{PT}$ symmetry is broken in the dynamic phase. We consider a wave solution $\rho_i = \rho_i^{avg} + \phi_i^{ow} \cos{(\theta_i^{ow} - v^{ow} t )} + \phi_i^{tw} \cos{(x-v^{tw}t + \theta_i^{tw})}$. Here, $\rho_i^{avg}$ is the time-integrated and space-integrated average value of $\rho_i$, and $w = \{tw, ow\}$ denotes the traveling wave and the temporal oscillating wave, respectively. The $\phi_i^{ow}$ and $\phi_i^{tw}$ characterize the wave amplitudes. Similarly, $v^{ow}$ and $v^{tw}$ are the wave velocities. Integrating \cref{eq:macroscopic_epr_non_reciprocal} over the periodic space and oscillation time period $\tau^{ow}$, the non-reciprocal EP for the system reads:
\begin{equation}\label{eq:macroscopic_epr_non_reciprocal_wave_ansatz}
\begin{split}
    \langle {\Sigma}^{nr} \rangle 
    & =
    \mathcal{V} \tau^{ow} \sum_{i,j,w} V_{ij}^{nr}  v^w \phi_i^w \phi_j^w \sin{( \theta_j^w - \theta_i^w )}.
\end{split}    
\end{equation}
\Cref{eq:macroscopic_epr_non_reciprocal_wave_ansatz} quantifies the non-reciprocal EP density over the total volume and time period. It has the physical meaning of the kinetic energy for a wave, where the mass is given by $v^w, V_{ij}^{nr}$ and $\sin{( \theta_j^w - \theta_i^w )}$, and the velocity is given by the wave amplitudes $\phi_i^w$. For $V_{ij}^{nr} \geq 0$, it physically implies that $\rho_i$ repels $\rho_j$ while $\rho_j$ is attracted to $\rho_i$, thus $\theta_j^{tw} \geq \theta_i^{tw}$ and $\theta_j^{ow} \geq \theta_i^{ow}$, which further implies $\langle {\Sigma}^{nr} \rangle \geq 0$. Importantly, the out-of-phase state $\theta_j^{tw} - \theta_i^{tw} = \pi/2$ maximizes $\langle {\Sigma}^{nr} \rangle$, while the in-phase state $\theta_j^{tw} = \theta_i^{tw}$ minimizes $\langle {\Sigma}^{nr} \rangle$. The in-phase to out-of-phase steady-state transition has been a key motif of non-reciprocal systems \cite{Fruchart2021}. Hence, it signifies that the transition from the in-phase to out-of-phase steady-state is equivalent to switching from the minimum to the maximum non-reciprocal EPR for steady-state selection. Moreover, $\langle \Sigma^{nr} \rangle$ for out-of-phase oscillation scales with the observation time $\tau$. This signifies a constant thermodynamic dissipation cost required to sustain the vorticity currents between the macrostates. Physically, it connects the dynamical phase transition to an analogy with equilibrium phase transitions, where the EPR for the non-equilibrium phase transition is analogous to the free energy for the equilibrium phase transition \cite{Ge_2009_prl}. It is characterized by different scaling regimes for $\langle \Sigma^{nr} \rangle$. The dissipative nature of the vorticity currents is realized only in the dynamic phase. Importantly, non-reciprocal phase transitions are equivalent to dynamical phase transitions and are characterized by non-analytical behavior of $\langle \Sigma^{nr} \rangle$.

The scaling of different contributions of $\Sigma$ with the observation time $\tau$ is summarized in \cref{table:EPR_scaling}. The self-propulsion and chemical driving EP follow scaling $O(\tau)$ for dissipative currents \cite{atm_2024_var_epr,atm_2025_var_epr_derivation}. However, a discontinuity (or a kink) in the self-propulsion or chemically driven EPR can be observed and is associated with a dynamical phase transition \cite{Ge_2009_prl,Vellela_2009,Ge_2011_JRSI,Nguyen_2018}. The non-reciprocal phase transitions are characterized by different scaling regimes for $\Sigma^{nr}$, which correspond to sustaining temporal or spatial vorticity currents.
\begin{table}[t!]
\begin{tabular}{|P{1.3cm}|P{1.3cm}|P{1.3cm}|P{1.3cm}|P{1.3cm}|}
     \hline
     \textbf{Phase}
     & $\Sigma^{c}$
     & $\Sigma^{nr}$ 
     & $\Sigma^{sp}$ 
     & $\Sigma^{ch}$ 
     \\
     \hline
     Static 
     & $O(1)$
     & $O(1)$
     & $O(\tau)$
     & $O(\tau)$
     \\
     \hline
     Dynamic
     & $O(1)$
     & $O(\tau)$
     & $O(\tau)$
     & $O(\tau)$ 
     \\
     \hline
\end{tabular}
\caption{Scaling of macroscopic mean EP as a function of the observation time $\tau$. }
\label{table:EPR_scaling}
\end{table}
\subsubsection{Orthogonal decomposition of the EPR}\label{sec:macroscopic_EPR_orthogonal}
We define the macroscopic relative state entropy: 
\begin{equation}\label{eq:macroscopic_state_entropy_orthogonal}
\begin{split}
    S^{state}_{E} = -\ln{\left( \frac{\mathcal{P} [\{\rho\}]}{\mathcal{P}^{E} [\{\rho\}]} \right)}.
\end{split}    
\end{equation}
Using the definition in \cref{eq:macroscopic_state_entropy_orthogonal}, $- d_t \langle {E} \rangle + d_t S^{gb}$ in \cref{eq:macroscopic_net_EPR_final} is further decomposed into the orthogonal form \cite{Qian_2001}:
\begin{equation}\label{eq:macroscopic_net_EPR_final_orthogonal}
\begin{split}
    \langle \dot{\Sigma} \rangle 
    &
    = - d_t\psi_{E} - d_t D^{KL}_{{E}}(\mathcal{P} [\{\rho\}])
    + \langle \dot{\Sigma}^{nr} \rangle
    + \langle \dot{\Sigma}^{ch} \rangle
    + \langle \dot{\Sigma}^{sp} \rangle. 
\end{split}
\end{equation}
$\psi_E$ is the macroscopic free energy defined as $\psi_E = - \ln{(\mathcal{Z}^E)}$. Here, $\mathcal{Z}^E = \int \mathbb{D}[\{\rho\}] e^{ - E }$ is the partition function, and the macroscopic Boltzmann distribution is ${\mathcal{P}^E [\{\rho\}]} = e^{ - E \: + \Psi^E }$. It holds that $\langle S^{state}_{E} \rangle = - D^{KL}_{{E}}(\mathcal{P} [\{\rho\}])$. Using the symmetry of the non-conservative external driving forces $F_{\gamma \gamma'}$ and $\vec{F}_i^{sp}$, namely the orthogonal decomposition of the transition affinities, \cref{eq:macroscopic_chemical_driving_EPR} is reduced to:
\begin{equation}\label{eq:macroscopic_driving_EPR_orthogonal}
\begin{split}
    \langle \dot{\Sigma}^{ch} \rangle 
    = \int_{\mathcal{V} } \sum_{\{ \Delta_{\gamma\gamma'} \}} 
    \langle
    J_{\gamma \gamma'}^{ch}
    \rangle
    \:
    F_{\gamma \gamma'}^{ch} 
    ,
    \hspace{0.3cm}
    \langle \dot{\Sigma}^{sp} \rangle
    = \int_{\mathcal{V}} \sum_{ i } \langle \vec{J}_i^{sp} \rangle \: \cdot \vec{F}_i^{sp} .
\end{split}    
\end{equation}
Here, $\vec{J}_i^{sp} = 2 d_i e^{\mu_i} \sinh{(\vec{F}_i^{sp} / 2)} = T_i^{\perp} \sinh{(\vec{F}_i^{sp} / 2)}$ is the anti-symmetric part of $\vec{J}_i^{\vec{\mathcal{D}} }$ under the adjoint transformation $\vec{F}_i^{sp} \to -\vec{F}_i^{sp}$. Similarly, $J_{\gamma \gamma'}^{ch} = d_{\gamma \gamma'} \left( e^{\mu_\gamma} + e^{\mu_{\gamma'}} \right) \sinh{ \left( {F_{\gamma \gamma'}^{ch}}/{2} \right) } = T^\perp_{\gamma \gamma'} \sinh{ \left( {F_{\gamma \gamma'}^{ch}}/{2} \right) }$ is the anti-symmetric part of $J_{\gamma \gamma'}$ under the adjoint transformation $F_{\gamma \gamma'}^{ch} \to - F_{\gamma \gamma'}^{ch}$. Importantly, $T_i^{\perp}$ and $T^\perp_{\gamma \gamma'}$ are the scaled (by $\Omega$) variances of currents (\cref{eq:eom_mft_variance_currents}) in the direction orthogonal to the external driving, which are obtained by plugging in $\vec{F}_i^{sp} = 0$ and $F_{\gamma \gamma'}^{ch} = 0$, respectively. Thus, in \cref{eq:macroscopic_driving_EPR_orthogonal}, the non-conservative EPR depends on a non-linear function of the external driving force and the current variance in the direction orthogonal to the driving. Importantly, $\langle \dot{\Sigma}^{ch} \rangle$ and $ \langle \dot{\Sigma}^{sp} \rangle$ depend on $e^{\mu_i}$ and are proportional to the macrostate mobility.

\Cref{eq:macroscopic_net_EPR_final_orthogonal} formulates the orthogonal decomposition of $\langle \dot{\Sigma} \rangle$. It decomposes $\langle \dot{\Sigma} \rangle$ into four linearly independent components. First, $- d_t\psi_{E}$ quantifies the rate of change of the free energy due to the external work needed to drive $E$ through the change in control parameters. Second, $ - d_t D^{KL}_{{E}}(\mathcal{P} [\{\rho\}])$ quantifies the EPR due to the relaxation of $E$. Third, $\langle\dot{\Sigma}^{nr} \rangle$ is the non-reciprocal EPR, whose case-specific simplifications for the $\mathcal{PT}$ preserving-breaking phases have been discussed before. Fourth, $\langle \dot{\Sigma}^{ch} \rangle + \langle \dot{\Sigma}^{sp} \rangle$ quantifies the EPR due to the non-conservative forces along the transitions, namely the self-propulsion and chemical driving for the diffusive and reactive systems, respectively.

The orthogonal decomposition for dissipation functions that are quadratic in the driving affinity, and thus a linear relationship between the current and the affinity, has been proven and rigorously studied \cite{Bertini_2002,Bertini_2010,Bertini_2015,Qian_2020}. In contrast, the non-linear relation between $J_{\gamma \gamma'}^{ch}$ and $F_{\gamma \gamma'}^{ch}$ (or  $\vec{J}_i^{sp}$ and $\vec{F}_i^{sp}$) leads to non-quadratic dissipation functions in \cref{eq:macroscopic_driving_EPR_orthogonal}. It gives exact and tighter bounds on the EPR. A more rigorous proof of the orthogonal decomposition for non-quadratic dissipation functions has been derived in Refs. \cite{Mielke_2017,Kaiser_2018,Peletier_2022,Renger_2021,Patterson_2024,Peletier_2023}, and its implications have been studied in Refs. \cite{kobayashi_2022_hessian_geometry,Kobayashi_2022,Sughiyama_2022,kobayashi_2023_information_graphs_hypergraphs,Loutchko_2023_geometry_tur,Renger_2023,duong_2023,Mizohata_2024}. The proof relies on the dynamical large deviation approach \cite{maes_2007,Maes_2008,Maes_2008_ldp_statistics}. In Ref.\cite{atm_cg_nr_2024}, we show that the large deviation functional of the non-reciprocal systems is the same as the one used in Refs. \cite{Mielke_2017,Kaiser_2018,Peletier_2022,Renger_2021,Patterson_2024,Peletier_2023}. This rigorously justifies the orthogonal decomposition for non-reciprocal systems. In contrast to previous works, the novelty of our approach lies in the proposal of orthogonal decomposition in state-space ($\mu_i$ and $\epsilon_i^\#$) and transition-space ($F_{\gamma \gamma'}^{ch}, \vec{F}_{i}^{sp}$ and $f_{\gamma \gamma'}^{ch}, \vec{f}_{i}^{sp}$) for both microscopic and macroscopic systems. The anti-symmetric part in state-space and transition-space gives $\langle \dot{\Sigma}^{nr} \rangle$ and $\langle \dot{\Sigma}^{ch} \rangle, \langle \dot{\Sigma}^{sp} \rangle$, respectively. The symmetric part gives the rate of change of the macroscopic stochastic Massieu potential $- d_t\psi_{E} - d_t D^{KL}_{{E}}(\mathcal{P} [\{\rho\}])$. Our formulation reveals similarities and differences in the underlying thermodynamic geometrical structure between the non-reciprocal and reciprocal systems. Importantly, the orthogonal decomposition of the thermodynamic cost in \cref{eq:macroscopic_net_EPR_final_orthogonal} is completely determined by physically measurable quantities, in particular, external driving forces and the corresponding current means and variances.

Importantly, the macroscopic self-propulsion EPR with $l=1$, $\langle \dot{\Sigma}_i^{sp} \rangle =  2 d_i^\parallel F_i^{sp} \tanh{(|F_i^{sp}|/2)}$, is bounded below by the continuum-space macroscopic self-propulsion EPR $ \dot{\Sigma}_i^{sp} = \tilde{d}_i^\parallel \left( f_i^{sp} \right)^2 $ in the continuum limit $l \to 0$ \cite{atm_cg_nr_2024}. This highlights the importance of the observation length scale for the coarse-grained macroscopic description. The macroscopic continuum description oversimplifies the coarse-grained description beyond the inherent diffusive length scale of the system, resulting in an underestimation of thermodynamic dissipation due to non-conservative self-propulsion forces  \cite{Pietzonka_2018,Speck_2023,bebon_2024_mips_thermodynamics}. This underestimation arises from the quadratic dissipation function in the continuum limit. Correctly identifying the discreteness/finiteness of the diffusive length scale restores the exact microscopic dissipation at the macroscale.
\subsubsection{Temporal cross-correlations between macrostates and relaxation EP}\label{sec:macroscopic_non_reciprocal_EP}
We define the temporal correlation $C_{ij}(t, \Delta t) = \rho_i( t + \Delta t) \rho_j(t)$ between macrostates and their anti-symmetric $C_{ij}^a(t, \Delta t) = C_{ji}(t, \Delta t) - C_{ij}(t, \Delta t)$ and symmetric $C_{ij}^s(t, \Delta t) = C_{ji}(t, \Delta t) + C_{ij}(t, \Delta t)$ decomposition \cite{Tomita_1974}. It can be trivially verified that $C_{ij}^a(t, \Delta t) = \Delta t \: \omega_{ij}$ for small $\Delta t$. Integrating \cref{eq:macroscopic_epr_non_reciprocal} from an initial time $t_i$ to a final time $t_f$, with $\tau = t_f - t_i$, leads to the following expression for the relaxation process:
\begin{equation}\label{eq:macroscopic_non_reciprocal_EP_cross_correlation}
\begin{split}
\langle \Sigma^{nr} \rangle  
& = \int_{\mathcal{V}} \sum_{ \{i > j\} }
{V}_{ij}^{nr} \left( \langle { {C}_{ij}^a(t_i, \tau) \rangle } \rangle - \langle { {C}_{ij}^a(t_i, 0) \rangle } \rangle
\right),
\\
\langle {E}^{int} \rangle
& = \frac 1 2 \int_{\mathcal{V}} \sum_{ \{i, j\} }
{V}_{ij}^{r} \left( \langle {C}_{ij}^s(t_i, \tau)  \rangle - \langle {C}_{ij}^s(t_i, 0) \rangle  \right). 
\end{split}    
\end{equation}
\Cref{eq:macroscopic_non_reciprocal_EP_cross_correlation} relates the EPR due to reciprocal $\langle E^{int} \rangle$ and non-reciprocal $\langle \dot{\Sigma}^{nr} \rangle$ interactions between the macrostates with the symmetric and anti-symmetric temporal correlations between the initial and final states. $\langle {C}_{ij}^s(t_i, \tau) \rangle$ and $\langle {C}_{ij}^a(t_i, \tau) \rangle$ are convenient to obtain experimentally. Using \cref{eq:macroscopic_non_reciprocal_EP_cross_correlation} in \cref{eq:macroscopic_net_EPR_final_orthogonal} yields a tighter bound on $\langle \dot{\Sigma} \rangle$ by using the relaxation of ${C}_{ij}^s(t_i, \tau)$ and ${C}_{ij}^a(t_i, \tau)$ from the initial state to the final state. This bound is similar to the TUR but is obtained using macrostate correlations instead of the precision of the transition currents \cite{Horowitz_2020}. Hence, it should be compared to Refs.~\cite{Ohga_2023,Kolchinsky_2023_spectral_bounds,Dechant_2023_correlation_bounds,Dechant_2023_power_spectrum_bounds,Shiraishi_2023,Liang_2023,vanvu_2023_dissipation}. $-\langle \dot{E}^{int} \rangle > 0$ ensures that attractive (repulsive) reciprocal interactions $V_{ij}^r<0$ ($V_{ij}^r>0$) increase (decrease) the symmetric temporal macrostate correlation $\langle {C}_{ij}^s(t_i, \tau) \rangle$ during the relaxation process. Similarly, $\langle \dot{\Sigma}^{nr} \rangle > 0$ implies that $ \langle { {C}_{ij}^a(t_i, \tau) \rangle } \rangle$ increases for $V_{ij}^{nr} > 0$.

Importantly, this highlights the physical correctness of `the orthogonal gauge'. The choice of any other gauge would incorrectly assign a part of the symmetric macrostate correlations to $\Sigma^{nr}$, which is physically contradictory. Consequently, $E^{int}$ and $\Sigma^{nr}$ are not linearly independent for other gauges, and a subsequent redefinition of the linearly independent contributions to the EPR leads to a formulation equivalent to the `orthogonal gauge' fixing.
\subsection{Mesoscopic, Macroscopic and Deterministic limits}
The parameter $\Omega$ dictates the scale of the coarse-grained description of the system. It consists of three important limits: the mesoscopic, macroscopic, and deterministic limits, characterized respectively by $\Omega$ being one, large, and infinite. Importantly, the intensive scaling of the microscopic Boltzmann weight $\epsilon_i^\#$ requires the constraint $v_{ij}^r, v_{ij}^{nr} \propto 1/\Omega$. A Taylor series expansion of \cref{eq:macroscopic_interaction_coefficeints} in $v_{ij}^r, v_{ij}^{nr}$ (or equivalently in $1/\Omega$) leads to the macroscopic interaction coefficients $V_{ij}^r = \beta \Omega v_{ij}^r + \frac{1}{2} \Omega \beta^2 ((v_{ij}^r)^2 + (v_{ij}^{nr})^2 ) + O( \Omega \beta^3 v_{ij}^3 )$ and $V_{ij}^{nr} = \beta \Omega v_{ij}^{nr} + \Omega \beta^2 v_{ij}^{nr} v_{ij}^r + O(\Omega \beta^3 v_{ij}^3)$. The macroscopic interaction coefficients satisfy the gauge-fixing conditions $V_{ij}^r = V_{ji}^r$ and $V_{ij}^{nr} = -V_{ji}^{nr}$. Taking the limit $\Omega \to \infty$ yields the deterministic limit, $\bar{V}_{ij}^r = \lim_{\Omega \to \infty} \beta \Omega v_{ij}^r$ and $\bar{V}_{ij}^{nr} = \lim_{\Omega \to \infty} \beta \Omega v_{ij}^{nr}$. Throughout this paper, we focus on the macroscopic coarse-grained description, though the system-specific scale $\Omega$ should be imposed depending on the physical context. In particular, when the mean number of particles per lattice site is small, a mesoscopic description is a better alternative for coarse-grained physical analysis.

\begin{table}[t!]
\begin{tabular}{|m{1.7cm}|m{2.1cm}|m{2.1cm}|m{2.1cm}|}
     \hline
     \textbf{Level of description}
     & \textbf{Mesoscopic}
     $\Omega = 1$
     \vspace{3pt}
     & \textbf{Macroscopic} 
     $\Omega = \mathcal{V} >> 1$
     \vspace{3pt}
     & \textbf{Deterministic}
     $\Omega = \mathcal{V} = \infty$
     \vspace{3pt}
     \\
     \hline
     State
     \vspace{5pt}
     & Fluctuating mean particle number
     & Fluctuating mean particle density
     & Deterministic mean particle density
     \\
     \hline
     Occupancy noise 
     \vspace{2pt}
     & Poissonian \cref{eq:macroscopic_interaction_coefficeints} \cite{atm_cg_nr_2024}
     \vspace{5pt}
     & Gaussian corrections around the mean field
     & Vanishes, recovering the mean-field limit
     \\
     \hline
     Transition noise
     & Poissonian$^\dagger$ 
     and $O(1)$ \cite{atm_2024_var_epr,atm_2025_var_epr_derivation}
     \vspace{1pt}
     & Gaussian 
     and $O(1/\Omega)$
     \vspace{1pt}
     & Vanishes
     \vspace{1pt}
     \\
     \hline
\end{tabular}
\caption{The summary of the impact and importance of fluctuations across different coarse-grained descriptions. }
\label{table:fluctuations}
\end{table}

\Cref{table:fluctuations} summarizes the implications of fluctuations at different observation scales: the mesoscopic, macroscopic, and deterministic scales. The relevant coarse-grained states are the number of particles at a lattice site, the particle density at a lattice point, and the particle density at a lattice point, respectively. The suitable physical models corresponding to these scales are the Lattice Gas Models (LGM), Macroscopic Fluctuation Theory (MFT), and Chemical Reaction Networks (CRN). CRNs are a special case where $\Omega = \mathcal{V}$. While structurally similar, these descriptions differ significantly in other physical aspects, particularly concerning the nature of the noise effects.

Noise plays a key role through two mechanisms: particle occupancy and the transitions between them. \Cref{eq:macroscopic_interaction_coefficeints} highlights that Poissonian occupancy noise renormalizes the mesoscopic $V_{ij}$, leading to a nonlinear dependence on $v_{ij}$. It is crucial for correctly predicting the microscopic phase diagram using coarse-grained mesoscopic EOMs \cite{atm_cg_nr_2024}. Moreover, the mean EPR correctly incorporates the microscopic noise effects of occupancy. $(\dagger)$ In comparison, the Gaussian/Langevin approximation \cref{eq:macroscopic_eom_mft} of the mesoscopic Poissonian transition noise is sufficient due to the van Kampen closure approximation and the correct identification of the fluctuation-response relation, as discussed in the subsequent section. The non-quadratic dissipation function leads to a non-quadratic Hamilton-Jacobi equation \cite{Freidlin_2012_random,Kubo_1973,Graham_1984,Graham_1977,Gang_1987}, and \cref{eq:macroscopic_eom_mft} incorporates this mesoscopic effect despite the Gaussian/Langevin formulation. A more systematic analysis of mesoscopic Poissonian transition fluctuations is detailed in \cite{atm_2024_var_epr,atm_2025_var_epr_derivation}. Importantly, we have not obtained the macroscopic EPR using the Langevin equation, which avoids the underestimation of the macroscopic EPR near equilibrium \cite{atm_2024_var_epr,atm_2025_var_epr_derivation}. In particular, the correct identification of effective transition affinities using the mean and fluctuations of the transition currents resolves this issue \cite{atm_2024_var_epr,atm_2025_var_epr_derivation}.
\section{Thermodynamic relations}\label{sec:generalized_thermodynamic_relation}
Here, we outline different thermodynamic relations using the macroscopic description. However, one could also use the microscopic description, which is more directly relatable in Stochastic Thermodynamics. We deliberately utilize the macroscopic description to extend and exhibit the applicability of Stochastic Thermodynamics to interacting many-body systems.
\subsection{Non-reciprocal Onsager relations}\label{sec:non_reciprocal_onsager_relation}
In this section, we demonstrate the non-reciprocal Onsager relation \cite{Onsager_1931,Onsager_1931_2}. Assuming vanishing external driving, $\vec{F}_i^{sp} = 0$ and $F_{\gamma \gamma'}^{ch} = 0$, the $J^{ \vec{\mathcal{D}} }_i$ in \cref{eq:macroscopic_eom_mean_current_mft} reads
\begin{equation}\label{eq:non-reciprocal_onsager_relation_diffusive}
\vec{J}^{ \vec{\mathcal{D}} }_i = - \sum_j \left( D_{ij}^r \left( \{ \rho \} \right) + D_{ij}^{nr} \left( \{ \rho \} \right) \right) \nabla^{ \vec{\mathcal{D}} } \rho_j,
\end{equation}
where $D_{ij}^r ( \{ \rho \} ) = D_i^{\vec{{\mathcal{D}}}} ( \{ \rho \} ) ( \beta V_{ij}^r + \delta_{ij}/\rho_i )$ and $D_{ij}^{nr} ( \{ \rho \} ) = \beta D_i( \{ \rho \} ) V_{ij}^{nr}$. The term $D_{ij}^r ( \{ \rho \} )$ satisfies Onsager's reciprocal relation \cite{Onsager_1931,Onsager_1931_2}, since $V_{ij}^{r} = {\partial^2 E}/{\partial \rho_j \partial \rho_i} = {\partial^2 E}/{\partial \rho_i \partial \rho_j}$. In contrast, $D_{ij}^{nr} ( \{ \rho \} )$ satisfies Onsager's anti-reciprocal relation, due to $V_{ij}^{nr} = -V_{ji}^{nr}$, or equivalently $\partial F_i^{nr}/\partial \rho_j = - \partial F_j^{nr}/\partial \rho_i$. We introduce the mobility, entropic, symmetric, and anti-symmetric interaction matrices $\mathbb{D}$, $\mathbb{S}$, $\mathbb{D}^r$, and $\mathbb{D}^{nr}$, with $ij^{th}$ element of the matrices being $\delta_{ij} D_i( \{ \rho \} )$, $\delta_{ij}/\rho_i$, $D_{ij}^r$, and $D_{ij}^{nr}$, respectively, and let $\boldsymbol{\nabla \rho}$ denote the column vector of gradients.

Thus, $\mathbf{J}^{ \vec{\mathcal{D}} } = (\mathbb{D}^r + \mathbb{D}^{nr}) \nabla \boldsymbol{\rho} = \mathbb{D} (\mathbb{S} + \beta(\mathbb{V}^r + \mathbb{V}^{nr})) \nabla \boldsymbol{\rho}$ with the corresponding $\mathbf{F}^{ \vec{\mathcal{D}}} = (\mathbb{F}^r + \mathbb{F}^{nr}) \nabla \boldsymbol{\rho} = \mathbb{D}^{-1} (\mathbb{D}^r + \mathbb{D}^{nr}) \nabla \boldsymbol{\rho}$. Hence, $\langle \dot{\Sigma}^{\mathcal{D}} \rangle = \langle \mathbf{F}^{ \vec{\mathcal{D}} } \cdot \mathbf{J}^{ \vec{\mathcal{D}} } \rangle = \boldsymbol{\nabla \rho}^T (\mathbb{D}^r + \mathbb{D}^{nr})^T \mathbb{D}^{-1} (\mathbb{D}^r + \mathbb{D}^{nr}) \boldsymbol{\nabla \rho}$. The square root of $\mathbb{D}^{-1}$, ${\mathbb{D}^{-\frac 1 2}}$, is also a diagonal matrix satisfying ${\mathbb{D}^{-\frac 1 2}} = {\mathbb{D}^{-\frac 1 2}}^T$, which reduces $\langle \dot{\Sigma}^{\mathcal{D}} \rangle = || \mathbb{D}^r + \mathbb{D}^{nr} ||^2$ to the norm obtained using $\mathbb{D}^{-1}$. In addition, the symmetric and skew-symmetric matrices satisfy ${\mathbb{D}^r}^T = \mathbb{D}^r$ and ${\mathbb{D}^{nr}}^T = -\mathbb{D}^{nr}$, such that $\langle \dot{\Sigma}^{\mathcal{D}} \rangle = || \mathbb{D}^r ||^2 + || \mathbb{D}^{nr} ||^2$. Thus, $\langle \dot{\Sigma}^{\mathcal{D}} \rangle = \langle \nabla \boldsymbol{\rho}^T ({\mathbb{D}^r}^T \mathbb{D}^{-1} \mathbb{D}^r + {\mathbb{D}^{nr}}^T \mathbb{D}^{-1} \mathbb{D}^{nr}) \nabla \boldsymbol{\rho} \rangle$, or equivalently, $\langle \dot{\Sigma}^{\mathcal{D}} \rangle = \langle \nabla \boldsymbol{\rho}^T ({\mathbb{F}^r}^T \mathbb{D}^{T} \mathbb{F}^r + {\mathbb{F}^{nr}}^T \mathbb{D}^{T} \mathbb{F}^{nr}) \nabla \boldsymbol{\rho} \rangle$. The term $\nabla \boldsymbol{\rho}^T {\mathbb{D}^r}^T \mathbb{D}^{-1} \mathbb{D}^r \nabla \boldsymbol{\rho}$ relates the symmetric response coefficients of $\mathbf{J}^{ \vec{\mathcal{D}} }$ to the mean EPR, analogous to Onsager's reciprocal relation near equilibrium \cite{Onsager_1931,Onsager_1931_2,Jaynes_1980}, while $\nabla \boldsymbol{\rho}^T {\mathbb{D}^{nr}}^T \mathbb{D}^{-1} \mathbb{D}^{nr} \nabla \boldsymbol{\rho}$ relates the anti-symmetric response coefficients to the mean EPR, representing the Onsager non-reciprocal relation. By construction, the norm is positive, leading to two independently positive terms for $\langle \dot{\Sigma}^{\mathcal{D}} \rangle$, interpretable as an orthogonal decomposition for non-reciprocal systems.

Using $\boldsymbol{\rho} = \boldsymbol{\rho}^{ss} + \delta \boldsymbol{\rho}$ and/or $\mathbf{J}^{ \vec{\mathcal{D}} } = \mathbf{J}^{ \vec{\mathcal{D}} }_{ss} + \delta \mathbf{J}^{ \vec{\mathcal{D}} }$, and exploiting the symmetry of fluctuations around the steady-state profile $\boldsymbol{\rho}^{ss}$ or $\mathbf{J}^{ \vec{\mathcal{D}} }_{ss}$, one can further decompose $\langle \dot{\Sigma}^{\mathcal{D}} \rangle$ into nonadiabatic and housekeeping contributions. Similarly, $\langle \dot{\Sigma}^{\mathcal{R}} \rangle$ can be incorporated by linearizing the reactive currents around equilibrium $\boldsymbol{\rho}^{eq}$ ($\mathbf{J}^{ \vec{\mathcal{R}} }_{eq}$) or steady-state $\boldsymbol{\rho}^{ss}$ ($\mathbf{J}^{ \vec{\mathcal{R}} }_{ss}$), requiring identification of analogous symmetric and anti-symmetric couplings $\mathbb{D}^r$ and $\mathbb{D}^{nr}$ in the discrete transition space \cite{Jaynes_1980}.
\subsection{Fluctuation response relation, higher order current cumulants and responses}\label{sec:fluctuation_responce_relation}
In statistical physics, the FRR is a fundamental principle that connects equilibrium fluctuations with the linear response of the system \cite{Einstein_1905,Sutherland_1905,Callen_1951,Green_1952,Green_1954,Kubo_1957,Yamamoto_1960,Zwanzig_1965,Kubo_1966,Helfand_1960,Kubo_1973}. The non-equilibrium analog of FRR has been postulated \cite{Kubo_1973,Cugliandolo_1993,Vlad_1994_1,Vlad_1994_2,Vlad_1994_3,Ross_1999,Bodineau_2004,Bertini_2001,Bertini_2002,Lippiello_2005,Speck_2006,Blickle_2007,Baiesi_2009,Speck_2009,Corberi_2010,Dechant_2018,Maes_2020_response_therory,Shiraishi_2021,Yang_2021}. We examine the FRR for non-reciprocal and driven systems in this section. We define the $n^{th}$ scaled cumulant $\langle \prescript{n}{}{J}_{\gamma \gamma'} \rangle_C $ of $J_{\gamma \gamma'}$. By construction, $\langle \prescript{1}{}{J}_{\gamma \gamma'} \rangle_C = {J}_{\gamma \gamma'}$ and $\langle \prescript{2}{}{J}_{\gamma \gamma'} \rangle_C = {T}_{\gamma \gamma'} = 2 \Omega B_{\gamma \gamma'}$. Here, $\Omega$ defines the scaling between the scaled cumulant and the cumulant, for example, the traffic ${T}_{\gamma \gamma'}$ and the variance $ B_{\gamma \gamma'} \left( \{ \rho \} \right)$. It satisfies the following hierarchical relationship \cite{atm_cg_nr_2024}:
\begin{equation}\label{eq:higher_order_current_cumulants}
\begin{split}
    & \langle \prescript{n}{}{J}_{\gamma \gamma'} \rangle_C =  \langle \prescript{n-2}{}{J}_{\gamma \gamma'} \rangle_C, 
    \\
    & \langle \prescript{n}{}{J}_{\gamma \gamma'} \rangle_C = {J}_{\gamma \gamma'}, \hspace{1cm} \text{n is odd},
    \\
    & \langle \prescript{n}{}{J}_{\gamma \gamma'} \rangle_C = {T}_{\gamma \gamma'}, \hspace{1cm} \text{n is even}.
\end{split}    
\end{equation}
\Cref{eq:higher_order_current_cumulants} reveals the recursive structure of the current cumulants. In particular, only the first and second cumulants are independent. This ensures that the van Kampen moments closure expansion is truncated up to the second order for the transition dynamics \cite{van_kampen,Onsager_1953,Onsager_1953_2,atm_2024_var_epr,atm_2025_var_epr_derivation}. Thus, \cref{eq:higher_order_current_cumulants} highlights the validity of the van Kampen closure for far-from-equilibrium, non-reciprocal, and externally driven systems. Physically, this important result enables the study of the macroscopic dynamics of non-reciprocal systems using the first two moments of the transition currents. A similar motif has previously been observed for MFT \cite{Bodineau_2004}. This ensures the correctness of the Langevin/Gaussian approximation of the macrostate stochastic dynamics formulated in the dynamics section.  

We define the response function $\prescript{A}{}{\zeta}_{\gamma \gamma'} = {\partial \: J_{\gamma \gamma'} }/{\partial A_{\gamma \gamma'} }$ and $\prescript{A}{}{\vec{\zeta}}_{i}^{\vec{\mathcal{D}}} = { \partial \: \vec{J}_{ i }^{ \vec{\mathcal{D}}}  }/{ \partial \vec{A}_i^{\vec{\mathcal{D}}} }$. The response function satisfies the FRR:
\begin{equation}\label{eq:frr_defination}
\begin{split}
    \prescript{A}{}{\zeta}_{\gamma \gamma'}
    = \Omega B_{ \gamma \gamma' },
    \hspace{0.5cm}
    \prescript{A}{}{\vec{\zeta}}_{i}^{\vec{\mathcal{D}}} 
    = \Omega B_{i}^{\vec{\mathcal{D}}}.
\end{split}    
\end{equation}
Here, $\Omega$ plays an analogous role to the inverse temperature for the macroscopic stochastic dynamics \cite{Freidlin_2012_random}. We use the convention of evaluating the response at the reference affinity $A_{\gamma \gamma'}^*$. Hence, it characterizes the reference state and the probability distribution around which the response is evaluated, for instance, a steady state or an equilibrium distribution. We define the response function for the currents with respect to the change in symmetric reactive transition affinity, $\prescript{U}{}\zeta_{\gamma \gamma'} = {\partial \: J_{\gamma \gamma'} }/{\partial U_{\gamma \gamma'} }$ and $\prescript{U}{}{\vec{\zeta}}_{i}^{\vec{\mathcal{D}}} = { \partial \: \vec{J}_{ i }^{ \vec{\mathcal{D}}}  }/{ \partial \vec{U}_i^{\vec{\mathcal{D}}} } $:
\begin{equation}\label{eq:frr_symmetric_defination}
\begin{split}
    \prescript{U}{}\zeta_{\gamma \gamma'} 
    = \frac 1 2 J_{ \gamma \gamma' }, \hspace{0.8cm} 
    \prescript{U}{}{\vec{\zeta}}_{i}^{\vec{\mathcal{D}}}  
    = \frac 1 2 \vec{J}_{i}^{\vec{\mathcal{D}}}.
\end{split}    
\end{equation}
The set of \cref{eq:frr_defination,eq:frr_symmetric_defination} satisfies the generic linear response relations between current and traffic \cite{Maes_2020,Maes_2020_response_therory}. 

We further delineate the underlying generic structure for the higher-order response function. The $n^{th}$ order response function for the $m^{th}$ current cumulant is defined as $\prescript{A}{n}{\zeta}_{\gamma \gamma'}^m = { \partial^n \: \langle \prescript{m}{}{J}_{\gamma \gamma'} \rangle_C }/{ \partial A_{\gamma \gamma'}^n }$ and  $\prescript{U}{n}{\zeta}_{\gamma \gamma'}^m = { \partial^n \: \langle \prescript{m}{}{J}_{\gamma \gamma'} \rangle_C }/{ \partial U_{\gamma \gamma'}^n }$:
\begin{equation}\label{eq:reactive_higher_order_frr_defination}
\begin{split}
    \prescript{A}{n}{\zeta}_{\gamma \gamma'}^m  
    = \frac{1}{2^n} \langle \prescript{m+n}{}{J}_{\gamma \gamma'} \rangle_C,
    \hspace{0.8cm}
    \prescript{U}{n}{\zeta}_{\gamma \gamma'}^m 
    = \frac{1}{2^n} \langle \prescript{m}{}{J}_{\gamma \gamma'} \rangle_C,
\end{split}    
\end{equation}
\cref{eq:reactive_higher_order_frr_defination} is the higher-order FRR that relates the non-linear response of any higher-order current cumulant to other current cumulants:
\begin{equation}\label{eq:reactive_higher_order_rrr_defination}
\begin{split}
    \prescript{A \:}{n_1}{\zeta}_{\gamma \gamma'}^{m_1}
    = 
    \delta(n_1 + m_1 - n_2 - m_2) 2^{n_2-n_1} \: \prescript{A\:}{n_2}{\zeta}_{\gamma \gamma'}^{m_2},
\end{split}    
\end{equation}
where $\delta(n_1-n_2)$ is the Kronecker delta function. \Cref{eq:reactive_higher_order_rrr_defination,eq:reactive_higher_order_frr_defination} generalizes the far-from-equilibrium higher-order FRR for non-reciprocal and driven systems \cite{Andrieux_2007_jsp_nonlin_responce,Saito_2008,Gaspard_2013}.

We use the FRR to infer the transition affinity $A_{\Delta}^*$ using the mean current $\langle J_\Delta \rangle$ and the current variance $2B_\Delta$ of the transition $\Delta$ \cite{atm_2024_var_epr,atm_2025_var_epr_derivation}:
\begin{equation}\label{eq:macroscopic_affinity_fdt_inference_relation}
\begin{split}
    A_{\Delta}^{*} = 2 \tanh^{-1}{ \left( \frac{J_{\Delta}}{2 \Omega B_{\Delta}} \right)}.
\end{split}    
\end{equation}
Using \cref{eq:macroscopic_affinity_fdt_inference_relation}, we formulate an inference-based inverse problem. The affinity $A^*_{\Delta^o}$ of an observable transition $\Delta_0$ is inferred using the observable mean current $J_{\Delta^o}$ and its variance $2B_{\Delta^o}$ \cite{atm_2024_var_epr,atm_2025_var_epr_derivation}:
\begin{equation}\label{eq:macroscopic_affinity_fdt_inference_relation_observable}
\begin{split}
    A_{\Delta^o}^{*} = 2 \tanh^{-1}{ \left( \frac{J_{\Delta^o}}{2 \Omega B_{\Delta^o}} \right)}.
\end{split}    
\end{equation}
$A_{\Delta^o}^{*}$ is the effective driving force corresponding to the observable current $J_{\Delta^o}$. This concludes the formulation of the FRR for the macroscopic fluctuating dynamics of non-reciprocal systems.
\subsection{Fluctuation relations}\label{sec:fluctuation_therorems}
The fluctuation relations (FR) stand as fundamental principles that illuminate the behavior of far-from-equilibrium fluctuating systems \cite{Bochkov_1977,Bochkov_1979,Evans_1993,Evans_1994,Jarzynski_1997,Jarzynski_1997_pre,Kurchan_1998,Crooks_1999,Maes_1999,Crooks_2000,Carberry_2004,Wang_2005}. They offer profound insight into the nature of fluctuations, shedding light on the asymmetry (symmetry) between the forward and backward physical processes at the microscopic level \cite{Gallavotti_1995,Maes_1999,Crooks_2000,Maes_2003,Lebowitz_1999,Kurchan_1998}. By examining the statistical properties of systems undergoing non-equilibrium stochastic dynamics, the FR unveil universal laws about the irreversible nature of thermodynamic processes. This bridges the gap between macroscopic irreversibility and the underlying microscopic dynamics, paving the way for a deeper understanding of the interplay between order and fluctuations in physical systems. The FR generalize the fluctuation-dissipation relation (FRR) and Onsager's regression hypothesis for systems operating far from equilibrium \cite{Bochkov_1981_1,Bochkov_1981_2,Gallavotti_1996,Andrieux_2004,Chetrite_2011}.

Here, we focus on a unified formalism of the FR based on measure theory \cite{Yang_2020,Hong_2020_measure_theory,Qian_2019,Jiang_2003} combined with the large deviation principle \cite{Touchette_2009}. In measure theory, the Radon-Nikodym derivative (RND) is defined as the transition probability measure between the process and the corresponding reference process. The LDB condition establishes the connection between the RND as a mathematical property, that is, a transition probability measure, and its physical interpretation as a stochastic transition EP. The measure-theoretical formalism of the FR based on the RND utilizes the contraction of the rate functional for the empirical microscopic transition currents to the corresponding rate functional for the EP \cite{atm_2024_var_epr,atm_2025_var_epr_derivation}. The orthogonal decomposition of the stochastic EP ensures that the operation and the reference operation chosen to evaluate the RND are spanned by operations that commute with each other. From the large deviation approach, this is equivalent to choosing the linearly independent empirical observable EP \cite{Talkner_2009,GarciaGarcia_2010,GarciaGarcia_2012,Lahiri_2015,GarciaGarcia_2016,atm_2025_var_epr_derivation}. The orthogonality condition delineates the linearly independent symmetry operations that act on the system \cite{Talkner_2009,GarciaGarcia_2010,GarciaGarcia_2012,Lahiri_2015,GarciaGarcia_2016,atm_2025_var_epr_derivation}.

We define a scaled-intensive observable $\hat{O} = \frac{1}{\Omega_{O}} \int_0^{\tau} O dt$ with a scaling factor $\Omega_{O}$. We consider the most general observable $\vec{\hat{O}} = \{ \Delta_0^\tau \hat{S}^{state}_E, \Delta_i^\tau \psi_E, \hat{W}, \{ \hat{\omega}_{ij}\}, \{\hat{J}_{\gamma \gamma'}^{ch} \}, \{ \vec{\hat{J}}_i^{sp} \} \}$ in vector representation with $\vec{\Omega}_{O} = \{ \ln{(\Omega)}, \Omega, \Omega, \{\Omega \tau\}, \{\Omega \tau\}, \{\Omega \tau\} \}$ and $\vec{\Omega}_{O}^{-1} = \{ (\ln{(\Omega}))^{-1}, \Omega^{-1}, \Omega^{-1}, (\Omega \tau)^{-1}, (\Omega \tau)^{-1}, (\Omega \tau)^{-1} \}$. Here, we have utilized the counting observable for the transition currents. The cumulant generating function $G_{\vec{{\mathbf{O}}}}(\vec{\chi}_{{O}})$ for $\vec{{O}}$ reads:
\begin{equation}\label{eq:observable_vector_orthogonal_generating_function_defination}
\begin{split}
    G_{\vec{{{O}}}}(\vec{\chi}_{{O}}) = \ln{ \langle e^{ \vec{\chi}_{{O}} \: \cdot \: \vec{\Omega} \: \odot \: \vec{\hat{O}} } \rangle },
\end{split}    
\end{equation}
with the observable conjugate vector $\vec{\chi}_{\vec{{O}}} = \{ \chi_{ \Delta \hat{S}^{state}_E }, \chi_{\Delta \hat{\Psi}_E}, \chi_{\hat{W}}, \{ \chi_{ \hat{\omega}_{ij} } \}, \{ \chi_{ \hat{J}_{\gamma \gamma'} } \},  \{ \chi_{ \hat{J}_{i}^{ \vec{\mathcal{D}} } } \} \}$. The scaling of $G_{\vec{{{O}}}}(\vec{\chi}_{{O}})$ is employed using $ \max\{ \vec{\Omega}_{O} \}$, which dominates the scaled $G_{\vec{{{O}}}}(\vec{\chi}_{{O}})$ (SCGF). The probability density measure $\mathcal{P}(\vec{ O })$ satisfies the following symmetry:
\begin{equation}\label{eq:observable_vector_orthogonal_probability_distributio_symmetry}
\begin{split}
    & \ln{ \left( \frac{ \mathcal{P}(\vec{ O }) }{\mathcal{P}( -\vec{ O } )} \right) } = \vec{A}_{\vec{{O}}} \cdot \vec{\Omega} \odot \vec{{O}}, 
    \\
    & G_{\vec{{{O}}}}(\vec{\chi}_{ \vec{{O}} }) = G_{ \vec{\hat{O}}}( - \vec{A}_{\vec{{O}}} - \vec{\chi}_{ \vec{{O}} } ).
\end{split}    
\end{equation}
Defined precisely, $\mathcal{P}(\vec{ O }) = \lim_{\delta \vec{O} \to 0} \mathcal{P}( \vec{O} < \vec{\hat{O}} < \vec{O} + \delta \vec{O} )$ and $\odot$ is the component-wise Hadamard product defined between two vectors. \Cref{eq:observable_vector_orthogonal_probability_distributio_symmetry} is the finite-$\vec{\Omega}_O$ Gallavotti-Cohen fluctuation relation symmetry, which requires identifying $\vec{A}_{\vec{\hat{O}}} = \{1,-1, 1, \{V^{nr}_{ij} \}, \{ F^{ch}_{\gamma \gamma'} \}, \{ \vec{F}^{sp}_{i} \} \}$, the forces conjugate to $\vec{\hat{O}}$ \cite{Gallavotti_1995,Gallavotti_1996,Jiang_2003,Gaspard_2004,Jarzynski_2004,Gaspard_2004_EPR_FT,Andrieux_2004,Andrieux_2006,Andrieux_2007,Andrieux_2007_single_current_FT,Saito_2008,Faggionato_2011,Hurtado_2011_symmetry,Gaspard_2013}.

We aim to identify the effective macroscopic symmetry of the coarse-grained observable denoted by $\vec{ O }^\parallel$, such that $\vec{ O }$ is decomposed into $\vec{ O } = \vec{ O }^\parallel \oplus \vec{ O }^\perp$. Using Bayes' theorem and \cref{eq:observable_vector_orthogonal_probability_distributio_symmetry}, the FR for $\vec{ O }^\parallel$ reads \cite{Talkner_2009,GarciaGarcia_2010,GarciaGarcia_2012,Lahiri_2015,GarciaGarcia_2016}:
\begin{equation}\label{eq:observable_vector_orthogonal_probability_distributio_symmetry_FR}
\begin{split}
    \frac{\mathcal{P}(\vec{ O }^\parallel)}{\mathcal{P}(-  \vec{{O}}^\parallel )} =  \frac{ e^{ \vec{A}_{ \vec{ O }^\parallel } \cdot \vec{\Omega}\odot \vec{ O }^\parallel }}{\langle e^{- \vec{ O }^\perp} | \vec{ O }^\parallel \rangle},
\end{split}    
\end{equation}
where $\langle e^{-\vec{ O }^\perp} | \vec{ O }^\parallel \rangle = \int \mathbb{D} \vec{O}^\perp \: {\mathcal{P}( \vec{ O }^\perp | \vec{ O }^\parallel )} e^{- \vec{A}_{\vec{\hat{O}}^\perp} \cdot \vec{\Omega}^\perp \odot \vec{ O }^\perp}$ quantifies the projection of the conditional probability measure of $\vec{ O }^\perp$ into the $\vec{ O }^\parallel$ space. $\langle e^{-\vec{ O }^\perp} | \vec{ O }^\parallel \rangle = 1$ for linearly independent observables $\vec{ O }^\parallel$ and $\vec{ O }^\perp$, which recovers the FR for $\vec{ O }^\parallel$. The proof follows trivially by the Taylor series expansion of $e^{-\vec{ O }^\perp}$ and the orthogonality relation $\langle \vec{ O }^\perp | \vec{ O }^\parallel \rangle = 0$. This emphasizes the importance of choosing linearly independent orthogonal observables. Physically, this generates a hierarchy of detailed fluctuation relations and delineates independent underlying symmetries of the EP.

By choosing $\vec{ O }^\parallel = \vec{ \hat{\Sigma} } = \{  \hat{W}, \Delta_0^\tau \hat{S}^{state}_E, \hat{\Sigma}^{nr}, \hat{\Sigma}^{ch}, \hat{\Sigma}^{sp} \}$ and $ \vec{ O }^\perp = \Delta_0^\tau \psi_E$, we obtain the detailed master FR for non-reciprocal systems:
\begin{equation}\label{eq:fluctuation_relation_generic}
\begin{split}
    \frac{\mathcal{P}( W , \Delta_0^\tau S^{state}_E, \Sigma^{nr}, \Sigma^{ch}, \Sigma^{sp} )}{\mathcal{P}(-W, -\Delta_0^\tau S^{state}_E, -\Sigma^{nr}, -\Sigma^{ch}, -\Sigma^{sp})} = e^{ W - \Delta_0^\tau \psi_E + \Delta_0^\tau S^{state}_E + \Sigma^{nr} + \Sigma^{ch}  + \Sigma^{sp} }.
\end{split}    
\end{equation}
We have implemented the contraction from the linearly independent empirical current observable \cref{eq:observable_vector_orthogonal_probability_distributio_symmetry} to the linearly independent EP \cref{eq:fluctuation_relation_generic} \cite{atm_2024_var_epr,atm_2025_var_epr_derivation}. \Cref{eq:fluctuation_relation_generic} satisfies the Lebowitz-Spohn symmetry, with the generating function satisfying $G_{ \vec{ \Sigma } }(\vec{\chi}_{ \vec{ \Sigma } }) = G_{ \vec{{\Sigma}} }( - 1 - \vec{\chi}_{ \vec{{\Sigma}} } )$ \cite{Lebowitz_1999,Kurchan_1998,Maes_1999}. The stochastic work $W$ is not a linearly independent variable; rather, the irreversible work $W - \Delta_0^\tau \psi_E$ is \cite{Lahiri_2015,GarciaGarcia_2016}. This is effectively realized by the factor of $W - \Delta_0^\tau \psi_E$ on the left-hand side of \cref{eq:fluctuation_relation_generic}. Hence, $\Delta_0^\tau \psi_E$ drops out, and its projection onto the work gives the factor $W - \Delta_0^\tau \psi_E$ on the right-hand side of \cref{eq:fluctuation_relation_generic}.

Assuming the specific case of $F_{\gamma \gamma'}^{ch} = 0$ and $\vec{F}_i^{sp} = 0$, which corresponds to non-reciprocity being the only source of the system's non-equilibrium-ness, \cref{eq:fluctuation_relation_generic} is reduced to  
the Crooks-Tasaki relation for non-reciprocal systems \cite{Tasaki_2000,Crooks_2000,Crooks_1999}:
\begin{equation}\label{eq:crooks_fluctuation_relation_nr}
\begin{split}
    \frac{\mathcal{P}( W , \Delta^\tau_0 S^{state}_E, \Sigma^{nr} )}{\mathcal{P}(-W, -\Delta_0^\tau S^{state}_E, -\Sigma^{nr})} = e^{ W - \Delta^\tau_0 \psi_E + \Delta_0^\tau S^{state}_E + \Sigma^{nr} }.
\end{split}    
\end{equation}
Assuming that the external driving control parameter changes instantly relax the probability distribution to the reference Boltzmann distribution (no error during driving, i.e., a quasistatic driving process \cite{Lahiri_2015,Sartori_2015,GarciaGarcia_2016}), we have $\Delta_0^\tau S^{state}_E = 0$. \Cref{eq:crooks_fluctuation_relation_nr} is then reduced to a familiar form of the Crooks-Tasaki fluctuation theorem, but defined here for non-reciprocal systems. By integrating \cref{eq:crooks_fluctuation_relation_nr}, we obtain the Jarzynski fluctuation relation for non-reciprocal systems:  
\begin{equation}\label{eq:jarzynski_fluctuation_relation_nr}
\begin{split}
    \langle e^{ - W - \Delta_0^\tau S^{state}_E - \Sigma^{nr} } \rangle = e^{-\Delta \psi_E }.
\end{split}    
\end{equation}
Using \cref{eq:macroscopic_non_reciprocal_EP_cross_correlation}, we relate the external driving work needed with the symmetric and anti-symmetric correlations between the initial and final macrostates. Using Jensen's inequality, the second law of thermodynamics (approximate law) \cref{eq:macroscopic_net_EPR_final_orthogonal} is recovered. However, using $C_{ij}^a(t,\tau)$ and the fluctuation theorem (exact law) \cref{eq:jarzynski_fluctuation_relation_nr} gives a tighter bound on the mean work. The work done on non-reciprocal systems requires extra thermodynamic cost associated with changing the anti-symmetric temporal cross-correlations between the macrostates. The validity of the non-reciprocal Crooks-Tasaki and Jarzynski relations \cref{eq:crooks_fluctuation_relation_nr,eq:jarzynski_fluctuation_relation_nr} is subject to not crossing the $\mathcal{PT}$ symmetry-breaking bifurcation manifold in the space of control parameters. Driving through a $\mathcal{PT}$ transition from a static to dynamic phase undergoes divergent fluctuations, which could lead to non-trivial FR symmetries and exponents. A more systematic study is needed for such physical problems. 

By choosing $\Delta_0^\tau S^{state}_E \neq 0$, the stochastic work incorporates the irreversible work contribution due to the non-quasistatic driving of the control parameters \cite{Kawai_2007,Saha_2009,Horowitz_2009,Deffner_2011,Sartori_2015,GarciaGarcia_2016,Lahiri_2012,Lahiri_2015}. Physically, it quantifies the mismatch between the instantaneous non-equilibrium probability distribution and the corresponding instantaneous Boltzmann probability distribution specified by the instantaneous control parameters of $E$. In the absence of external work $(W=0)$, $\Delta_0^\tau S^{state}_E \neq 0$ quantifies the relaxation EP and restores the finite-time FR.  

Assuming $W=0$ and combining the boundary term $\Delta \psi_E$ with the non-conservative EP $\Sigma^{nr} + \Sigma^{sp} + \Sigma^{ch}$, \cref{eq:fluctuation_relation_generic} is reduced to the following equation:
\begin{equation}\label{eq:fluctuation_relation_generic_housekeepin_excess_decomposition_orthogonal}
\begin{split}
    \frac{\mathcal{P}( \Delta_0^\tau S^{state}_E, \Sigma^{hk}_E ) }{ \mathcal{P}( -\Delta_0^\tau S^{state}_E, - \Sigma^{hk}_E ) } = e^{ \Delta_0^\tau S^{state}_E + \Sigma^{hk}_E }.
\end{split}    
\end{equation}
\Cref{eq:fluctuation_relation_generic_housekeepin_excess_decomposition_orthogonal} represents generic detailed fluctuation relations for the non-adiabatic, housekeeping, and total EP, applicable to non-reciprocal systems \cite{Espigares_2012}. It is the effective version of the underlying FR symmetry \cref{eq:fluctuation_relation_generic}. The detailed FR for the non-adiabatic, housekeeping, and total EP are obtained by using $E=ss$ \cite{Saha_2009,Esposito_2010_3dft,Espigares_2012}, the Hatano-Sasa relation for the excess EP for a transition from an initial to a final steady state \cite{hatano_2001,Espigares_2012}, or the Speck-Seifert relation for the housekeeping EP for a transition from an initial to a final steady state \cite{Speck_2005}.  

The steady-state condition is not required to formulate the orthogonal decomposition, making it applicable in a broader context, with a suitable gauge-fixing choice of $E$ according to the required physical or experimental constraints \cite{atm_2025_var_epr_derivation,maes_2007,dechant_2022_geometric_epr,dechant_2022_geometric_epr_cpl,yoshimura_2023_epr_decomposition_nld,nagayama_2023_geometric_epr_rd,Ding_2022_st_ft,kobayashi_2022_hessian_geometry,Kobayashi_2022,Sughiyama_2022,kobayashi_2023_information_graphs_hypergraphs,Loutchko_2023_geometry_tur,Renger_2023,duong_2023,Mizohata_2024,Espigares_2012}. 
\subsection{Thermodynamic-Kinetic Uncertainty Relation}\label{sec:thermodynamic_uncertainity_relations}
The thermodynamic uncertainty relation (TUR) delineates the trade-off between the precision of an observable current and the minimum EP required to sustain it \cite{Horowitz_2020,Barato_2015,Gingrich_2016,Horowitz_2017}. Here, the precision is the ratio between the square of the mean observable transition current and the variance of the transition current. The TUR gives a tighter bound on the EP than the second law of thermodynamics. In other words, the precision of an observable current comes at a minimum EP required to sustain it. The bounds on thermodynamic dissipation can be improved by choosing other appropriate physical observables and constraints \cite{Gingrich_2017,Garrahan_2017,Skinner_2021,Skinner_2021_EPR,Manikandan_2020,Otsubo_2020,Van_vu_2020,Kim_2020,Otsubo_2022,Dechant_2021,Shiraishi_2021,Garrahan_2017,Di_Terlizzi_2019,Dieball_2023} and by its unification into the thermodynamic-kinetic uncertainty relation (TKUR) \cite{Vo_2022}. The TKUR's connection to Speed Limits (SL) \cite{Shiraishi_2018_classical_SL,Ito_2018,Nicholson_2018,Nicholson_2020,Vo_2020,Ito_2020,Yoshimura_2021,Vo_2022,Van_vu_2023} and to the FR symmetry \cite{Merhav_2010,Timpanaro_2019,Hasegawa_2019,Francica_2022,Francica_2024} has been established. Here, we demonstrate the TKUR for non-reciprocal systems, while a more fundamental analysis of the TKUR and non-quadratic speed limits is outlined in Refs.~\cite{atm_2024_var_epr,atm_2025_var_epr_derivation}.
\subsubsection{Coarse-grained observable transition currents}
We consider a set of macroscopic observable transition currents denoted by $O^J = \{\Delta^o\}$ and the corresponding mean inferred EPR $\langle \dot{\Sigma}_{O^J} \rangle$. The mapping $\{\{ \Delta_{\gamma\gamma'} \},\{ \Delta_i^{\vec{\mathcal{D}}}\} \} = \{ \Delta \} \to \{\Delta^o\}$ is considered to be many-to-one, so that each microscopic transition contributes to one macroscopic observable transition current. This mapping is represented using an observable matrix $\mathbb{O}^{J}$, with the row and column indices corresponding to the observable and microscopic currents, respectively. Mathematically, the mapping $\{\{ \Delta_{\gamma\gamma'} \},\{ \Delta_i^{\vec{\mathcal{D}}}\} \} \to \{\Delta^o\}$ implies $\mathbb{O}_{o'i'}^J = 1 \implies \mathbb{O}_{o''i'}^J = 0, \forall o'' \in \{\Delta^o\} - o'$. The observable current and traffic vectors satisfy $\vec{J}_{\Delta^o} = \mathbb{O}^J \vec{J}_{\Delta}$ and $\vec{T}_{\Delta^o} = \mathbb{O}^J \vec{T}_{\Delta}$. In particular, using \cref{eq:macroscopic_eom_mean_current_mft,eq:eom_mft_variance_currents}, one can equivalently obtain the same expressions for the mean and variance of the observable current. Then, the mean inferred EPR $\langle \dot{\Sigma}_{O^J} \rangle$ reads \cite{atm_2024_var_epr,atm_2025_var_epr_derivation}:
\begin{equation}\label{eq:mean_EPR_observable_transition}
\begin{split}
    \langle \dot{\Sigma}_{O^J} \rangle 
    = \sum_{ \{\Delta^o\} } 2 J_{ \Delta^o } \tanh^{-1}{ \left( \frac{ J_{ \Delta^o } }{ T_ {\Delta^o } } \right) }.
\end{split}
\end{equation}
The inequality $\langle \dot{\Sigma} \rangle \geq \langle \dot{\Sigma}_{O^J} \rangle$ is proven using the generalized log-sum inequality and the definitions $\vec{J}_{\Delta^o} = \mathbb{O}^J \vec{J}_{\Delta}$ and $\vec{T}_{\Delta^o} = \mathbb{O}^J \vec{T}_{\Delta}$ \cite{atm_2024_var_epr,atm_2025_var_epr_derivation,Dannan_2014_log_sum_generalization}. The non-quadratic formulation gives a tighter bound on the EPR than the traditional quadratic TUR \cite{Yoshimura_2021} and is closely related to \cite{Lee_2022,Vo_2020,Vo_2022,Van_vu_2023}. The mean observable inferred EP $\langle {\Sigma}_{O^J} \rangle$ satisfies the inequality \cite{atm_2024_var_epr,atm_2025_var_epr_derivation}:
\begin{equation}\label{eq:mean_EP_observable_transition_current}
\begin{split}
    \langle \Sigma_{O^J} \rangle 
    \geq \sum_{ \{\Delta_o \}} 2 J_o^{\tau}  \tanh^{-1}{ \left( \frac{ J_o^\tau}{ T_o^\tau } \right) }, 
\end{split}     
\end{equation}
where $J_o^{\tau} = \int_{0}^\tau J_o dt$ and $T_o^{\tau} = \int_{0}^\tau T_o dt$ are the time-integrated current and traffic. We have utilized Jensen's inequality to obtain the inequality in \cref{eq:mean_EP_observable_transition_current} from the equality in \cref{eq:mean_EPR_observable_transition}. The non-quadratic dissipation function gives a tighter bound than the quadratic TKUR, and its detailed physical and technical analysis is summarized in Refs.\cite{atm_2024_var_epr,atm_2025_var_epr_derivation}. 
\subsubsection{Coarse-grained vorticities and state-correlations}
Here, we aim to examine the EPR inference using state-space observables. We choose a specific sub-case of the set of vorticity currents between the observable macrostate $\rho_o = \mathbb{O}^{\rho} \vec{\rho}$, where $\mathbb{O}^{\rho}$ quantifies the many-to-one mapping from $\{\rho_i\} \to \{\rho_o\}$. $O^{\omega} =\{\omega_{oo'} \}$ is quantified by the observable vorticity operator $\mathbb{O}^{\omega}$, which connects microscopic vorticities $\{\omega_{ij}\}$ to observable vorticities through $\vec{\omega}_{O} = \mathbb{O}^{\omega} \vec{\omega}$. The anti-symmetric (vorticity) and symmetric (traffic) components are denoted by $\langle \omega_{o o'} \rangle = \langle \rho_o \partial_t \rho_{o'} -  \rho_{o'} \partial_t \rho_o \rangle$ and  $ \langle \omega_{o o'}^s \rangle  = \langle \rho_o \partial_t \rho_{o'} +  \rho_{o'} \partial_t \rho_o \rangle$, respectively. The mean inferred EPR using the observable vorticities reads \cite{atm_2024_var_epr,atm_2025_var_epr_derivation}:
\begin{equation}\label{eq:mean_EPR_observable_vorticity}
\begin{split}
    \langle \dot{\Sigma}_{O^\omega} \rangle = \sum_{ \{\omega_{oo'}\} } 2 \langle \omega_{o o'} \rangle \tanh^{-1}{\left( \frac{ \langle \omega_{o o'} \rangle }{ | \langle \omega_{o o'}^s \rangle | } \right)}.
\end{split}    
\end{equation}
The inferred mean EP $\langle \dot{\Sigma}_{O^\omega} \rangle$ provides a lower bound on $\langle \dot{\Sigma} \rangle$. Integrating \cref{eq:mean_EPR_observable_vorticity} and using Jensen's inequality leads to a bound on the inferred EP using the state correlations:
\begin{equation}\label{eq:mean_EP_observable_vorticity}
\begin{split}
    \langle {\Sigma}_{O^\omega} \rangle 
    \geq 
    \sum_{ \{\omega_{oo'}\} } 2 \langle \Delta_0^\tau C_{o o'}^a \rangle \tanh^{-1}{\left( \frac{ \langle \Delta_0^\tau C_{o o'}^a \rangle }{ \langle | \Delta_0^\tau C_{o o'}^s | \rangle } \right)}.
\end{split}    
\end{equation}
Here, $\mathcal{C}_{oo'}^a(t, \Delta t) = \rho_o( t + \Delta t) \rho_{o'}(t) - \rho_{o'}( t + \Delta t) \rho_{o}(t)$ and $\mathcal{C}_{oo'}^s(t, \Delta t) = \rho_o( t + \Delta t) \rho_{o'}(t) + \rho_{o'}( t + \Delta t) \rho_{o}(t)$ characterize the anti-symmetric and symmetric parts of the observable state correlation. Furthermore, $\Delta_0^\tau C_{o o'}^a = C_{o o'}^a(t_i,\tau) - C_{o o'}^a(t_i,0)$ and $\Delta_0^\tau C_{o o'}^s = C_{o o'}^s(t_i,\tau) - C_{o o'}^s(t_i,0)$ denote the differences between the initial and final times.

Using $\{\omega_{oo'} \} = \{\omega_{ij} \}$, the tightest possible bound on $\langle \dot{\Sigma} \rangle$ is obtained using state-space correlations. Physically, this signifies that considering the temporal cross-correlations between all linearly independent macrostates yields the tightest possible bound on the EP using the state-space TUKR. The set of \cref{eq:mean_EPR_observable_vorticity,eq:mean_EP_observable_vorticity} provides the formulation of the short-time and finite-time TKUR using temporal correlations between the observable states \cite{atm_2024_var_epr,atm_2025_var_epr_derivation}. This should be compared to the quadratic \cite{Shiraishi_2023} and non-quadratic \cite{Kolchinsky_2023_spectral_bounds,Liang_2023} counterparts. 
\subsection{Information thermodynamics}\label{sec:information_thermodynamics}
The framework of information thermodynamics takes into account the thermodynamic implications of statistical information. Statistical information on a system's state is encoded in Gibbs's entropy \cref{eq:macroscopic_diffusive_reactive_gibbs_EPR,eq:macroscopic_state_entropy}. An information gain can be achieved through a feedback-controlled operation or measurement \cite{Sagawa_2010,Sagawa_2012,Lahiri_2012,Sartori_2014,Parrondo_2015}. This information gain reduces the statistical uncertainty associated with the system, resulting in a lower EP, signifying the reduced statistical uncertainty.

We define the hidden variable as $Y$, which does not affect the transition dynamics (bulk term) of the relevant macrostates $ \{\rho\}$. Then, the stochastic mutual information $I_{ \{\rho\}, Y }$ between them is defined as:
\begin{equation}\label{eq:mutual_information}
\begin{split}
    I_{ \{\rho\}, Y } = - \ln{\left( \frac{\mathcal{P}_{\{\rho\}, Y}}{\mathcal{P}_{\{\rho\}} \: \mathcal{P}_{Y}} \right) } 
    & = - \ln{(\mathcal{P}_{ \{ \rho \}|Y })} + \ln{(\mathcal{P}_{\{ \rho \}})}. 
\end{split}    
\end{equation}
Here, Bayes' theorem is used to reach the second equality in \cref{eq:mutual_information}. Mutual information quantifies the statistical correlation (or dependence) between $\{ \rho \}$ and the hidden variable $Y$. It vanishes if the macrostate of the system is statistically independent of the hidden variable, that is, $\mathcal{P}_{ \{ \rho \}|Y } = \mathcal{P}_{ \{ \rho \} }, \forall Y \in \mathcal{V}_Y, \{ \rho \} \in \mathcal{V}_{\{ \rho \}}$. Here, $\mathcal{V}_Y$ and $\mathcal{V}_{\{ \rho \} }$ denote the probability measure spaces for the hidden variable $Y$ and macrostates $\{ \rho \}$, respectively. The mean mutual information is defined as $\langle I_{\{ \rho \}, Y } \rangle = \int_{\mathcal{V}_Y} \int_{ \mathcal{V}_{\{ \rho \}} } I_{\{ \rho \}, Y } \: \mathcal{P}_{\{\rho\}, Y} $. Here, $ \int \mathbb{D}[ \{ \rho \} ] \equiv \int_{ \mathcal{V}_{\{ \rho \}} }$ is the shorthand notation used for summation over the probability measure space of the macrostate $\{\rho\}$. Then, the mutual information EPR is defined as $\langle \dot{I}_{\{ \rho \}, Y } \rangle = \int_{\mathcal{V}_Y} \int_{ \mathcal{V}_{\{ \rho \}} } I_{\{ \rho \}, Y } \: d_t \mathcal{P}_{\{\rho\}, Y} $, whose exact expressions for the mutual information EP and EPR are: 
\begin{equation}\label{eq:mean_mutual_information_EPR}
\begin{split}
    \langle I_{ \{\rho\}, Y } \rangle 
    & = \mathcal{S}^{gb}(\mathcal{P}_{ \{ \rho \}|Y }) - \mathcal{S}^{gb}(\mathcal{P}_{ \{ \rho \} }), 
    \\
    \langle \dot{I}_{ \{\rho\}, Y } \rangle 
    & = d_t \mathcal{S}^{gb}(\mathcal{P}_{ \{ \rho \}|Y }) - d_t \mathcal{S}^{gb}(\mathcal{P}_{ \{ \rho \} }) 
\end{split}.    
\end{equation}
Incorporating the mutual information EP and EPR, the second law of thermodynamics \cref{eq:macroscopic_net_EPR_final_orthogonal} is modified. Thus, the Gibbs EPR is replaced by the conditional Gibbs EPR. The conditional Gibbs entropy satisfies the inequality $\mathcal{S}^{gb}(\mathcal{P}_{ \{ \rho \} }) \geq \mathcal{S}^{gb}(\mathcal{P}_{ \{ \rho \}|Y })$. Physically, this corresponds to the information gained due to the hidden variable $Y$, which reduces the statistical ignorance (Gibbs entropy) of the macrostate $\{ \rho \}$ by encapsulating its correlations with the hidden variable $Y$. $\langle \dot{I}_{ \{\rho\}, Y } \rangle \leq 0$ holds, which gives the sign convention for the information gain. \Cref{eq:mean_mutual_information_EPR} satisfies the equality, $\langle \dot{I}_{ \{\rho\}, Y } \rangle = d_t \mathcal{S}^{gb}(\mathcal{P}_{ \{ \rho \}, Y }) - d_t \mathcal{S}^{gb}(\mathcal{P}_{ \{ \rho \} }) - d_t \mathcal{S}^{gb}(\mathcal{P}_{Y})$. Similarly, the fluctuation theorem follows trivially by modifying \cref{eq:fluctuation_relation_generic} to incorporate stochastic mutual information \cref{eq:mutual_information}, or equivalently, by replacing the reference state entropy with the reference conditional state entropy $S_E^{state} \to S_E^{state|Y} = - \ln{ ( \mathcal{P}_{ \{ \rho \}|Y }/ \mathcal{P}^E_{ \{ \rho \}|Y } ) } $. This is a manifestation of the physical symmetry of measurement (certainty) and uncertainty. We have briefly highlighted the procedure to incorporate information thermodynamics for non-reciprocal systems. The rest is rather trivial. 
\section{Illustrative examples} \label{sec:example}
\subsection{Diffusive systems}
We consider systems without reactive transitions, for which $\{\gamma\gamma'\} = \emptyset$ is an empty set. 
\subsubsection{Generalized Kawasaki-Dean equation}
Considering $\vec{F}_{sp} = 0$, \cref{eq:macroscopic_eom_mft} is reduced to \cite{atm_cg_nr_2024}: 
\begin{equation}\label{eq:macroscopic_eom_mft_dean}
\begin{split}
    \partial_t \rho_i 
    & = \tilde{d}_i \nabla \cdot
    \left( e^{\mu_i} \nabla \mu_i  
    \right) 
    + \frac{1}{\sqrt{\Omega}} \nabla \cdot \left( \sqrt{2 \tilde{d}_i e^{\mu_i}} \vec{ \hat{\xi}}_i^{\mathcal{D}} \right).
\end{split}    
\end{equation}
\Cref{eq:macroscopic_eom_mft_dean} generalizes the Kawasaki-Dean equation for interacting systems \cite{Kawasaki_1994,Dean_1996}, valid in the regime of small interaction parameters. In contrast, \cref{eq:macroscopic_eom_mft_dean} encapsulates the effect of the interaction on the amplitude of the noise. For an ideal non-interacting reciprocal particle, $\mu_i = \ln{\rho_i}$, one recovers the Kawasaki-Dean equation for the ideal particles. 
\subsubsection{Macroscopic fluctuation theory}\label{sec:mft}
We consider systems without chemical reaction changes between the particle types. In this case, the generalized macroscopic fluctuation theory EOM for the macrostate reads:
\begin{equation}\label{eq:macroscopic_eom_mft_diffusive}
\begin{split}
    \partial_t \rho_i 
    & = \nabla^{\vec{\mathcal{D}}} \cdot
    \left\{ 
    D_i^{\vec{\mathcal{D}}}
    \left( \nabla^{\vec{\mathcal{D}}} \mu_i + \nabla^{\vec{\mathcal{D}}} F_i^{nr} - \vec{F}_i^{sp} \right)
    \right \} 
    \\
    & \hspace{1cm} + \frac{1}{\sqrt{\Omega}} \nabla^{\vec{\mathcal{D}}} \cdot \left( \sqrt{ {2D_i^{\vec{\mathcal{D}}}}  } \: \vec{\hat{\xi}}_i \right).
\end{split}
\end{equation}
\Cref{eq:macroscopic_eom_mft_diffusive} is the MFT formulation for non-reciprocal systems \cite{Bertini_2015}. In contrast to Ref.\cite{Bertini_2015}, we have systematically derived the coarse-grained EOM with the density-dependent mobility $D_i^{\vec{\mathcal{D}}} = d_i^{\vec{\mathcal{D}}} e^{\mu_i + F_i^{nr}}$ \cite{atm_cg_nr_2024}. The second law of thermodynamics for this system is $\langle \dot{\Sigma} \rangle = - d_t D^{KL}_{E}(\mathcal{P}[\{\rho\}]) + \langle \dot{\Sigma}^{nr} \rangle + \langle \dot{\Sigma}^{sp} \rangle$. Here, $\langle \dot{\Sigma}^{sp} \rangle$ is a non-quadratic dissipation function, in contrast to the quadratic one obtained in MFT \cite{Bertini_2015}, therefore avoiding the underestimation of the dissipation for far-from-equilibrium systems \cite{atm_2024_var_epr,atm_2025_var_epr_derivation}.  
\subsection{Non-reciprocal two species}\label{sec:exmaple_nr_2_species}
\subsubsection{Chemo sensing bacteria}\label{sec:example_chemo_sensing_bacteria}
We consider a prototypical system of bacteria (b) attracted toward a chemical (c). Thus, the microscopic interaction rules for this physical system read $v_{bc} = -1$, $v_{cb} = 0$, $v_{cc} = 0$, $v_{bb} = 0$. This leads to $v_{bc}^r = -\frac 1 2$ and $v_{cb}^{nr} = -v_{bc}^{nr} = \frac 1 2$. We choose this model with a minimal motif of non-reciprocity and its implications for diffusive dynamics; therefore, the self-propulsion forces vanish, $\vec{f}_c^{sp} = \vec{f}_b^{sp} = 0$. The macroscopic energy functional $E$ for the density of the bacteria ($\rho_b$) and the chemical ($\rho_c$) reads:  
\begin{equation}\label{eq:energy_functional_chemotaxis}
\begin{split}
    E &= \int_{\mathcal{V}} \left[ V_{bc} \rho_b \rho_c + \rho_b \ln{\left(\frac{\rho_b}{e}\right)} + \rho_c \ln{\left(\frac{\rho_c}{e}\right)} \right],
\end{split}
\end{equation}
where $V_{bc}^r = \beta \left(-\frac 1 2 + \frac{\beta}{4 \Omega} \right)$ and $V_{cb}^{nr} = \beta \left( \frac 1 2 - \frac{\beta}{4 \Omega} \right)$. Using $\mu_b^r = \ln{(\rho_b)} + \rho_c V_{bc}^r$ and $\mu_c^r = \ln{(\rho_c)} + \rho_b V_{bc}^r$, and $F_b^{nr} = - V_{cb}^{nr} \rho_c$ and $F_c^{nr} = \frac{1}{2} V_{cb}^{nr} \rho_c$. The mobility of the bacterial density and chemical density are $D_\rho = \tilde{d}_\rho \rho_b e^{ \rho_c  (V_{bc}^{r} + V_{bc}^{nr}) } = \tilde{d}_\rho \rho_b e^{ - \beta \rho_c  (1 - \frac{\beta}{2 \Omega} ) }$ and $D_c = \tilde{d}_c \rho_c$, respectively. Thus, $\langle \dot{\Sigma}^{sp} \rangle = \langle \dot{\Sigma}^{ch} \rangle = 0$. The expression for $\langle \dot{\Sigma}^{nr} \rangle$, which quantifies the thermodynamic cost of maintaining the vorticity currents between the bacteria and the chemical, reads:
\begin{equation}\label{eq:epr_chemotaxis_nr_diffusive}
\begin{split}
    \langle \dot{\Sigma}^{nr} \rangle 
    & = V^{nr}_{cb} \int_{\mathcal{V}} \langle \rho_c \partial_t \rho_b - \rho_b \partial_t \rho_c \rangle. 
\end{split}    
\end{equation}
The second law of thermodynamics states $\langle \dot{\Sigma} \rangle = - d_t \langle E \rangle + d_t S^{gb} + \langle \dot{\Sigma}^{nr} \rangle$. Or equivalently, $\langle \dot{\Sigma} \rangle = - d_t D^{KL}_{E}(\mathcal{P}[\{\rho\}]) + \langle \dot{\Sigma}^{nr} \rangle$. Here, $d_t D^{KL}_{E}(\mathcal{P}[\{\rho\}])$ quantifies the EPR due to the relaxation of $E$, which equivalently quantifies the relaxation of the symmetric macrostate correlations. $\langle W \rangle = 0$, since the control parameters of the system are fixed. 
\subsubsection{Predator-Prey Model}\label{sec:exmaple_nr_sheep_dog}
We consider a predator-prey system with a minimalist attraction-repulsion mechanism, such as dogs-sheep, without any self-propulsion. Thus, the microscopic interaction rules are $v_{ds} = -1$, $v_{sd} = 1$, $v_{dd} = 0$, $v_{ss} = -1$, which implies $v_{ds}^r = -1$, $v_{sd}^r = 1$, $v_{dd}^r = 0$, $v_{ss}^r = -1$. This leads to the macroscopic interaction rules $V_{sd}^r = \frac{\beta^2}{2 \Omega}$, $V_{sd}^{nr} = \beta$, $V_{ds}^{nr} = -\beta$, and $V_{ss}^r = -\beta + \frac{\beta^2}{2 \Omega}$. We have intentionally chosen $v_{ss} = -1$, which models an attractive interaction between the sheep, physically corresponding to the flocking of the sheep herd. In addition, $f_{ds}^{ch} = \vec{f}_s^{sp} = \vec{f}_d^{sp} = 0$. Then, the macroscopic coarse-grained energy functional for this model obtained using the density of the sheep ($\rho_s$) and the dogs ($\rho_d$) reads:  
\begin{equation}\label{eq:energy_functional_sheep_dog}
\begin{split}
    E &= \int_{\mathcal{V}} \left[ \frac 1 2 V_{ss}^r \rho_s^2 + V_{sd}^r \rho_s \rho_d + \rho_d \ln{\left(\frac{\rho_d}{e}\right)} + \rho_s \ln{\left(\frac{\rho_s}{e}\right)} \right],
\end{split}
\end{equation}
where the reciprocal macroscopic Boltzmann weights are $\mu_{d}^r = \ln{\rho_d} + V_{sd}^r \rho_s$ and $\mu_{s}^r = \ln{\rho_s} + V_{sd}^r \rho_d + V_{ss}^r \rho_s$, and the non-reciprocal macroscopic Boltzmann weights are $F_{d}^{nr} = V_{ds}^{nr} \rho_s$, $F_{s}^{nr} = V_{sd}^{nr} \rho_d$. The diffusive mobilities are then $D_{d} = \tilde{d}_{d} \rho_d e^{ (V_{ds}^r + V_{ds}^{nr}) \rho_s }$ and $D_{s} = \tilde{d}_{s} \rho_s e^{ \left( V_{ss}^r \rho_s + (V_{sd}^r + V_{sd}^{nr}) \rho_d \right) }$. In addition, $\langle \dot{\Sigma}^{sp} \rangle = \langle \dot{\Sigma}_\mathcal{R}^{nr} \rangle = \langle \dot{\Sigma}^{ch} \rangle = 0$, since $F_{ds}^{ch} = \vec{F}_s^{sp} = \vec{F}_d^{sp} = 0$. The expression for $\langle \dot{\Sigma}^{nr} \rangle$ reads:
\begin{equation}\label{eq:epr_sheep_dog_nr_diffusive}
\begin{split}
    \langle \dot{\Sigma}^{nr} \rangle 
    & = V^{nr}_{sd} \int_{\mathcal{V}} \langle \rho_s \partial_t \rho_d - \rho_d \partial_t \rho_s \rangle. 
\end{split}    
\end{equation}
Assuming $\langle W \rangle = 0$, the second law of thermodynamics for this model is $\langle \dot{\Sigma} \rangle = - d_t \langle E \rangle + d_t S^{gb} + \langle \dot{\Sigma}^{nr} \rangle$. Or equivalently, $\langle \dot{\Sigma} \rangle = - d_t D^{KL}_{E}(\mathcal{P}[\{\rho\}]) + \langle \dot{\Sigma}^{nr} \rangle$. 

The change in the symmetric part of the macrostate correlations is $O(1/\Omega)$, as $V_{sd}^{nr} \propto O(1/\Omega)$, compared to the chemo-sensing bacteria in \cref{sec:example_chemo_sensing_bacteria}, where $V_{sd}^{nr} \propto O(1)$, which highlights a very important difference between the two models. Importantly, despite the similarity of $\langle \dot{\Sigma}^{nr} \rangle$ for both models, the symmetric part of the macrostate correlations distinguishes between the underlying different physical mechanisms that could generate similar vorticity currents. In the chemo-sensing bacteria section, it is one-way attraction, and here it is a mutual attraction-repulsion mechanism between the dogs and sheep that leads to non-reciprocal vorticity currents.  
\subsection{Thermodynamically consistent Active Ising Model}\label{sec:example_aim}
\subsubsection{Single species}
Here, we consider the thermodynamically consistent Active Ising Model (AIM). The AIM consists of two different types of self-propelled particles, positive and negative \cite{Solon2013}. The particles of the same type attract each other, otherwise repel. Hence, the microscopic interaction rules are $v_{++} = -1$, $v_{--} = -1$, and $v_{-+} = 1$. This leads to the macroscopic interaction coefficients $V_{--}^r = V_{++}^r = -1 + \frac{\beta}{2 \Omega}$ and $V_{-+}^r = 1 + \frac{\beta}{2 \Omega}$. The macroscopic reciprocal Boltzmann weights are $\mu_{+}^r = \ln{\rho_+} + \beta \left( V_{++}^r \rho_+ + V_{+-}^r \rho_- \right)$ and $\mu_{-}^r = \ln{\rho_-} + \beta \left( V_{--}^r \rho_- + V_{-+}^r \rho_+ \right)$. The diffusive mobility for the density fields $\rho_+$ and $\rho_-$ reads $D_+ = \tilde{d} e^{\mu_+^r}$ and $D_- = \tilde{d} e^{\mu_-^r}$, respectively. We report a detailed study in Ref. \cite{atm_2024_flocking_thermo,atm_cg_nr_2024} and briefly outline the thermodynamic consequences here.  

The macroscopic energy functional $E$ is trivially obtained using \cref{eq:macro_energy_functional} and $V^r_{ij}$. The only non-conservative force for the thermodynamically consistent AIM is the self-propulsion. Thus, $\langle \dot{\Sigma}^{nr} \rangle = \langle \dot{\Sigma}^{ch} \rangle = 0$. The macroscopic mean EPR due to the self-propulsion reads:
\begin{equation}\label{eq:epr_sp_maim}
\begin{split}
\langle \dot{\Sigma}^{sp} \rangle = \int_{\mathcal{V}} 2 \left( \langle D_+ + D_- \rangle \right) F^{sp} \sinh{ \left( \frac{F^{sp}}{2} \right) }.
\end{split}
\end{equation}
The macroscopic second law of thermodynamics reads $\langle \dot{\Sigma} \rangle = - d_t \langle E\rangle + \langle \dot{\Sigma}^{sp} \rangle + d_t S^{gb}$, or equivalently, $\langle \dot{\Sigma} \rangle = - d_t D^{KL}_{E}(\mathcal{P}[\{\rho\}]) + \langle \dot{\Sigma}^{sp} \rangle$. We report other important implications for the correctness of the macroscopic phase diagram in Ref.\cite{atm_cg_nr_2024}.
\subsubsection{Non-reciprocal two species}\label{sec:example_aim_nrmaim}
\begin{table*}[t!]
\begin{tabular}{ |m{2.2cm}|m{2.4cm}|m{2.2cm}|m{1.0cm}|m{1.6cm}|m{1.8cm}|m{1.8cm}|m{1.8cm}|m{1.8cm}| }
     \hline
     \centering
     \textbf{Level}
     & 
     \multicolumn{2}{|c|}{\textbf{Dynamics}} 
     &
     \centering
     {\textbf{LDB}}
     & \multicolumn{5}{|c|}{\textbf{Thermodynamics}}
     \\
     \hline
     \textbf{} 
     &
     \centering
     \textbf{State}
     \vspace{5pt}
     &
     \centering
     \textbf{Equations of Motion}
     \vspace{2pt}
     & 
     \centering
     & 
     \centering
     \textbf{Gibbs entropic}
     \vspace{2pt}
     & 
     \centering
     \textbf{Boltzmann entropic}
     \vspace{2pt}
     & 
     \centering
     \textbf{Reciprocal interaction} 
     \vspace{2pt}
     &\centering
     \textbf{Non-reciprocal interaction}
     & 
     \textbf{External driving}
     \vspace{2pt}
     \\
     \hline
     \centering
     \textbf{Microscopic description}
     & 
     \centering
     Multiparticle state probability $P_{ \{ N \} }$
     \vspace{2pt}
     & 
     \centering
     Master equation
     \cref{eq:microscopic_master_equation}
     \vspace{2pt}
     &
     \centering
     \cref{eq:microscopic_multi_particle_local_detailed_balance_final} 
     \vspace{2pt}
     & \centering
     $s_{state}$ in \cref{eq:microscopic_state_entropy}
     \vspace{2pt}
     & \centering
     $s_b$ in \cref{eq:microscoipic_boltzmann_entropy}
     \vspace{2pt}
     & \centering
     $\epsilon_i^r$ in \cref{eq:microscopic_reciprocal_interaction_energy_particle}
     \vspace{2pt}
     & \centering 
     $f_i^{nr}$ in \cref{eq:microscopic_reciprocal_interaction_energy_particle}
     \vspace{2pt}
     &
     $f_{\gamma\gamma'}^{ch}$ and $\vec{f}_{i}^{sp}$ in \cref{eq:microscopic_multi_particle_local_detailed_balance_final}
     \\
     \hline
     \centering
     \textbf{Macroscopic fluctuating description} 
     \vspace{5pt}
     & \centering
     Fluctuating macrostate density $\rho_i$
     \vspace{4pt}
     & \centering
     Generalized macroscopic fluctuation theory \cref{eq:macroscopic_eom_mft}
     & \centering
     \cref{eq:macroscopic_local_detailed_balance_reactive}
     \vspace{4pt}
     & \centering
     $S_{state}$ in \cref{eq:macroscopic_state_entropy}
     \vspace{4pt}
     & \centering
     $\mu_i^{id} = \ln{(\rho_i)}$ 
     \vspace{4pt}
     &
     $\mu_i^r$ in \cref{eq:macroscpic_boltzmann_weight}
     \vspace{4pt}
     &
     $F_i^{nr}$ in \cref{eq:macroscpic_boltzmann_weight}
     \vspace{4pt}
     &
     $F_{\gamma \gamma'}^{ch}$ and $\vec{F}_i^{sp}$ in \cref{eq:macroscopic_local_detailed_balance_reactive}
     \vspace{4pt}
     \\
     \hline
     \centering
     \textbf{Hydrodynamic order parameters} \cite{Hohenberg_1977,Cross_1993}
     \vspace{2pt}
     & \centering
     Phenomenological order parameters $\phi = g(\rho_i)$
     \vspace{2pt}
     & \centering
     Model A, B, etc. 
     \vspace{17pt}
     & \centering
     
     \vspace{4pt}
     & \centering
     Absent
     \vspace{17pt}
     & \centering
     $ \mu_{\phi_i}^{id} = \frac{\delta \mathcal{F}_{GL}^{id}}{\delta \phi_i}$
     \vspace{14pt}
     & \centering
     $ \mu_{\phi_i} = \frac{\delta \mathcal{F}_{GL}}{\delta \phi_i}$
     \vspace{14pt}
     & \centering
     Usually absent but considered in      \cite{You_2020,Saha_2020,saha_2022effervescent,Frohoff_2023,Frohoff_2021,Frohoff_2021_nonreciprocal,brauns_2023_nonreciprocal,Suchanek_2023_prl,Alston_2023,Suchanek_2023_pre_epr,Suchanek_2023_pre_pt}
     &
     Usually absent but considered in \cite{Bertini_2015,Bertini_2002,Bertini_2010}
     \vspace{3pt}
     \\
     \hline
\end{tabular}
\caption{Coarse-graining diagram: where $\mathcal{F}_{GL}$ and $\mathcal{F}_{GL}^{id}$ are the corresponding Ginzburg-Landau energy functional derived using phenomenological arguments and the ideal particle counterpart that incorporates the Boltzmann entropic contributions only. }
\label{table:summary}
\end{table*}
We aim to study a more sophisticated model that combines both non-reciprocity and self-propulsion simultaneously. The bird-flocking phenomena for a predator-prey bird species with two different preferred flying directions are denoted by $+$ and $-$. The predator (prey) is a Falcon (Starling). The microscopic interaction parameters for this model are $v_{s^+s^+} = v_{s^-s^-} = v_{f^+f^+} = v_{f^-f^-} = -1$ and 
$v_{s^+s^-} = v_{f^+f^-} = 1$ and 
$v_{s^+f^+} = v_{s^-f^-} = v_{s^+f^-} = v_{s^-f^+} = 1$ and 
$v_{f^+s^+} = v_{f^-s^-} = v_{f^+s^-} = v_{f^-s^+} = -1$. It leads to the macroscopic interaction coefficients 
$V_{s^+s^+}^{r} = V_{s^-s^-}^{r} = V_{f^+f^+}^{r} = V_{f^-f^-}^{r} = - \beta + \frac{\beta^2}{2\Omega}$, 
$V_{s^+s^-}^{r} = V_{f^+s^-}^{r} = \beta + \frac{\beta^2}{2 \Omega}$,
$V_{s^+f^+}^{r} = V_{s^+f^-}^{r} = V_{s^-f^+}^{r} = V_{s^-f^-}^{r} = \frac{\beta^2}{2 \Omega}$, 
$V_{s^+f^+}^{nr} = V_{s^+f^-}^{nr} = V_{s^-f^+}^{nr} = V_{s^-f^-}^{nr} = \beta$. The macroscopic interaction energy functional $E$ is trivially obtained using \cref{eq:macro_energy_functional} and $V^r_{ij}$. The macroscopic self-propulsion EPR reads:
\begin{equation}\label{eq:epr_sp_nrmaim}
\begin{split}
    \langle \dot{\Sigma}^{sp} \rangle 
    & = \int_{\mathcal{V}} 2 \bigg[ 
     F_s^{sp} \sinh{ \left( \frac{F_s^{sp}}{2} \right) } \left( D_{s^+} + D_{s^-} \right) 
    \\
    & \hspace{1.5cm} +  F_f^{sp} \sinh{ \left( \frac{F_f^{sp}}{2} \right) } \left( D_{f^+} + D_{f^-} \right)  
    \bigg],
\end{split}    
\end{equation}
and the macroscopic non-reciprocal EPR reads: 
\begin{equation}\label{eq:epr_nr_nrmaim}
\begin{split}
    \langle \dot{\Sigma}^{nr} \rangle
    & = V_{s^+f^+}^{nr} \int_{\mathcal{V}} \langle \rho_{s} \partial_t \rho_{f} - \rho_f \partial_t \rho_s \rangle.
\end{split}    
\end{equation}
Here, $\rho_s = \rho_{s^-} + \rho_{s^+}$ and $\rho_f = \rho_{f^-} + \rho_{f^+}$ quantify the total density of the Starling and Falcon, irrespective of their flying direction. The set of \cref{eq:epr_sp_nrmaim,eq:epr_nr_nrmaim} gives the orthogonal decomposition of the macroscopic second law of thermodynamics, $\langle \dot{\Sigma} \rangle = - d_t \langle E\rangle + \langle \dot{\Sigma}^{nr} \rangle + \langle \dot{\Sigma}^{sp} \rangle + d_t S^{gb}$. Or equivalently, $\langle \dot{\Sigma} \rangle = - d_t D^{KL}_{E}(\mathcal{P}[\{\rho\}]) + \langle \dot{\Sigma}^{nr} \rangle + \langle \dot{\Sigma}^{sp} \rangle$, in the absence of the external driving of $E$, the driving work is $\langle W \rangle = 0$.
\section{Conclusion and Outlook}\label{sec:conclusion}
We have formulated a generic novel framework of stochastic thermodynamics for non-reciprocal systems that relies on systematic thermodynamically consistent coarse-graining approaches. The novel formulation of the Local Detailed Balance condition for non-reciprocal systems provides a thermodynamically consistent formulation across the scales. Hence, our framework broadens the applicability of stochastic thermodynamics to different observation scales of the system. We further decompose the mean EPR into four orthogonal contributions, namely the conservative, the non-reciprocal, the external chemical/self-propulsion driving, and the rate of change in the free energy (driving work). They correspond to the entropy production cost associated with the relaxation (towards the reference Boltzmann distribution), sustaining the vorticity currents (due to non-reciprocity), sustaining the dissipative transition currents, and the quasistatic mean stochastic work, respectively. Importantly, the systematic coarse-graining using the Doi-Peliti field theory ensures the equivalence of the systems' dynamics and thermodynamics across the observation scale.

We compute the dynamic equations of motion for the macrostate using the Langevin approximation, which exhibits the multiplicative demographic noise. This generalizes the Macroscopic Fluctuation Theory to non-reciprocal and externally driven systems. We demonstrated that the microscopic non-reciprocal interactions lead to the manifestation of Onsager's non-reciprocal relations on the macroscale. We formulate the fluctuation-response relation and its generalizations, namely the higher-order current cumulant response relations. In the spirit of stochastic thermodynamics, we formulate the generic fluctuation relations for non-reciprocal systems. We obtain the tightest Thermodynamic Kinetic Uncertainty Relation for the non-reciprocal systems. It relies on using the observable transition currents and the observable vorticity that lies in the transition and state space, respectively. We briefly highlight how to incorporate the information thermodynamics for non-reciprocal systems. Our framework opens up a plethora of model-specific extensions of non-reciprocal systems across a different observable scale. Different aspects of this work are summarized in \cref{table:summary}.

\textit{Multi-body microscopic interactions}. \textemdash \:
One can explore the consequences of the nonlinear dependence of $\epsilon_i$ on $N_i$, which exhibits a richer phenomenology \cite{saha_2022effervescent}. The structure of our framework (both dynamic and thermodynamic) is robust to such a modification to the non-linear dependence of the Boltzmann weights on the particle number. Hence, a straightforward generalization is obtained \cite{atm_cg_nr_2024}. Importantly, the higher-order multi-body interaction plays a key role in ensuring that the interaction energy functional is bounded from below.

\textit{Linear cyclic CRN}. \textemdash \: A recent development in the interacting (non-ideal) CRN has led to the applicability of the methodology developed for the non-interacting CRN \cite{schnakenberg_1976,Jiang_2004_book,Andrieux_2007,Kalpazidou_2007_cycle_book,Faggionato_2011,Altaner_2012,Ge_2016} to the interacting CRN under certain conditions \cite{Miangolarra_2022}. We intentionally avoided the interplay of the interacting systems and the topological properties of the CRN encapsulating the underlying reactive transitions. Nevertheless, the interacting CRNs are fundamentally similar to the non-interacting CRNs, provided that certain constraints on the topological properties of the CRN are met. An important implication is the vanishing of diffusive currents \cite{Miangolarra_2022}. However, the interplay of diffusive and reactive currents has been the key mechanism of pattern formation \cite{Brauns_2020}. Therefore, this opens up an interesting avenue to explore, under a more generic condition of a spatially extended interacting CRN that does not satisfy the constraints in Ref.\cite{Miangolarra_2022}. The study of the interplay between diffusive fluxes and the reactive fluxes modeled by interacting CRNs and their thermodynamic and dynamic implications for pattern formation remains an open problem. We expect that the spatially extended non-reciprocal systems might exhibit a richer phenomenology.

\textit{Nonlinear CRN}. \textemdash \:
Our framework can be straightforwardly incorporated into the non-linear deterministic CRN \cite{kobayashi_2023_information_graphs_hypergraphs,Dal_chengio_2023} by incorporating the topological structure of the CRN. Moreover, the stochastic non-linear CRN exhibits more sophisticated effects arising due to the interplay of non-linearity and stochasticity \cite{Feinberg_1972,Horn_1972,Feinberg_1995,Gunawardena_2003,Anderson_2010,Anderson_2016}. It is an interesting avenue to explore the non-linear stochastic CRN by incorporating non-reciprocal interactions. Since our framework systematically incorporates the microscopic occupancy and transition fluctuations on the macroscale, it is an inherently better starting point than the deterministic CRN to study such effects.

\textit{The role of demographic noise}. \textemdash \:
Demographic noise plays a key role in understanding the physics of ecological models
\cite{Baras_1996,Domokos_2004,Traulsen_2005,Galla_2009,Berr_2009,Melbinger_2010,Rogers_2012,Pigolotti_2012,Kessler_2015,Biroli_2018,Weissmann_2018,West_2020,Altieri_2021}. For instance, the steady-state selection, the stability of the attractors for the deterministic and stochastic dynamics \cite{Domokos_2004,Traulsen_2005,Galla_2009,Berr_2009,Melbinger_2010,Rogers_2012,Pigolotti_2012,Kessler_2015,Biroli_2018,Weissmann_2018,West_2020,Altieri_2021}, and the mismatch between the deterministic and stochastic dynamics \cite{Kurtz_1971,Kurtz_1972,Vellela_2007,Vellela_2009}. It is an interesting avenue to explore the model-specific implications of demographic noise and potentially richer novel phenomena using the exact \cref{eq:macroscopic_eom_mft} for their thermodynamically consistent dynamics. Importantly, our framework enables us to compute the thermodynamic dissipation cost associated with the different phases as well.

{ \textit{The finite-time optimal control of non-reciprocal systems} \textemdash \: We have omitted the time-dependent change of the control parameters of the non-reciprocal models, except for the slow driving of the control parameters of the conservative energy functional $E$ that quantifies the quasistatic work. However, a novel framework for the finite-time optimal control of any control parameters of far-from-equilibrium systems has been recently developed in Ref.\cite{atm_2025_gftoc}. It has revealed the importance of the exact non-quadratic dissipation functions and the discontinuous endpoint jumps in optimal driving protocols, which have been shown to be of paramount importance for far-from-equilibrium systems \cite{atm_2024_var_epr,atm_2025_var_epr_derivation}. It is an interesting avenue to explore model-specific phenomena for the optimal control of non-reciprocal systems in a finite driving time. }
%
%
%
%
%
\appendix
\section{Microscopic excess and house-keeping decomposition of the EPR}\label{sec:microscopic_epr_excess_housekeeping_decomosition}
The total mean microscopic bulk EPR can be further divided using another physical aspect of the system. For instance, a state variable is often easier to track compared to the underlying thermodynamic energy functional. One particular case arises when the temporal dynamics of the microstate lead to a steady-state probability distribution $P_{\{N\}}^{ss}$ for the microstate $\{N\}$. The mean bulk EPR ($\langle \dot{\sigma}^{\mathcal{R}} \rangle + \langle \dot{\sigma}^{\mathcal{D}} \rangle$) can be decomposed into two components, called excess and housekeeping EPR, using the steady state of the dynamics \cite{Oono_1998,hatano_2001,Speck_2005,seifert_2012,Ge_2009,Ge_2010,Esposito_2010_3dft}. This decomposition is commonly referred to as the Hatano-Sasa (HS) decomposition of the total mean EPR. The mean excess HS EPR is denoted by $\langle \dot{\sigma}^{ex}_{hs} \rangle$, with the exact expression given by:        
\begin{equation}\label{eq:microscopic_HS_excess_epr}
\begin{split}
    \langle \dot{\sigma}^{ex}_{hs} \rangle 
    & = \sum_{ \{ \Delta_{\gamma \gamma' }^\# \} }^{\{N\}} j_{\gamma \gamma'}^\# ( \{ N \} )  \ln{\left(
    \frac
    {
    P_{ \{ N + \Delta_{\gamma \gamma'}^\# \} }^{ss} (t)
    }
    { 
    P^{ss}_{ \{ N \} } (t)
    }
    \right)} 
    \\
    & \hspace{1cm} + \sum_{ \{ \Delta_i^{\Vec{\mathcal{D}} \#} \} }^{\{N\}}
    j_i^{{ \vec{\mathcal{D}} \# }} ( \{ N \} )
    \ln{
    \left(
    \frac
    {
    P^{ss}_{ \{ N + \Delta_i^{\Vec{\mathcal{D}} \#}  \} } (t)
    }
    { 
    P^{ss}_{ \{ N \} } (t)
    }
    \right)}.
\end{split}    
\end{equation}
Using the master equation \cref{eq:microscopic_master_equation}, the mean microscopic excess HS EPR satisfies $ \langle \dot{\sigma}^{ex}_{hs} \rangle  = \sum_{\{N\}} d_t P_{\{ N \}}(t) \ln{ \left( P^{ss}_{ \{ N \} } (t) \right)}$. The time dependence of the steady state indicates the validity of the definition in \cref{eq:microscopic_HS_excess_epr} for non-autonomous dynamics implemented through thermodynamic work.  

The exact expression for the microscopic housekeeping EPR reads:
\begin{equation}\label{eq:microscopic_HS_housekeeping_EPR}
\begin{split}
    \langle \dot{\sigma}^{hk}_{hs} \rangle
    = \langle \dot{\sigma} \rangle - \langle \dot{\sigma}^{ex}_{hs} \rangle - d_t s^{gb}.
\end{split}    
\end{equation}
Positivity of excess and housekeeping EPR has been demonstrated \cite{hatano_2001,Speck_2005,Ge_2009,Ge_2010,Esposito_2010_3dft}. By combining the rate of change of the Gibbs entropy term $d_t s^{gb}$ with the mean excess HS EPR $ \langle \dot{\sigma}^{ex}_{hs} \rangle $, one defines the non-adiabatic mean EPR $ \langle \dot{\sigma}^{na} \rangle $. Its closed-form expression, $\langle \dot{\sigma}^{na} \rangle = \langle \dot{\sigma}^{ex}_{hs} \rangle + d_t s^{gb}$, reads:
\begin{equation}\label{eq:microscopic_non_adiabatic_EPR}
\begin{split}
    \langle \dot{\sigma}^{na} \rangle = - \sum_{\{N\}} d_t P_{\{ N \}} (t) \ln{ \left( \frac {P_{ \{ N \} } (t)}{ P^{ss}_{ \{ N \} } (t) } \right)}.
\end{split}    
\end{equation}
The non-adiabatic mean EPR $ \langle \dot{\sigma}^{na} \rangle $ quantifies microscopic EPR due to a mismatch between the instantaneous probability distribution $\{ P_{ \{ N \} } (t) \}$ and the steady-state distribution $\{ P^{ss}_{ \{ N \} } (t) \}$. In steady state, the instantaneous probability distribution satisfies $\{ P_{ \{ N \} } (t) \} = \{ P^{ss}_{ \{ N \} } (t) \}$, so that the non-adiabatic mean EPR $ \langle \dot{\sigma}^{na} \rangle $ vanishes. Consequently, in steady state, the total mean microscopic EPR $\langle \dot{\sigma} \rangle$ is completely determined by the mean housekeeping EPR, $\langle \dot{\sigma}^{hk}_{hs} \rangle = \langle \dot{\sigma} \rangle$.  

Reorganizing the definitions of the mean microscopic excess HS EPR and the microscopic non-adiabatic EPR shows that they are related to the rate of change of the stochastic Massieu potential. In particular, $ \langle \dot{\sigma}^{ex}_{hs} \rangle  = d_t \langle \ln{(P^{ss}_{ \{ N \} } (t))} \rangle $ and  $ \langle \dot{\sigma}^{na} \rangle  = - d_t \langle \ln{  (P_{ \{ N \} } (t)) / (P^{ss}_{ \{ N \} } (t)) } \rangle $, where the time derivative is defined as $d_t\langle * \rangle = \sum_{\{N\}} (*) d_t P_{\{N\}}$. The same convention was previously used to define the second law of thermodynamics. Thus, $-\ln{(P^{ss}_{ \{ N \} } (t))}$ and $\ln{  (P_{ \{ N \} } (t)) / (P^{ss}_{ \{ N \} } (t)) }$ are stochastic Massieu potentials whose temporal variations are $\langle \dot{\sigma}^{ex}_{hs} \rangle $  and $\langle \dot{\sigma}^{na} \rangle$, respectively.

The Kullback-Leibler divergence is defined as $D^{KL}_{ss} (P(t) ||P^{ss}(t)) = \sum_{ \{N\} } P_{\{N\}}(t) \ln{ ( P_{\{N\}}(t) / P^{ss}_{\{N\}}(t) )} $ \cite{Bergman_1955,Lebowitz_1957,Schloegl_1971,Amari_2000_book,Qian_2001,Ge_2009,Ge_2010,Shiraishi_2019}. We introduce a shorthand notation $D^{KL}_{ss}(t) = D^{KL}(P(t) || P^{ss}(t))$, where the superscript indicates the reference distribution. Thus, $\langle \dot{\sigma}^{na} \rangle = - d_t D^{KL}_{ss} (P(t))$. Physically, this relates the non-adiabatic EPR to the relaxation of the Kullback-Leibler divergence between the instantaneous and steady-state distributions. Plugging this into \cref{eq:microscopic_HS_housekeeping_EPR} gives $ \langle \dot{\sigma}^{hk}_{hs} \rangle =  \langle \dot{\sigma} \rangle|_{P^{ss}}$, showing that the housekeeping EPR equals the total microscopic EPR in steady state, while the microscopic adiabatic EPR vanishes. For a system without external non-conservative or non-reciprocal forces, $ \langle \dot{\sigma}^{hk}_{hs} \rangle = 0$ trivially holds. The microscopic EPR due to relaxation is fully quantified by the rate of change of the Kullback-Leibler divergence, $\langle \dot{\sigma} \rangle =  - d_t D^{KL}_{ss} (t) $.  

A stronger bound on the microscopic entropy production is obtained by exploiting the positivity of $\langle \dot{\sigma}^{na} \rangle$ and its Kullback-Leibler representation \cite{Shiraishi_2019}. In particular, $\langle \sigma^{na} \rangle = D^{KL}_{ss} (0) - D^{KL}_{ss} (\tau)$, where $\tau$ is the relaxation time. Using the triangle inequality \cite{Shiraishi_2019,Amari_2000_book}, this can be further simplified as $\langle \sigma^{na} \rangle \geq D^{KL} (P(0) || P(\tau)) $, which is a tighter bound than the second law of thermodynamics due to the non-negativity of the KL divergence.
%
%
%
%
%
%
\bibliography{reference}
%
%
%
%
%
%
%
\end{document}